\documentclass[letterpaper]{JHEP3} 
\usepackage{epsfig,array,cite,latexsym,amsmath,oldgerm,hhline}
\preprint{INT-PUB 04-04, BUHEP 05-05}
\newcommand\beq{\begin{eqnarray}}
\newcommand\eeq{\end{eqnarray}}
\newcommand\bal{ \begin{align}}
\newcommand\eal{\end{align} }

\newcommand\eqn[1]{\label{eq:#1}} 
\newcommand\Eq[1]{eq.~\eqref{eq:#1}} 
\newcommand\eq[1]{eq.~\eqref{eq:#1}} 
\newcommand\half{{\textstyle{\frac{1}{2}}}} 
\newcommand\fourth{{\textstyle{\frac{1}{4}}}}

\newcommand\ket[1]{\vert #1 \rangle}

\newcommand\bfz{\mathbf{z}}
\newcommand\bfZ{\mathbf{Z}}
\newcommand\bft{\mathbf{T}}

\newcommand\bfXi{\boldsymbol{\Xi}}

\newcommand\bfmu{\boldsymbol{\mu}}

\newcommand{\CC}{{\cal C}}
\newcommand{\CD}{{\cal D}}
\newcommand{\CE}{{\cal E}}

\newcommand{\CN}{{\cal N}}
\newcommand{\CP}{{\cal P}}
\newcommand{\CG}{{\cal G}}
\newcommand{\CQ}{{\cal Q}}

\newcommand{\bfn}{{\bf n}}
\newcommand{\bfr}{{\bf r}}
\newcommand{\bfR}{{\bf R}}
\newcommand{\bfw}{{\bf w}}
\newcommand{\bfm}{{\bf m}}
\newcommand{\bfe}{{\bf  e}}
\newcommand{\bfnab}{\mathbf{\nabla}}

\DeclareMathOperator{\Tr}{Tr\,}

\newcommand{\sla}[1]%
        {\kern .25em\raise.18ex\hbox{$/$}\kern-.60em #1}

\newcommand{\vslash}{\hbox{\sla $v$}}%
\newcommand{\mybar}[1]%
        {\kern 0.6pt\overline{\kern -0.6pt#1\kern -0.6pt}\kern 0.6pt}
\newcommand{\dig}{\kern-1.5pt \raisebox{.9ex}{$\cdot$}  \kern1.5pt
  \raisebox{0ex}{${\mathbf\cdot}$}\kern1.5pt \raisebox{-.9ex}{$\cdot$}} 
\newcommand{\digb}{\kern-1.5pt \raisebox{.75ex}{$\cdot$}  \kern1.5pt
  \raisebox{0ex}{${\mathbf\cdot}$}\kern1.5pt \raisebox{-.75ex}{$\cdot$}} 
\newcommand{\digc}{\kern-1.5pt \raisebox{1.05ex}{$\cdot$}  \kern1.5pt
  \raisebox{0ex}{${\mathbf\cdot}$}\kern1.5pt \raisebox{-1.05ex}{$\cdot$}} 
\newcommand{\drawsquare}[2]{\hbox{%
\rule{#2pt}{#1pt}\hskip-#2pt
\rule{#1pt}{#2pt}\hskip-#1pt
\rule[#1pt]{#1pt}{#2pt}}\rule[#1pt]{#2pt}{#2pt}\hskip-#2pt
\rule{#2pt}{#1pt}}
\newcommand{\Yfund}{\raisebox{-.5pt}{\drawsquare{6.5}{0.4}}}
\newcommand\fverb{\setbox\pippobox=\hbox\bgroup\verb}
\newcommand\fverbdo{\egroup\medskip\noindent%
                        \fbox{\unhbox\pippobox}\ }
\newcommand\fverbit{\egroup\item[\fbox{\unhbox\pippobox}]}
\newbox\pippobox

\title{A Euclidean Lattice Construction of  Supersymmetric Yang-Mills
  Theories with Sixteen Supercharges}

\author{David B. Kaplan \\ Institute for Nuclear Theory, University of
  Washington, 
  Seattle, WA 98195-1550 \\Email: 
\email{dbkaplan@phys.washington.edu}}

\author{Mithat {\"U}nsal\\ Dept. of Physics, Boston University, 590 Commonwealth Ave, Boston, MA 
02215\\Email:
\email{unsal@buphy.bu.edu}
  }

\abstract{We formulate supersymmetric Euclidean spacetime $A_d^*$ lattices whose classical continuum 
limits are $U(N)$ supersymmetric Yang-Mills theories with sixteen
supercharges in $d=1,2,3$ and $4$ dimensions. This family  
  includes the especially interesting  $\CN=4$ supersymmetry in four
  dimensions, as well as a Euclidean path integral formulation of
  Matrix Theory on a one dimensional lattice.  
}

\begin{document} 

\section{Introduction}
\label{sec:1}

Sixteen is the maximal number of supercharges  that can be 
accommodated in a theory with particle multiplets of spin $s\le 1$. 
Such
 theories are extremely 
constrained and much  is known or surmised about
them. The  $\CQ=16$  $SU(N)$ gauge theory  in $d=4$ 
dimensions, known as $\CN=4$ supersymmetry, is believed to be  finite
and superconformal, possessing monopole and dyon   
excitations \cite{Witten:1978mh,Osborn:1979tq}, as well as a discrete
$SL(2,Z)$ symmetry generalizing   electric-magnetic duality   
\cite{Montonen:1977sn,Goddard:1977qe} which
exchanges  weak and 
strong coupling. In the large $N$
limit, the theory is conjectured to be equivalent to supergravity in $AdS_5$
space \cite{Maldacena:1998re,Gubser:1998bc,Witten:1998qj}.  The $d=3$
theory with $\CQ=16$ supercharges is expected to have a nontrivial
infrared fixed point \cite{Seiberg:1998ax}, while 
in $d=1$ and $d=0$
dimensions, the respective  quantum mechanical and matrix theories are
conjectured to be related in the large-$N$ limit to
$M$-theory\cite{Banks:1997vh,Ishibashi:1997xs}.  
Despite the intense interest in these theories, and the obvious need
for a nonperturbative definition, none existed until their construction
on a spatial lattice in ref.~\cite{Kaplan:2002wv}. In this paper we
continue the program of refs.~\cite{Cohen:2003aa,Cohen:2003qw} and
show how the $\CQ=16$ supercharge theories may be constructed on
Euclidean spacetime lattices;  as we shall show,  the $A_d^*$ structure of the
lattices we construct have a particularly elegant structure\footnote{Some of the
  results appearing in this paper  were presented quite some time ago in conference proceedings
\cite{Kaplan:2003uh} and public lectures  \cite{bern},
\cite{seminars}, however the details of the construction have not been
presented before.}.

  The challenge  confronting attempts
to put these and other supersymmetric Yang-Mills (SYM) theories on the
lattice has been how to maintain enough supersymmetry in the absence of
continuous translations in order to forbid the numerous relevant
operators
which violate the symmetries of the desired continuum
theory.  Obvious and egregious examples of such unwanted operators
are mass terms for the scalar partners of the gauge bosons; only some
sort of residual supersymmetry can forbid such operators. However,
early attempts to construct supersymmetric lattices failed to yield
Lorentz invariant continuum field theories \cite{Banks:1982ut}.
Recently two approaches have been developed for constructing lattices respecting
exact supersymmetries, which yield Lorentz invariant supersymmetric
theories in the continuum with either no or little fine tuning.  The
approach pioneered by Catterall and collaborators and followed up by Sugino starts with
nilpotent charges which form a subset of the supercharges
of the target
theory\cite{Catterall:2001fr,Catterall:2001wx,Catterall:2003uf,Catterall:2003wd,Sugino:2003yb,Giedt:2004qs,Sugino:2004uv,Catterall:2004np}.
The approach followed here creates the lattice theories by performing
an orbifold projection on supersymmetric matrix models obtained by
dimensionally reducing SYM theories in various
dimensions\cite{Kaplan:2002wv,Cohen:2003aa,Cohen:2003qw}; for a review
see \cite{Kaplan:2003uh}. The projection creates a lattice action while
preserving some of the supersymmetries of the matrix model. The
theories have a degenerate manifold of ground states in the infinite
volume limit  (the moduli
space), where the distance from the origin in moduli space is
identified as the inverse lattice spacing of the theory.  The
continuum limit is thus defined as a trajectory out to infinity in the
moduli space, and the result is a SYM field theory.  Again, the exact
supersymmetries of the lattice guarantee that the continuum limit can
be achieved with little or no fine tuning.  This method for
constructing supersymmetric lattice theories has its origins in 
orbifold projection methods of string theory
\cite{Douglas:1996sw} and 
deconstruction
\cite{ArkaniHamed:2001ca,ArkaniHamed:2001ie,Rothstein:2001tu}. In the
next section we address generalities of the construction, and then we
discuss each case in turn, from dimension $d=4$ down to $d=1$.

\section{The  mother theory and the orbifold projection}
\label{sec:2}

\subsection{The mother theory}
\label{sec:2a}

Our starting point is the  $\CQ=16$  mother theory, which is the   dimensional reduction 
of   $\hbox{\CN=1}$ SYM with  gauge group $\CG$ from ten Euclidean dimensions
down to zero dimensions.  The mother theory is a theory of matrices
--- ten bosonic and sixteen Grassmann --- and inherits the sixteen
supersymmetries as well as the $\CG\times SO(10)$ symmetry of its
ten-dimensional precursor. Each of the bosons and fermions transform
as an 
adjoint under $\CG$, while under the  $SO(10)$ symmetry they transform
as the  $\boldsymbol{10}$ and  $\boldsymbol{16}$
representations respectively.   We choose a Hermitean chiral basis for
the ten, 32-dimensional gamma matrices $\Gamma_\alpha$ of $SO(10)$, and the chirality
matrix $\Gamma_{11}$ satisfying
\beq
\{\Gamma_\alpha,\Gamma_\beta\} = 2 \delta_{\alpha\beta}\ ,\qquad
\Gamma_\alpha=\Gamma_\alpha^\dagger\ ,\qquad
\Gamma_{11} =-i\prod_{\alpha=1}^{10} \Gamma_\alpha \ .
\eeq
The generators of $SO(10)$ transformations are given by
\beq
M_{\alpha\beta} = \frac{1}{4i}\left[\Gamma_\alpha,\Gamma_\beta\right]\ ,
\eqn{gen}\eeq
and the charge conjugation matrix $C$ 
satisfies
\beq
C^{-1} \Gamma_\alpha C = -\Gamma_\alpha^T\ ,\qquad
C^\dagger=C^{-1}=C\ .
\eeq

For greater ease in comparing lattice and continuum theories, we will
choose the convention
\beq
\Gamma_m^T = \begin{cases} -\Gamma_m & m=1,\ldots,5\cr +\Gamma_m &
  m=6,\ldots,10 \end{cases}
\eqn{gamsym}\eeq
allowing us to define the charge conjugation matrix as
\beq
C = \prod_{m=1}^5 \, \Gamma_m\ .
\eqn{ccdef}
\eeq
For an explicit $\Gamma$ matrix basis worked out in detail, see Appendix~\ref{sec:ap}.

We define a   left-handed Grassmann spinor $\omega = + \Gamma_{11}\omega$ which
is written as a 32-component Dirac spinor, but which only has 16
independent components and transforms as the irreducible ${\bf 16}$
representation of $SO(10)$.  We  also introduce a real bosonic variable
$v_\alpha$ transforming as the ${\bf 10}$ representation of $SO(10)$.
 Then the action of the mother  
theory  may be written  as
\beq
S = \frac{1}{g^2}\left(\frac{1}{4} \Tr v_{\alpha\beta} v_{\alpha\beta} + \frac{i}{2} \Tr
 \omega^{T}\,C \,
\Gamma_\alpha\, [ v_\alpha, \omega]\  \right)
\label{eq:act1}
\eeq
where 
$v_{\alpha\beta} =
i[v_\alpha,v_\beta]$. In
this expression $\omega=\omega^a T_a$ and $v_\alpha = v_\alpha^a T_a$ are
matrices, where $T_a$ are the hermitean generators of $\CG$ in the
defining representation of $\CG$, normalized so that
\beq
\Tr T_a T_b = \delta_{ab}\ .
\eeq

The global  $G_R=SO(10)$ symmetry of the above action  
 is just the  ten dimensional Lorentz symmetry 
transmuted to a global symmetry of the zero dimensional mother theory. Explicitly, 
the fermionic and bosonic fields transform under $SO(10)$ as
\beq
\omega \to \Omega \omega ,\qquad \vslash \to \Omega\, \vslash \,\Omega^{-1}\ ,\qquad
\vslash\equiv  \Gamma_\alpha v_\alpha\ ,\qquad\Omega \in SO(10)\ .
\eeq

One can also show that the action  \Eq{act1}  is invariant under the 
 supersymmetry   transformation, 
\beq
\delta v_\alpha= - \kappa^{T}C\Gamma_\alpha \omega \ , \qquad 
\delta \omega = i v_{\alpha\beta} M_{\alpha\beta} \kappa  \ , 
\eqn{mothersusy}
\eeq
where $\kappa$ is a
chiral Grassmann spinor satisfying $\Gamma_{11}\kappa=+\kappa$  with
16 independent components  parameterizing the $\CQ=16$ supersymmetry
of the mother theory.  Note that
 the supersymmetry transformations do not commute with $SO(10)$, and
 that $\kappa$ (and hence the supercharges) transform as a $\boldsymbol{16}$.

\subsection{Specifying the orbifold projection}
\label{sec:2b}

The procedure of creating lattices by using the orbifold projection technique 
 has been 
explained in detail  in  \cite{Cohen:2003aa, Cohen:2003qw}, and we
summarize it briefly here.  To construct the a $d$-dimensional
lattice with $N$ sites in each direction for a target theory possessing a $U(k)$ gauge symmetry
and in the continuum, we choose the group $\CG$ of the mother
theory to be  $\CG= U(kN^d)$. 
Each of the variables of
the mother theory are therefore $kN^d$-dimension matrices. We then
project out all variables in the mother theory 
which are charged under a certain $Z_N^d$ subgroup of the
$U(kN^d)\times SO(10)$ symmetry of the mother theory, and the action
written in terms of the surviving variables has a natural lattice
interpretation.  The structure and symmetries of the lattice depend on
the embedding of the discrete  $Z_N^d$ subgroup. 

The embedding of
the discrete  $Z_N^d$ subgroup within  $U(kN^d)$ is given by the
natural decomposition $U(k)^{Nd}\ltimes Z_N^d$. It is
convenient to consider each  one of the bosonic or fermionic matrices of the mother theory, which we will
refer to generically as $\Phi$,  to be  a matrix consisting  of $N^{2d}$ independent $k\times k$
blocks. These blocks can be written as $\Phi_{\bfm,\bfn}$, where
$\bfm$ and $\bfn$ are  two independent $d$-dimensional vectors with
integer components, each of which run from $1$ to 
$N$.   This action of the mother theory \eq{act1} can be considered as
an extremely nonlocal lattice action in 
$d$-dimensions, where each site is labeled by a $d$-dimensional
integer vector $\bfm$, and each nonzero block $\Phi_{\bfm,\bfn}$ is a a $k\times k$ matrix valued
lattice variable living on the link between sites $\bfm$ and
$\bfn$. In the case where $\bfm=\bfn$,  the diagonal $\Phi_{\bfm,\bfm}$
block is a variable sitting at the site $\bfm$.  

The orbifold projection sets most of these lattice blocks to zero.  In
particular, each $\Phi$ variable is assigned a $Z_N^d$ charge $\bfr$
according to its weight vector in $SO(10)$, where $\bfr$ is a
$d$-component vector with integer coefficients running from $1$ to
$N$.  The exact relation between $\bfr$ and the $SO(10)$ weights will
be discussed further below.  The orbifold projection then sets to zero
all blocks $\Phi_{\bfm,\bfn}$ not satisfying  $\bfn = \bfm + \bfr$.
If all components of the $\bfr$ vectors equal zero or $\pm1$, then the
action \eq{act1} written in terms of the projected variables will look
like a very local lattice action.

 The projection breaks the original $\CG=U(kN^d)$ symmetry of the mother
theory down to an independent $U(k)$ symmetry associated with
each lattice site, which constitutes the lattice version of a $U(k)$
gauge symmetry. Each site variable  transforms as an adjoint under
the local $U(k)$ symmetry, while each link variable  transforms as
a bifundamental $(\Yfund,\mybar\Yfund)$ or its conjugate under the
$U(k)\times U(k)$ symmetries associated with the endpoints of the
links.

The orbifold projection breaks some or all
of the supersymmetries of the mother theory.  That is because the
supercharges are transform as a spinor under $SO(10)$, but are
$\CG$-invariant.  Thus only supercharges with $\bfr = 0$ 
survive the orbifold projection. 

It remains to specify how the $\bfr$ charges are related to the
$SO(10)$ symmetry. Our choice of how to embed the  $Z_N^d$ symmetry
into $ SO(10)$ is 
guided by three principles:
\begin{enumerate}
\item 
Since we will eventually take $N\to\infty$, the embedding must take
the form $Z_N^d\in U(1)^d\in SO(10)$;

\item
Supersymmetry generators $Q$ are $\CG$-invariant and transform as a
$\boldsymbol{16}$ of $SO(10)$. Therefore,
in order to break as few supersymmetries as possible, for any lattice
dimension $d$ we
will want to maximize the number of elements in the spinor of $SO(10)$
which are singlets under the $Z_N^d$ symmetry ({\it e.g.} which have
$\bfr=0$);

\item
Lattice variables associated with the surviving $(\bfm,\bfn) = (\bfm,\bfm+\bfr)$
block of a mother theory variable $\Phi$ resides on the link between
sites $\bfm$ and $\bfm+\bfr$. Therefore,  in order to keep the lattice action as local as possible, we want the
components of $\bfr$  to all be $0$ or $\pm 1$, to avoid having link
variables connecting distant sites.

\end{enumerate}
The first point implies that the $Z_N^d$ orbifold group should be
embedded within the $U(1)^5$ Cartan subgroup of $SO(10)$. It
immediately follows that the maximum lattice dimension we could
construct in this manner is $d=5$.  (We will see that requiring that the lattice be
supersymmetric will actually constrain the maximum dimension to be
$d=4$).  We can take the five generators $q_m$ of this $U(1)^5$ symmetry to
be 
  \beq
q_m =  M_{m,m+5}\ ,\quad m=1,\ldots,5
\eqn{qdef}
\eeq
corresponding to rotations in the $x_m-x_{m+5}$ plane in a ten dimensional space. 

The  five complex bosonic 
fields 
\beq
z_m =i  (v_{m} - i v_{m+5})/\sqrt{2}\ ,\qquad
\mybar z_m = -i  (v_{m} + i v_{m+5})/\sqrt{2}
\eqn{zdef}\eeq
 are  eigenstates of the $U(1)^5$ symmetry generated by the $q_m$,
where $z_m$ has charge $q_n = \delta_{mn}$, and $\mybar
z_m$ has  charge  $q_n =- \delta_{mn}$.

The charges of the fermions are determined by defining 
the anticommuting raising and 
lowering operators
\begin{equation}  
\hat A^m= \half \left( \Gamma_{m} -i \Gamma_{m+5}\right), \qquad
\hat A_m^{\dagger}= \half \left( \Gamma_{m} +i \Gamma_{m+5}\right),
\qquad m=1,\ldots,5,
\eqn{adef}\end{equation}
which satisfy the
relations 
\begin{equation}
\{\hat A^m, \hat A^n\}=\{\hat A_m^{\dagger}, \hat A_n^{\dagger}\}=0, \qquad 
\{\hat A^m, \hat A_n^{\dagger}\}= \delta^m_n,
\eqn{acom}
\end{equation}
 familiar as the algebra of fermionic ladder operators.  The
spinor 
representation is then constructed in a Fock space of five different
species of one-component fermion, each of which has occupation number
zero or one.
The operators forming Cartan subalgebra of $SO(10)$ can be expressed in terms 
of these fermionic raising and lowering operators as 
\beq
 q_m = M_{m, m+5}=\left( \hat A_m^{\dagger}\hat A^m - \half\right) \qquad
 \text{(no sum on $m$).}
\eqn{qmat}
\eeq 
The 32-dimensional reducible spinor representation thus consists
of the states with $q_m$ charges 
\beq
\ket{\pm\half,\pm\half,\pm\half,\pm\half, \pm\half} 
\eqn{qcharges}\eeq 
The 16-dimensional irreducible representations are found by projecting out those states with $\Gamma_{11}=\pm1$.
As $\Gamma_{11} =
-i\prod_{\alpha=1}^{10}\Gamma_\alpha =
\prod_{m=1}^5 (2 q_m)$,  the 
${\bf 16}$ of $SO(10)$ consists of states with $q_m$ 
charges given by the 32 states in \eq{qcharges}, subject to the
additional constraint 
\beq 
\prod_{m=1}^5\, 2q_m = +1  
\ . 
\eqn{chicon}\eeq

The $Z_N^d$ orbifold symmetry is then embedded in $SO(10)$ by defining
the $\bfr$ charges in terms of the five $q_m$.  The three guidelines above
may be  economically satisfied if we define
\beq
r_\mu = q_\mu-q_{d+1}\ ,\qquad  \mu= 1\,\ldots,d\le 4\ .
\eqn{rcharges}
\eeq
With this definition we see that each of the components of $\bfr$ only assumes the values
$\pm1$  or $0$ for any of the bosonic or fermionic variables of the
mother theory. This fulfills our requirement that the orbifold
projection be chosen so that there are only neighbor interactions on
the resulting lattice.  Furthermore, it follows that $2^{4-d}$ of the
sixteen fermions have 
vanishing $\bfr$, which is the maximum possible.  With the above
definition of $\bfr$ we can construct $d$-dimensional local lattices
with  $2^{4-d}$ unbroken supercharges. We will consider below all the 
lattices with $d\le 4$, and will display the $\bfr$ charges explicitly
in each case.

The $d$
$\bfr$ charges generate the Cartan subgroup of an $SU(d+1)$ subgroup of
the original $SO(10)$, an
observation which proves useful in constructing the lattice theories,
as it implies that the position on the lattice assigned to each
variable is determined by its $SU(d+1)$ weight. 
To see this, note that the anticommutation
relations \eq{acom} imply that the operators
\beq
\hat T_a \equiv \hat A^\dagger_m (T_a)^m_{\ n} \hat A^n\ ,\qquad m,n=1,\ldots,5\ ,
\eqn{tdef}\eeq
satisfy the commutation relations
\beq
\left[\hat T_a,\,\hat T_b\right] = \hat A^\dagger_m \left[ T_a,\,
  T_b\right]^m_{\ n}  \hat A^n\ .
\eeq
The components of $\bfr$ defined in \eq{rcharges} may be written as
\beq
r_\mu = \hat A^\dagger_m (R_\mu)^m_{\ n} \hat A^n ,\qquad (R_\mu)^m_{\ n}\equiv 
(\delta^m_\mu\delta_{n\mu} - \delta^m_5\delta_n^5)\ ,\qquad \mu=1,\ldots,d.
\eqn{rsu}
\eeq
Since the  $R_\mu$ matrices are linearly independent, real,
diagonal and traceless, it follows that the four $\bfr$ charges generate  the Cartan subalgebra
of an $SU(d+1)$ subgroup of $SO(10)$.

\subsection{From orbifold projection to spacetime lattice}
\label{sec:2c}

As described above, the orbifold projection gives rise to an action 
which can be conveniently described  as a lattice action, assigning
variables to links and sites of a $d$-dimensional lattice as
determined by their $\bfr$ charges. However, at this point the lattice
does not resemble a spacetime lattice (for example, the action derived
from \Eq{act1} has only   cubic and quartic terms, and nothing
resembling a kinetic ``hopping term''.  Furthermore, there is no
intrinsic dimensionful scale in the action, and hence no metric or
definition of distance given to our lattice links.  All that is defined by
the action of the orbifold theory is a connectivity. This can be made
precise by recognizing that the $\bfm$ and $\bfn = \bfm + \bfr$
vectors are not themselves lattice vectors.  Instead, the lattice
point $\bfm$ resides at spacetime point $\vec x = \sum_a m_a \vec e_a
\equiv \bfm\cdot \vec \bfe$, where the $\vec e_a$ are a complete set
of lattice vectors, to be specified\footnote{The integer valued
  $d$-vectors such as $\bfm$ and $\bfr$ are represented in bold-face;
  $d$-dimensional spacetime vectors such as $\vec x$ or $\vec e_a$ are
  denoted with a vector.  The complete set of $d$, \hbox{$d$-dimensional} lattice vectors is $\vec
  \bfe$.}. Similarly, link variables
associated with the vector $\bfr$ reside on links corresponding to the
spacetime vector $\bfr\cdot \vec\bfe$.

As seen in earlier examples
\cite{Kaplan:2002wv,Cohen:2003aa,Cohen:2003qw,Kaplan:2003uh}, turning
the orbifold lattice into a spacetime lattice is accomplished by
expanding  the orbifold
projected action about a certain point in moduli space.  The vacuum
expectation values of the bosonic link variables at this point in
moduli space are interpreted as the inverse lengths of the
corresponding links of the spacetime lattice, and the continuum limit
is taken by moving out to infinity in moduli space \footnote{The
  moduli space is the set of orbifold projected matrices for which the lattice action
  \Eq{act1} vanishes.}.  Expanding about different points in moduli
space can leave intact different symmetries, and correspond to
different spacetime lattice structures.  For example, if we were to
choose the point where the $z_m$ and $\mybar z^m$ variables were equal
and proportional to the identity matrix for those with $r_a =
\delta_{ai}$ for $i=1,\ldots,d$, with all other bosonic variables
vanishing, the resultant $d$-dimensional lattice would be hypercubic, with various
diagonal link variables.  However, as we will show below, the most
symmetric choice corresponds to not to an $A_d^*$ lattice, rather than
a hypercubic one.

We now turn to the explicit construction of the supersymmetric
lattices in dimensions $d=1,\ldots,4$.  In particular, we write down
the orbifold action for dimension $d$, and make explicit the exact
supersymmetry retained on the lattice. We then show how the desired
target theory is obtained at the classical level  as one travels along
the $A_d^*$ trajectory in moduli space, described in the previous
section, making explicit the peculiar way in which the lattice
variables assemble themselves into the continuum variables of the
$\CQ=16$ target theory, which typically form large multiplets of both
supersymmetry and a chiral $R$-symmetry.  For more information about
these target theories, we refer the reader to
ref.~\cite{Seiberg:1998ax}.

\section{Construction of the lattice in four dimensions }
\label{sec:3}

\subsection{The target theory}

\label{sec:3a}

For $d=4$ the continuum  target theory with $\CQ=16$ supercharges is $\CN=4$ SYM
theory with a $U(k)$ gauge group. The action for this theory can be
simply obtained  
by dimensional reduction of the simple $\CN=1$ SYM theory in ten
dimensions down to four dimensions.  The action  therefore 
possesses a global $SO(4) \times SO(6) $  symmetry  inherited 
by dimensional reduction (where the $SO(4)\simeq SU(2)\times SU(2)$  is the Euclidean
version of the Lorentz symmetry, and $SO(6)\simeq SU(4)$ is called the
$R$-symmetry of the theory). The gauge fields $v_\alpha$ of the
ten-dimensional  theory reduce to four gauge fields $V_\mu$ in four
dimensions, plus six scalar fields $S_a$.  The sixteen component
gaugino $\tilde\omega$ of the ten-dimensional theory reduces to four complex Weyl
doublet fermions.  The transformations of these fields is
\beq
\begin{array}{c|c}
\quad & SO(6)\times SU(2)\times SU(2)\cr
\hline
V_\mu & (1,2,2)\cr
S_a & (6,1,1)\cr
\tilde\omega & (4,2,1)\oplus (\mybar 4,1,2) 
\end{array}
\eqn{targarray}\eeq
In order to make a direct connection between the lattice and the continuum
theory, it is simplest to express the target action in
four dimensions in a notation that retains some of the structure
inherited from ten dimensions.  Therefore we write the action for
$\CN=4$ SYM in four dimensions as\footnote{In the above equation, and throughout this section we adopt the
convention that repeated indices are summed, where $\alpha,\beta\ldots$
are $SO(10)$ indices summed over $1,\ldots,10$; the indices
$\mu,\nu,\dots$ are $SO(4)$ spacetime indices summed over $1,\ldots,4$; the
indices $a,b,\ldots$ are $SO(6)$ $R$-symmetry indices summed over $1,\ldots,6$; and
$m,n,\ldots$ are $SU(5)$ indices summed $1,\ldots,5$.}
\beq
S_{\text{target}} &=& \frac{1}{g_4^2} \,\int d^4\bfR\, \Tr
\Bigl(\frac{1}{4} V_{\mu\nu} V_{\mu\nu} + \frac{1}{2} (D_\mu S_a)^2 -\frac{1}{4}\left[S_a,S_b\right]^2\cr
 &&\qquad\qquad\qquad\qquad
+\frac{1}{2}\tilde \omega^T\,\tilde{C}\, D_\mu \, \tilde \Gamma_\mu \,\tilde \omega +
\frac{i}{2}\, \tilde\omega^T\, \tilde{C}\, \tilde\Gamma_{4+a}\, [S_a,\tilde \omega]\Bigr)
\eqn{target4}\eeq
Here we have introduced $SO(10)$ gamma matrices $\tilde\Gamma_\alpha$
and charge conjugation matrix $\tilde C$ \footnote{ These matrices
  appear with tildes  to
indicate a difference in basis from that chosen for the mother theory
in \eq{act1}.  In fact,  when we establish the correspondence between lattice and
continuum fermion variables, we will identify the similarity
transformation between  the two bases $\Gamma$ and $\tilde \Gamma$.}.  The
$SO(4)\times SO(6)$ invariance of the above theory is manifest.

\subsection{The mother theory in $SU(5)\times U(1)$ multiplets }
\label{sec:3b}

To create a four dimensional lattice from the $\CQ=16$ mother theory, we 
orbifold by $Z_N^4$, where the four $Z_N$ transformations are determined by 
the four-vector $\bfr$ charges defined in \eq{rcharges} with $d=4$.  As we showed in
\S\ref{sec:2b}, the $\bfr$ charges generate the Cartan subgroup of the
$SU(5)$ subgroup of the original $SO(10)$ symmetry of the mother
theory, and that the assignment of variables of the mother theory
onto links and sites of the lattice follows from their $SU(5)$ weights.  It is
convenient therefore to decompose the variables 
of the mother theory under the subgroup  $SU(5)\times U(1)\in SO(10)$, where the
$U(1)$ is generated by 
\beq
Q_0\equiv\sum_{m=1}^5 q_m\qquad \text{($U(1)$ generator)}\ ,
\eqn{u1}
\eeq 
where the $q_m$ are defined in \eq{qdef};  $Q_0$ generates a  rotation in all of the $m$, $m+5$
planes of the ten dimensional space simultaneously.

The bosons $v_\alpha$
transform as a {\bf 10} of $SO(10)$ and decompose under as the $SU(5)\times
U(1)$ subgroup as
\beq
v \sim{\bf 10} \ \longrightarrow\  z\oplus \mybar z \sim {\bf 5_1}\oplus {\bf \mybar  5_{-1}}\ .
\eeq
It should be evident that  the $z_m$ and $\mybar z_m$ variables
defined in \eq{zdef}  have $U(1)$ charges $Q_0=+1$ and $Q_0=-1$
respectively, and so it must be that $z_m\sim{\bf 5_1}$ and $\mybar
z_m\sim {\bf \mybar 5_{-1}}$. As  for
the fermions, the variable $\omega$ of the mother theory, transforming
as a {\bf 16} of $SO(10)$, decomposes under $SU(5)\times
U(1)$ as
\beq
\omega\sim {\bf 16} \ \longrightarrow\  \lambda\oplus \psi \oplus \xi \sim{\bf 1_{\frac{5}{2}}}\oplus {\bf 5_{-\frac{3}{2}}}\oplus
  {\bf \mybar {10}_{\frac{1}{2} }}\ .
\eeq
To perform this decomposition of $\omega$ explicitly we use the
fermionic ladder operators  $A^m$ and  $A_m^\dagger$ defined in
\eq{adef}.  The $SU(5)$ generators are given by $\hat
T_a=(\hat A^\dagger T^a A)$ in \eq{tdef}, and the  $U(1)$ generator is given by
$Q_0=(-\frac{5}{2} +\sum_m \hat A^\dagger_m \hat A^m)$, combining \eq{u1} and
\eq{qmat}. For any particular basis for the $SO(10)$ $\Gamma$ matrices
one can then find a normalized spinor $\nu_+$ annihilated by all of the $\hat A^\dagger_m$:
\beq
\hat A^\dagger_m \nu_+ = 0\ ,\qquad \nu_+^\dagger \nu_+ = 1\ .
\eqn{nudef}\eeq
Note that $Q_0\, \nu_+ = +\frac{5}{2}$, and that applying to $\nu_+$ a
lowering operator
$\hat A_m$ decreases the $Q_0$ charge by one unit.
The variable $\omega$ may then be
expanded as
\beq
\omega = \left(\lambda +   \xi_{mn}\,\frac{1}{2}
\hat A^m \hat A^n 
-\psi^m \,\frac{\epsilon_{mnpqr}}{24} \hat A^n \hat A^p \hat A^q \hat A^r
  \right)\nu_+
\eqn{omexp}
\eeq
with  $\lambda$, $\xi_{mn}$ and $\psi^m$ transforming under
$SU(5)\times U(1)$ as  the  ${\bf 1_{\frac{5}{2}}}$,  ${\bf \mybar{10}_{\half }}$
and ${\bf
  5_{-\frac{3}{2}}}$
respectively. Written in terms of this $SU(5)\times U(1)$ decomposition, the action
of the mother theory \eq{act1} becomes
\beq
S &=& \frac{1}{g^2} \Tr \Bigl[ \sum_{m,n} \left( \frac{1}{2}[\mybar z_m, z^m][\mybar z_n, z^n] + 
[z^m, z^n][\mybar z_n,\mybar z_m]\right)\Bigr.\cr
&&\qquad\quad\Bigl. +  
{\sqrt 2}\left( \lambda [\mybar z_m, \psi^m] - 
\xi_{mn}[z^m, \psi^n] + \frac{1}{8} \epsilon^{mnpqr}
\xi_{mn}[\mybar z_p, \xi_{qr}] \right) \Bigr]\ .
\eqn{act2}
\eeq
The $q_m$ and $\bfr$ charges of each of these variables is easily
computed, using \eq{qmat} and \eq{rcharges};  the results are shown
below in Table~\ref{tab1}.

 The correspondence between the $\bfr$
charges and the $SU(5)$ tensor notation is made explicit by defining
the five vectors
\beq
\bfmu_1 &=& \{1,0,0,0\}\ ,\cr
\bfmu_2 &=& \{0,1,0,0\}\, \cr
\bfmu_3 &=& \{0,0,1,0\}\, \cr
\bfmu_4 &=& \{0,0,0,1\}\, \cr
\bfmu_5 &=& \{-1,-1,-1,-1\}\ .
\eqn{mudef}\eeq
The utility of the $\bfmu_m$ vectors is that
they specify the $\bfr$ charge directly in terms of the $SU(5)$ tensor
indices: for each variable the $\bfr$ charge is given by a sum of
$\bfmu_m$ for each
upper  $SU(5)$ index $m$, and $-\bfmu_m$ for each lower index
$m$. Thus, as seen in Table~\ref{tab1},   $z^m$ and $\psi^m$ have
$\bfr=\bfmu_m$; while $\mybar z_m$ has
$\bfr=-\bfmu_m$, $\xi_{mn}$ has $\bfr=-(\bfmu_m + \bfmu_n)$ and 
$\lambda$ has $\bfr=0$.

\begin{table}[t]
\begin{tabular}[t]
{|r||r||r|r|r|r|r||rcl|}
\hline
& $Q_0$\ & $q_1\ $ & $q_2\ $ & $q_3\ $ & $q_4\ $ & $q_5\ $ &
&\bfr&
\\ \hline
$z_1$ &  $\, \ 1$ &$\,\ 1$ & $\, \ 0$ & $\, \ 0$ & $\, \ 0$ & $\, \ 0$ & 
$\{1,0,0,0\}$&=&$\ \ \bfmu_1$\\ 
$ z_2$&  $\, \ 1$ &$\, \ 0$ & $\,\ 1$ & $\, \ 0$ & $\, \ 0$ &$\, \ 0$ & 
$\{0,1,0,0\}$&=&$\ \ \bfmu_2$\\
 $z_3$ &$\, \ 1$  & $\, \ 0$ & $\, \ 0$ & $\,\ 1$ & $\, \ 0$ & $\, \
 0$ & %
$\{0,0,1,0\}$&=&$\ \ \bfmu_3$\\
$z_4$ & $\, \ 1$  & $\, \ 0$ & $\, \ 0$ & $\, \ 0$ & $\,\ 1$ & $\, \ 0$&
$\{0,0,0,1\}$&=&$\ \ \bfmu_4 $\\
$z_5$ &  $\,\ 1$ & $\, \ 0$ & $\, \ 0$ & $\, \ 0$ & $\, \ 0$ &$\, \ 1$ &
$\{-1,-1,-1,-1\}$&=&$\ \ \bfmu_5$\\ \hline
$\mybar z_1 $ & $-1$ &  $\,\ -1$  & $\, \ 0$ & $\, \ 0$ & $\, \ 0$ & $\, \ 0$ &
$\{-1,0,0,0\}$&=&$-\bfmu_1$\\ 
$\mybar z_2$& $-1$  & $\, \ 0$ & $\, \ -1$ & $\, \ 0$ & $\, \ 0$ & $\, \ 0$&
$\{0,-1,0,0\}$&=&$-\bfmu_2$ \\
$\mybar z_3$ &  $-1$ &$\, \ 0$ & $\, \ 0$ &  $\,\ -1$ & $\, \ 0$ & $\, \ 0$ &
$\{0,0,-1,0\}$&=&$-\bfmu_3$ \\
$\mybar z_4$ &  $-1$ &$\, \ 0$ & $\, \ 0$ & $\, \ 0$ & $-1$ & $\, \ 0$ &
$\{0,0,0,-1\}$&=&$-\bfmu_4$ \\
$ \mybar z_5$ &  $-1$ &$\, \ 0$ & $\, \ 0$ & $\, \ 0$ & $\, \ 0$ & $\,\ -1$ &
$\{1,1,1,1\}$&=&$-\bfmu_5$\\   \hline   \hline 
$\lambda$ & \, \ $\frac{5}{2}$ & $\ \half$ & $\ \half$ & $\ \half$ &
$\ \half$ & $\ \half$ &
$\{0,0,0,0\}$&=&\,\ \ {\bf 0}\\ \hline 
$\psi_1$ &  $-\frac{3}{2}$ & $\ \half$ & $-\half$ & $-\half$ & $-\half$ & $-\half$ &
$\{1,0,0,0\}$&=&$\ \ \bfmu_1$
\\
$\psi_2$  & $-\frac{3}{2}$ & $-\half$ & $\ \half$ & $-\half$ & $-\half$ & $-\half$ &
$\{0,1,0,0\}$&=&$\ \ \bfmu_2$ \\ 
$\psi_3$& $-\frac{3}{2}$ & $-\half$ & $-\half$ & $\ \half$ & $-\half$ &
$-\half$ &
$\{0,0,1,0\}$&=&$\ \ \bfmu_3$\\
$\psi_4$ & $-\frac{3}{2}$ & $-\half$ & $-\half$ & $-\half$ & $\ \half$ & $-\half$ &
$\{0,0,0,1\}$&=&$\ \ \bfmu_4 $\\
$\psi_5$ & $-\frac{3}{2}$ & $-\half$ & $-\half$ & $-\half$ & $-\half$ & $\ \half$ &
$\{-1,-1,-1,-1\}$&=&$\ \ \bfmu_5$\\  \hline 
$\xi_{12}$& \, \ $\frac{1}{2}$ & $-\half$ & $-\half$ & $\ \half$ & $\ \half$ & $\ \half$ &
$\{-1,-1,0,0\}$&=&$-\bfmu_1 - \bfmu_2$\\
$\xi_{13}$ & \, \ $\frac{1}{2}$ & $-\half$ & $\ \half$ & $-\half$ & $\ \half$ & $\ \half$ &
$\{-1,0,-1,0\}$&=&$-\bfmu_1-\bfmu_3$\\
$\xi_{14}$& \, \ $\frac{1}{2}$ & $-\half$ & $\ \half$ & $\ \half$ &
$-\half$ & $\ \half$ &
$\{-1,0,0,-1\}$&=&$-\bfmu_1-\bfmu_4$ \\
$\xi_{23}$ & \, \ $\frac{1}{2}$ & $\ \half$ & $-\half$ & $-\half$ &
$\ \half$ & $\ \half$ &
$\{0,-1,-1,0\}$&=&$-\bfmu_2-\bfmu_3$\\
$\xi_{24}$& \, \ $\frac{1}{2}$ & $\ \half$ & $-\half$ & $\ \half$ & $-\half$ & $\ \half$ &
$\{0,-1,0,-1\}$&=&$-\bfmu_2-\bfmu_4$  \\
$\xi_{34}$ & \, \ $\frac{1}{2}$ & $\ \half$ & $\ \half$ & $-\half$ & $-\half$ & $\ \half$ & 
$\{0,0,-1,-1\}$&=&$-\bfmu_3-\bfmu_4$ \\
$\xi_{15}$& \, \ $\frac{1}{2}$ & $-\half$ & $\ \half$ & $\ \half$ &
$\ \half$ & $-\half$ &
$\{0,1,1,1\}$&=&$-\bfmu_1-\bfmu_5$ \\
$\xi_{25}$& \, \ $\frac{1}{2}$ & $\ \half$ & $-\half$ & $\ \half$ &
$\ \half$ & $-\half$ &
$\{1,0,1,1\}$&=&$-\bfmu_2 - \bfmu_5$ \\ 
$\xi_{35}$ & \, \ $\frac{1}{2}$ & $\ \half$ & $\ \half$ & $-\half$ & $\ \half$ & $-\half$ &
$\{1,1,0,1\}$&=&$-\bfmu_3-\bfmu_5$\\
$\xi_{45}$ & \, \ $\frac{1}{2}$ & $\ \half$ & $\ \half$ & $\ \half$ &
$-\half$ & $-\half$ &
$\{1,1,1,0\}$&=&$-\bfmu_4-\bfmu_5$\\  \hline
 \end{tabular} 
\caption{ The $Q_0$, ${q_m}$ and $r_\mu = (q_\mu-q_5)$ charges of the
  bosonic variables $v$
  and fermionic variables $\omega$  of the $\CQ=16$ mother  theory under the
  $ SO(10)\supset SU(5)$ decomposition 
$v=10\to 5\oplus\mybar 5 = z^m   \oplus \mybar z_m $, and 
$\omega= 16 \to 1\oplus 5\oplus  \mybar {10} = \lambda \oplus \psi^m \oplus \xi_{mn}$. 
 }

\label{tab1}
\end{table}

\subsection{ Manifest $\CQ=1$ supersymmetry of the mother theory}
\label{sec:3c}

The above action \eq{act2} is just a rewriting of the mother theory
\eq{act1} in terms of new variables, and so it respects the full
sixteen independent supersymmetry transformations of
\eq{mothersusy}, which are parametrized by the constant Grassmann
spinor $\kappa$, transforming as a ${\bf 16}$ under $SO(10)$. After
the orbifold projection by $(Z_N)^4$, only a 
single supersymmetry remains intact, corresponding to  that
component of $\kappa$ which has $\bfr=0$.  This surviving supersymmetry
corresponds to
\beq
 \kappa = \eta \nu_+\ ,
\eeq
where $\eta$ is a Grassmann number, and $\nu_+$ is the constant $SO(10)$ spinor
defined in \eq{nudef}.
This sole surviving supersymmetry transformation, in terms of our new variables, is
\beq
 \delta z^m &=&  i \sqrt 2 \, \eta\, \psi^m, \cr
 \delta \mybar z_m &=& 0  \cr
\delta \lambda &=&  -i \eta \,  [\mybar z_m, z^m]  \cr
\delta\psi^m&=&0\cr 
\delta \xi_{mn} &=& -2 i \eta \, [\mybar z_m, \mybar z_n] \qquad
\eqn{q1susy}
\eeq
where $m, n = 1\ldots 5$ and repeated indices are summed.

We now rewrite the mother theory action \Eq{act2} in a superfield formalism which makes this
$\CQ=1$ supersymmetry manifest. By doing this before the orbifold
projection, it makes it quite easy to see the exact supersymmetry of
the lattice theory after the orbifold projection.
We introduce a Grassmann valued coordinate $\theta$ and nilpotent
supersymmetry charge $Q$ which generates the above 
supersymmetry transformations 
\beq
\delta= i\eta\, Q\ , \qquad Q=\frac{\partial}{\partial \theta}
\eeq
The  transformations \eq{q1susy} can then be realized in terms of following 
superfields 
\beq
{\bf Z}^m  &=&  z^m + \sqrt 2  \theta \psi^m \cr
{\bf \Lambda} &=& \lambda  -\theta  ( [\mybar z_m, z^m] + id ) \cr
{\bf \Xi}_{mn} &=& \xi_{mn} -2 \theta  [\mybar z_m, \mybar z_n]\ , \qquad
\eeq
along with the singlet $\mybar z_m$, where we have introduced an auxiliary field $d$ and modified the
transformation of $\lambda$ such that
\beq
\delta \lambda &=&  -i \eta \, ( [\mybar z_m, z^m] + i\,d ) \cr
\delta d &=& i\delta\left([\mybar z_m, z^m]\right) = -\sqrt{2} \eta\,
[\mybar z_m, \psi^m]\ ,
\eqn{ddef}\eeq
allowing the supersymmetry to close off-shell (this is necessary,
since $\delta$ as defined in \eq{q1susy} does not satisfy $\delta^2=0$
without invoking the equations of motion, as we see below).

In terms of these superfields, the action for mother theory \Eq{act2} 
is written in manifestly $\CQ=1$ invariant form  as 
\beq
S &=& \frac{1}{g^2} \Tr  \int \;  d \theta \left( -\frac{1}{2} {\bf \Lambda} 
{\partial}
_{\theta} {\bf \Lambda}  - {\bf \Lambda} [\mybar z_m, {\bf Z}^m]  + \frac{1}{2}
{\bf \Xi}_{mn}[{\bf Z}^m,{\bf Z}^n] \right) + \frac{\sqrt 2 }{8} 
 \epsilon^{mnpqr} {\bf \Xi}_{mn}
[\mybar z_p , {\bf \Xi}_{qr}] \qquad \qquad
\eqn{act3}
\eeq
The last term in the action is not integrated over $\theta$;  that it
is supersymmetric may be shown by means of  the Jacobi identity of the
Lie algebra which implies
\beq
\frac{\partial}{\partial \theta}  
\epsilon^{mnprq} {\bf \Xi}_{mn}
[\mybar z_p , {\bf \Xi}_{qr}]=0\ .
\eeq
Thus this term is  $\theta$  independent and hence supersymmetric. 

One can readily verify that the action \eq{act3} in component form is
equivalent to \eq{act2}, except for the addition of a new term
involving the auxiliary field, 
$\frac{1}{2g^2} \Tr d^2$.  By differentiating $S$ by $d$ and by
$\lambda$ one finds the equations of motion $d=0$ and $[\mybar z_m,
\psi^m]=0$ respectively.  The latter equals $\delta d$ by \eq{ddef}, and so that the off-shell supersymmetry
transformations \eq{ddef} are consistent with the supersymmetry of the
mother theory \eq{q1susy} after invoking the equations of
motion. The auxiliary field $d$ fulfills here an analogous role to that played by
auxiliary fields in the more familiar four dimensional supersymmetric
field theories.

\subsection{The $D=4$, $\CQ=1$ lattice action and its symmetries}
\label{sec:3d}

The charges given in Table~\ref{tab1} make it simple to write down
the action of the lattice theory that results from the orbifold
projection. In component form, the result is

\begin{equation}
\begin{aligned} 
S =
  \frac{1}{g^2}\sum_{\bfn} \Tr \biggl[ & \half\Bigl(\sum_{m=1}^5 \left(\mybar z_m(\bfn-\bfmu_m)
    z^m(\bfn-\bfmu_m) - z^m(\bfn)\mybar z_m(\bfn)\right)
\Bigr)^2\\&
  +\sum_{m,n=1}^5
\Bigl\vert\, z^m(\bfn) z^n(\bfn + \bfmu_m)-  z^n(\bfn) z^m(\bfn + \bfmu_n)
\Bigr\vert^2
  \\ & 
-\sqrt{2}\Bigl( \Delta_\bfn(\lambda,\mybar z_m, \psi^m)+
  \Delta_\bfn(\xi_{mn}, z^m,\psi^n)
  +\frac{1}{8}\epsilon^{mnpqr}\Delta_\bfn(\xi_{mn},\mybar z_p,\xi_{qr})
\Bigr) 
\biggr] 
\\ &
\eqn{d4lat}
\end{aligned} 
\end{equation} 
 We have introduced the labeling convention that
$z^m(\bfn)$, $\psi^m(\bfn)$ and $\mybar z_m(\bfn)$ live on the same link, running
between site $\bfn$ and site $(\bfn+\bfmu_m)$; similarly
$\xi_{mn}(\bfn)$ lives on the link between sites $\bfn$ and
$(\bfn+\bfmu_m + \bfmu_n)$, while $\lambda(\bfn)$ resides at the site
$\bfn$.  The site vector $\bfn$, a four-vector with integer-valued
components,  should be distinguished from $SU(5)$
indices $n$. 

We have introduced the triangular plaquette function $\Delta_\bfn$
defined as:
\begin{equation}
\begin{aligned} 
 \Delta_\bfn(\lambda,\mybar z_m, \psi^m)=&-
\lambda(\bfn) \Bigl(\mybar z_m(\bfn-\bfmu_m)  \psi^m(\bfn-\bfmu_m) -
\psi^m(\bfn)\mybar z_m(\bfn)\Bigr)\ ,\\ 
 \Delta_\bfn(\xi_{mn}, z^m,\psi^n)=&   \xi_{mn}(\bfn)\Bigl(
   z^m(\bfn) \psi^n(\bfn+\bfmu_m) - \psi^n(\bfn)
   z^m(\bfn+\bfmu_n)\Bigr)\ ,\\
\Delta_\bfn(\xi_{mn},\mybar z_p,\xi_{qr})=&-\xi_{mn}(\bfn)\Bigl(\mybar
z_p(\bfn-\bfmu_p) \xi_{qr}(\bfn+\bfmu_m+\bfmu_n) \\&\qquad\qquad- 
\xi_{qr}(\bfn-\bfmu_q-\bfmu_r)\mybar z_p(\bfn+\bfmu_m+\bfmu_n)\Bigr)
\end{aligned}
\eqn{Deltadef}
\end{equation}
Note that $\Delta$ corresponds to the signed sum of two terms, each of
which is a string of three
variables along a closed and oriented path on the lattice, with the
sign determined by the orientation of the path.
 As discussed in \S~\ref{sec:2}, there is a $U(k)$ gauge
symmetry associated with each site, with $\lambda(\bfn)$ transforming
as an adjoint, while the oriented link variables transform as bifundamentals
under the two $U(k)$ groups associated with the originating and
destination sites of the link. A  string of variables along any closed
path on the lattice, such as we see in the definition of $\Delta$, is
gauge invariant. In the continuum limit, the $\Delta$ terms will form
the gaugino hopping terms and Yukawa couplings of the $\CQ=16$ SYM theory.

It is now simple to write down the action for the lattice theory that
results from the orbifold projection, in a form which is manifestly
$\CQ=1$ supersymmetric.

    After orbifold projection, there are
superfields associated with each lattice site $\bfn$, where $\bfn$ is a four component
vector of integers, each component ranging from 1 to $N$:
\beq
{\bf Z}^m({\bf n})  &=&  z^m(\bfn) + \sqrt 2  \theta \psi^m(\bfn) 
\cr
{\bf \Lambda}(\bfn) &=& \lambda(\bfn)  -\theta \left( 
\left[ \mybar z_m (\bfn - \bfmu_m) z^m(\bfn -\bfmu_m)- z^m(\bfn) 
\mybar z_m(\bfn)\right] + id(\bfn) \right) \cr
{\bf \Xi}_{mn}(\bfn) &=& \xi_{mn}(\bfn) -2 \theta  
\left[\, \mybar z_m(\bfn+ \bfmu_n) \mybar z_n(\bfn) - \mybar z_n(\bfn + \bfmu_m ) 
\mybar z_m(\bfn)\right]
\eeq
In addition there is the singlet field $\mybar z_m(\bfn)$. 
In the above expressions, subscripts and superscripts $m,n=1,\ldots,5$ and repeated indices are
summed over.  Note that the superfields are not
entirely local, and that in the continuum they will depend on
derivatives of fields as well as the fields themselves.

The lattice action we obtained may be written in manifestly  $\CQ=1$ 
supersymmetric form as 
\beq
S &=& \frac{1}{g^2}\Tr \sum_{\bfn}  \int \;  d \theta \left( 
-\frac{1}{2} {\bf \Lambda}(\bfn) {\partial}_{\theta} {\bf \Lambda}(\bfn)  
- {\bf\Lambda}(\bfn)\Bigl[
\mybar z_m(\bfn - \bfmu_m) 
{\bf Z}^m(\bfn - \bfmu_m)
 - {\bf Z}^m(\bfn) \mybar z_m(\bfn) \Bigr]\right. \cr &&\qquad\qquad\qquad\qquad
 + \left.
\frac{ 1}{2}
{\bf \Xi}_{mn}(\bfn)\Bigl[{\bf Z}^m(\bfn) {\bf Z}^n (\bfn + \bfmu_m)- {\bf
 Z}^n(\bfn) 
 {\bf Z}^m(\bfn + \bfmu_n)\Bigr]
\right) \cr
&&
 +
  \frac{\sqrt 2 }{8}  \epsilon^{mnpqr}{\bf \Xi}_{mn} (\bfn)
 \Bigl[\mybar z_p (\bfn-  \bfmu_p)  {\bf \Xi}_{qr}( \bfn +  \bfmu_m + 
\bfmu_n )
-{\bf \Xi}_{qr}(\bfn -  \bfmu_q - \bfmu_r) \mybar z_p(\bfn +  
\bfmu_m + \bfmu_n) \Bigr]
\cr&&
\eqn{d4latss}
\eeq
The auxiliary field $d(\bfn)$ has no hopping term, and after
eliminating it by the equations of motion on can show that the above
action in terms of superfields is equivalent to the lattice action
given in component form in \eq{d4lat}.

The purpose for formulating the action in the supersymmetric form is
to facilitate analysis of allowed operators and the continuum limit of
the lattice theory.

\subsection{The continuum limit for $d=4$ lattice: tree level  }
\label{sec:3e}

The lattice defined by the orbifold projection cannot be directly
considered to be a spacetime lattice, as all terms in the lattice
action \eq{d4latss} are trilinear and  conventional
hopping terms are absent. To generate a spacetime lattice and take the continuum
limit one must follow the example of deconstruction
\cite{ArkaniHamed:2001ca} and follow a particular trajectory  out to
infinity in the
moduli space of the theory, interpreting the distance
from the origin of moduli space as the inverse lattice spacing.

As can be seen in \eq{d4lat}, the moduli space in the present theory  corresponds to all values for
the bosonic $z$ variables such that
\begin{equation}
\begin{aligned} 
0=\sum_{\bfn} \Tr \biggl[ & \half\Bigl(\sum_m \left(\mybar z_m(\bfn-\bfmu_m)
    z^m(\bfn-\bfmu_m) - z^m(\bfn)\mybar z_m(\bfn)\right)
\Bigr)^2\\&
  +
\sum_{m,n}
\Bigl\vert\, z^m(\bfn) z^n(\bfn + \bfmu_m)-  z^n(\bfn) z^m(\bfn + \bfmu_n)
\Bigr\vert^2\biggr] .
\end{aligned}
\end{equation}

\subsubsection{A hypercubic lattice}

  There are clearly a large class of solutions to these equations.
One possibility is
\beq
z^m(\bfn) &=& \mybar z_m(\bfn) = \frac{1}{a\sqrt{2} }   {\bf 1}_k
,\qquad  m= 1, \ldots, 4,\cr
z^5(\bfn) &=& \mybar z_5(\bfn) = 0\ ,
\eeq
where $a$ is the length scale associated with the lattice spacing,
interpreted as the physical length (up to a factor of $4/5$) of the links on which $z_m$ and
$\mybar z^m$ variables reside, for $m=1,\ldots,4$. Such a  lattice can be interpreted as a hypercubic lattice
of length $a$ on an edge, since the $\bfr$ charges for these variables
correspond to Cartesian unit vectors,
as seen  in Table~\ref{tab1}.  In this case, the physical location of
site $\bfn$ is simply the four-vector  $\bfR = a \bfn$.
Because the $\xi_{mn}$, $z^5$, $\mybar z_5$ and $\psi^5$ variables reside on various diagonal
links of this hypercubic lattice, and all links are oriented, the
symmetry of the lattice action is $S_4$,  much
smaller than the hypercubic group.

\subsubsection{The $A_4^*$ lattice}

Instead of the above trajectory, we choose to examine the most symmetric solution, in the theory
that the greater the symmetry of the spacetime lattice, the fewer
relevant or marginal operators will exist.  A solution which
treats all five $z^m$ symmetrically ({\it e.g.} which preserves an
$S_5$ permutation symmetry) is to have the five links on which they
reside correspond to the vectors connecting the center of a 4-simplex to
its corners. The lattice generated by such vectors is known to mathematicians as $A_4^*$. \footnote{The $A_4$
  lattice is generated by the 
simple roots of $SU(5)=A_4$; then $A_4^*$ is the dual lattice,
generated by the fundamental weights of $SU(5)$, or equivalently, by
the weights of the defining representation of $SU(5)$. Lower dimension
analogues are  $A_2^*$, the triangular lattice, and $A_3^*$, the
body-centered cubic lattice. For further discussion, see
\cite{Conway:1991}}; for a picture of $A_3^*$, see
Fig.~\ref{fig:lat3dBCC} below.  The point in moduli space about which we expand is the
symmetric point:
\beq
z^m(\bfn) &= \mybar z_m(\bfn) = \frac{1}{a\sqrt{2}}    {\bf 1}_k
,\qquad  m= 1, \ldots, 5\ .
\eqn{s5}\eeq
Once again  $a$ is interpreted as the spacetime length of the link
that each $z^m$ resides upon.    The symmetry of our
lattice is $S_5$, corresponding to permutations of the $SU(5)$ indices
in the mother theory action \eq{act2}, accompanied by fermion phase
redefinitions  $\xi\to i\xi$, $\psi\to -i\psi$, and $\lambda\to
i\lambda$ in the case of odd permutations.  The symmetry of the action is not the full
symmetry of the $A_4^*$ lattice, as reflection symmetries which
exchange $z$ and $\mybar z$ are not symmetries of the action.

To relate the lattice site $\bfn$ with a physical location in
spacetime, we introduce a specific basis, in the form of  five,
four-dimensional lattice vectors 
\beq
    {\bf e_1} &=& (\frac{1}{\sqrt 2},  \frac{1}{\sqrt 6}, 
\frac{1}{\sqrt{12}}, \frac{1}{\sqrt{20}})  \cr
{\bf e_2} &=& (-\frac{1}{\sqrt 2},  \frac{1}{\sqrt 6},
\frac{1}{\sqrt{12}}, \frac{1}{\sqrt{20}})  \cr
{\bf e_3} &=& (0,  -\frac{2}{\sqrt 6},
\frac{1}{\sqrt{12}}, \frac{1}{\sqrt{20}})  \cr
{\bf e_4} &=& (0, 0,
-\frac{3}{\sqrt{12}}, \frac{1}{\sqrt{20}})  \cr
{\bf e_5} &=& (0, 0,
0, -\frac{4}{\sqrt{20}}).  
\eqn{latvec}\eeq 
These vectors satisfy the relations
\beq
\sum_{m=1}^5 \bfe_m = 0\ ,\qquad  \bfe_m\cdot \bfe_n =
\left(\delta_{mn}-\frac{1}{5}\right)\ ,\qquad \sum_{m=1}^5
(\bfe_m)_\mu (\bfe_m)_\nu = \delta_{\mu\nu}\ .
\eqn{eprop}
\eeq
The lattice vectors \eq{latvec} are simply related to the $SU(5)$
weights of the {\bf 5} representation, and  the  $5\times 5$ matrix
$\bfe_m\cdot \bfe_n$ can be recognized as the 
Gram matrix for $A_4^*$ \cite{Conway:1991}.

The site $\bfn$ on our
lattice is then defined to be at the spacetime location 
\beq
\bfR = a \sum_{\nu=1}^4 (\bfmu_\nu\cdot \bfn)\, \bfe_\nu = a
\sum_{\nu=1}^4 \, n_\nu \,\bfe_\nu \ ,
\eqn{ra4}\eeq
where $a$ is the lattice spacing introduced in \eq{s5}, and the
vectors $\bfmu_\nu$ (which have integer components) were defined in
\eq{mudef}.  By making use of the fact that $\sum_m\bfe_m=0$, it is
easy to show that a small lattice displacement of the form $d\bfn =
\bfmu_m$ corresponds to a spacetime translation by $(a\,\bfe_m)$:
\beq
d\bfR =  a \sum_{\nu=1}^4 (\bfmu_\nu\cdot d\bfn)\, \bfe_\nu =  a
\sum_{\nu=1}^4 (\bfmu_\nu\cdot \bfmu_m)\, \bfe_\nu =a\,\bfe_m\ .
\eqn{dra4}\eeq
Thus from the last column in Table~\ref{tab1} one can read
off the physical location of each of the variables.  For example, at
the site $\bfn={\bf 0}$, $z^1({\bf 0})$ lies on the link directed from $\bfR={\bf 0}$ to
$\bfR = a \,\bfe_1$, while $\xi_{45}({\bf 0})$  lies on the link
directed from the site
$\bfR= a \,(\bfe_4 + \bfe_5)$ to the site $\bfR={\bf 0}$. From the
relation \eq{dra4} we see that each of the five links occupied by the
five $z^m$ variables has length $|a\,\bfe_m|=\sqrt{\frac{4}{5}}\, a$, unlike
the case of the hypercubic lattice mentioned above, where $z^5$
resided on a  link twice as long as the links occupied by the other four $z^m$ variables.

To relate the lattice \eq{d4lat} to the continuum target theory, we
expand about the point \eq{s5} in inverse powers of the lattice
spacing $a$. The procedure is somewhat awkward, as the lattice
structure is related to the $SU(5)\times U(1)$ subgroup of $SO(10)$,
while the target theory has a structure determined by the $SO(4)\times
SO(6)$ subgroup of $SO(10)$. The effort is  facilitated by introducing
the $5\times 5$ real orthogonal
matrix $\CE$ defined as 
\beq
\CE = \begin{pmatrix}
 \ \ \frac{1}{\sqrt{2}} &  \ \ \frac{1}{\sqrt{6}} & 
\ \ \frac{1}{\sqrt{12}} &  \ \ \frac{1}{\sqrt{20}}& \ \ \frac{1}{\sqrt{5}} \cr
-\frac{1}{\sqrt 2} &  \ \ \frac{1}{\sqrt{6}} & \ \ \frac{1}{\sqrt{12}} & \ \ \frac{1}{\sqrt{20}} & \ \ \frac{1}{\sqrt{5}}  \cr
\ \ 0 &  -\frac{2}{\sqrt{6}} & \ \ \frac{1}{\sqrt{12}} & \ \ \frac{1}{\sqrt{20}} & \ \ \frac{1}{\sqrt{5}}  \cr
\ \ 0 & \ \ 0 & -\frac{3}{\sqrt{12}} & \ \ \frac{1}{\sqrt{20}} & \ \ \frac{1}{\sqrt{5}}  \cr
\ \ 0 & \ \ 0 & \ \ 0 & -\frac{4}{\sqrt{20}} & \ \ \frac{1}{\sqrt{5}} 
\end{pmatrix} = \left(\CE^T\right)^{-1}
\eqn{cedef}\eeq
Note that  $\CE_{m\mu}$, with $\mu=1,\ldots 4$, are the components of the vectors
$\bfe_m$ of \eq{latvec}. This matrix has the property that
\beq
\left(\sum_{m=1}^{5} \bfe_m \,\CE_{mn}\right)_\mu
=\begin{cases}\delta_{ n\mu} & n=1,\ldots,4\cr 0 & n=5 \ .\end{cases}
\eqn{ceprop}
\eeq
which serves as a bridge between the $SU(5)$ tensors of the lattice
construction, and the $SO(4)$ representations of the continuum theory.
In terms of this matrix we then define the
expansion of $z^m$ about the point in moduli space \eq{s5} to be
\beq
z^m = \frac{1}{a\sqrt{2}} + \sum_{m=1}^5 \CE_{mn} \Phi_n =
\frac{1}{a\sqrt{2}} + \sum_{\mu=1}^4 (\bfe_m)_\mu \Phi_\mu +
\frac{1}{\sqrt{5}} \Phi_5\ ,
\eqn{zexp4}\eeq
with
\beq
\Phi_\mu \equiv \left(\frac{S_\mu + i V_\mu}{\sqrt{2}}\right)\ ,\quad
\mu=1,\ldots,4\ ,\qquad \Phi_5 \equiv \left(\frac{S_5 + i
    S_6}{\sqrt{2}}\right)\ ,
\eeq
where $V_\mu$ and $S_a$ are hermitean $k\times k$ matrices, and
$\mybar \Phi = \Phi^\dagger$. Recall also that $\mybar z_m = (z^m)^\dagger$.

We now will expand the action \eq{d4lat} to leading order in powers of
the lattice spacing $a$, with the goal to show the equivalence in the
continuum limit  at tree
level between our lattice action and the target theory action,
\eq{target4}.  Since the Jacobian of the transformation 
between lattice coordinates $\bfn$ and spacetime coordinates $\bfR$ in
\eq{ra4} equals $a^4/ \sqrt{5}$, we first must rescale our coupling $g$ such
that
\beq
\frac{1}{g^2} = \frac{a^4}{\sqrt{5}\,g_4^2}  \ ,\qquad
\lim_{a\to 0} \frac{1}{g^2} \sum_{\bfn} = \frac{1}{g_4^2} \int d^4R\ .
\eeq

Next we consider in turn the terms in the bosonic part of \eq{d4lat}.
The relation \eq{dra4} dictates that we Taylor expand  shifted
variables such as $z^m(\bfn + \bfmu_p)$ as
\beq
z^m(\bfn + \bfmu_p) = \frac{1}{a\sqrt{2}} + \left(1 +
  a\,\bfe_p\cdot{\mathbf\nabla}\right)\CE_{mn}\Phi_n(\bfR) +  O(a^2)\
. 
\eeq
With this relation, we find for the first term in \eq{d4lat}
\footnote{We remind the reader of  our convention that repeated indices are summed.
  $SU(5)$ tensor indices are denoted by  $m,n$ and are summed
from 1 to 5; $SO(6)$ vector indices are denoted by $a,b$ and are summed from
1 to 6; and $SO(4)$ vector indices are denoted by $\mu$,$\nu$ and are summed from
1 to 4.}
\begin{equation}
\begin{aligned}
&
\frac{1}{g^2}\sum_{\bfn}\Tr\,\half\left[\sum_m \left(\mybar z_m(\bfn-\bfmu_m)
    z^m(\bfn-\bfmu_m) - z^m(\bfn)\mybar z_m(\bfn)\right)\right]^2 \cr 
=&
\frac{1}{g_4^2}\int d^4\bfR\, \Tr\,\half\Biggl[
 \left( \frac{1}{a\sqrt{2}} +(1-
   a\,\bfe_m\cdot\bfnab)\CE_{mp}\Phi_p^\dagger+O(a^2)\right) \cr &\qquad\qquad\qquad\qquad\times\left(
   \frac{1}{a\sqrt{2}} + (1-
   a\,\bfe_m\cdot\bfnab)\CE_{mq}\Phi_q+O(a^2)\right) 
\cr &\qquad\qquad\qquad\qquad 
 -   \left(\frac{1}{a\sqrt{2}}
  + \CE_{mq}\Phi_q\right) \left(\frac{1}{a\sqrt{2}}
  + \CE_{mp}\Phi_p^\dagger\right)\Biggr]^2\cr
=&
\frac{1}{g_4^2}\int d^4\bfR\, \Tr\,\half\left(-
  \frac{1}{\sqrt{2}}(\bfe_m\,\CE_{mn}\cdot\bfnab)(\Phi_n+\Phi_n^\dagger)
  +[\Phi_m^\dagger,\Phi_m] + O(a)\right)^2\cr
=& \frac{1}{g_4^2}\int d^4\bfR\, \Tr\,\half\left( -D_\mu
  S_\mu + i\,[S_5,S_6]+O(a)\right)^2
\end{aligned}
\eqn{bose1}\end{equation}
where $D_\mu$ is the covariant derivative of the target theory,
\beq
D_\mu = \partial_\mu + i \left[ V_\mu,\, \cdot\, \right]
\eeq

The second term in \eq{d4lat} has the expansion
\begin{equation}
\begin{aligned}
&\frac{1}{g^2}\sum_{\bfn}\Tr\,
\Bigl\vert\, z^m(\bfn) z^n(\bfn + \bfmu_m)- m\leftrightarrow n
\Bigr\vert^2\cr
&=
\frac{1}{g_4^2}\int d^4\bfR\, \Tr\,
\Biggl\vert\,\left( \frac{1}{a\sqrt{2}} +
  \CE_{mp}\Phi_p\right)\left(\frac{1}{a\sqrt{2}} +(1 +  a\,\bfe_m\cdot\bfnab) \CE_{nq}\Phi_q
+O(a^2)\right)- m\leftrightarrow n
\Biggr\vert^2\cr
&=  \frac{1}{g_4^2}\int d^4\bfR\ \frac{1}{4}
\Tr\,\Biggl\vert (D_\mu S_\nu
  - D_\nu S_\mu) + i (V_{\mu\nu}-i [S_\mu, S_\nu]) \Biggr\vert^2\cr
  &\qquad\qquad\qquad\quad
 + 
2 \Biggl\vert (-D_\mu S_5  -i[S_\mu, S_6]) -i (D_\mu S_6  -i[S_\mu, S_5]) 
\Biggr\vert^2 \cr
&=  \frac{1}{g_4^2}\int d^4\bfR\ \frac{1}{4}\Tr\,\Bigl(V_{\mu\nu}^2 + (D_\mu S_\nu
  - D_\nu S_\mu)^2 + 2(D_\mu S_5)^2 +
  2 (D_\mu S_6)^2 -2i V_{\mu\nu}[S_\mu,S_\nu]
  \cr &\qquad\qquad\qquad\quad
 +4i[S_\mu,S_6] D_\mu S_5-4i[S_\mu,S_5]D_\mu S_6- [S_\mu,S_\nu]^2 - 
2 [S_\mu,S_5]^2  -2 [S_\mu,S_6]^2\Bigr)\cr&
\end{aligned}
\eqn{bose2}\end{equation}
where $V_{\mu\nu}=-i[D_\mu,D_\nu]$ is the nonabelian field strength. In the 
penultimate line, the expressions inside modulus are split into hermitean 
and antihermitean parts for convenience. 

Note that neither of the bosonic terms \eq{bose1} nor \eq{bose2} are
individually $SO(4)\times SO(6)$ invariant.  However, upon adding them
one gets the bosonic part of the target theory action,
\beq
S_\text{boson}= \frac{1}{g_4^2}\int d^4\bfR\, \Tr\,\left[
\frac{1}{2} (D_\mu S_a)^2 +\frac{1}{4} V_{\mu\nu}^2 -\frac{1}{4}
[S_a,S_b]^2\right] +O(a)
\eeq
This should seem rather miraculous: in this theory the six  $S_a$ fields arise from
link variables transforming nontrivially under the lattice symmetries,
yet they become in scalars under the $SO(4)$ spacetime rotations,
transforming instead under the independent $SO(6)$ global $R$-symmetry that
emerges in the continuum.

We now turn to the fermionic part of the action
\beq
S_{\text{fermion}}= -\frac{\sqrt{2}}{g^2}\sum_\bfn \left(
\Delta_\bfn(\lambda,\mybar z_m, \psi^m) +  \Delta_\bfn(\xi_{mn},
z^m,\psi^n)+\frac{1}{8}\epsilon^{mnpqr}\Delta_\bfn(\xi_{mn},\mybar
z_p,\xi_{qr}) \right)\ ,\cr
\eqn{sfermionlat}\eeq
where  the three triangular
plaquette $\Delta$ 
functions were introduced in \eq{Deltadef}.  They  have the expansions
 \begin{equation}
\begin{aligned} 
-\sqrt{2}\, \Delta_\bfn(\lambda,\mybar z_m, \psi^m)=&
-\lambda \Bigl[(\bfe_m\cdot\bfnab)\psi^m -
\CE_{mn}[S_n,\psi^m] + i\CE_{m5}[S_6,\psi^m]\Bigr] +O(a)\ ,\\ 
-\sqrt{2} \Delta_\bfn(\xi_{mn}, z^m,\psi^n)=&  -
 \xi_{mn}\Bigl[(\bfe_m\cdot\bfnab)\psi^n +\CE_{mp}[S_p,\psi^n] +i\CE_{m5}
  [S_6,\psi^n]\Bigr] +O(a) \ ,\\
-\frac{\sqrt{2}}{8}\epsilon^{mnpqr}\Delta_\bfn(\xi_{mn},\mybar z_p,\xi_{qr})=&
-\frac{1}{8}\epsilon^{mnpqr}\xi_{mn}\Bigl[(\bfe_p\cdot\bfnab)\xi_{qr}
- \CE_{pt}[S_t,\xi_{qr}] \\&\qquad\qquad\qquad
+ i \CE_{p5} [S_6,\xi_{qr}]\Bigr] +O(a)\ .
\end{aligned}
\eqn{Deltalim}
\end{equation}
As before, repeated Latin indices are summed $1,\ldots,5$, while bold
dot products are between four-vectors. The vectors $\bfe$ and matrix
$\CE$ were defined in \eq{latvec} and \eq{cedef} respectively.

In order to make the $SO(4)\times SO(6)$ symmetry manifest, it is
convenient to reassemble the fermions in a sixteen component spinor $\tilde\omega$,
similar to \eq{omexp}, except for now $\tilde\omega(\bfR)$ is s
spinorial field in four dimensions:
\beq
\tilde\omega(\bfR) = \left(\lambda(\bfR) +   \xi_{mn}(\bfR)\,\frac{1}{2}
\hat A^m \hat A^n 
-\psi^m(\bfR) \,\frac{\epsilon_{mnpqr}}{24} \hat A^n \hat A^p \hat A^q \hat A^r
  \right)\nu_+\ .
\eqn{omexpii}
\eeq
Then by
making use of the expansions \eq{Deltalim} and extensive use of Mathematica, we can express the
continuum limit of the fermion action \eq{sfermionlat} in terms of $\tilde\omega$ as
\beq
S_\text{fermion}&=&\frac{1}{g_4^2}\int d^4\bfR\, \Tr\,\frac{1}{2}\,
\tilde\omega^T C\, \bigl(\Gamma_m\, \CE_{m\mu} D_\mu \tilde\omega +
  i\,\Gamma_{m+5}\, \CE_{mn}[S_n,\tilde \omega]+ i \, \Gamma_{m}\,
  \CE_{m5}[S_6,\tilde \omega] \bigr)\cr&&
\eqn{sfc}\eeq
where the $\Gamma_\alpha$ are $SO(10)$ gamma matrices in the basis
used to define the mother theory, \eq{act1}.We can define a new gamma matrix basis for $SO(10)$
\beq
\tilde \Gamma_\mu = \Gamma_m\, \CE_{m\mu}\ ,\qquad
\tilde \Gamma_{n+4} = \Gamma_{m+5}\, \CE_{mn}\ ,\qquad 
\tilde \Gamma_{10} = \Gamma_m\, \CE_{m5}\ .
\eeq
where $ \mu=1,\ldots4$, $n=1,\ldots 5$, and the sum $\sum_{m=1}^5$ is
implied in each of the above expressions. From the orthogonality of
the matrix $\CE$, it follows that
$\{\tilde\Gamma_\alpha,\tilde\Gamma_\beta\}=2\delta_{\alpha\beta}$, for $\alpha,\beta=1,\ldots,10$.
In the new basis the charge conjugation matrix is unchanged,
$\tilde C = C$.  Therefore, the above continuum limit of the lattice
fermion action \eq{sfc} may be written as
\beq
S_\text{fermion}&=&\frac{1}{g_4^2}\int d^4\bfR\  \frac{1}{2}\, \Tr\,\Bigl(
\tilde\omega^T \tilde C\, \tilde \Gamma_\mu D_\mu \tilde\omega +
i\,\tilde\omega^T \,\tilde C \,\tilde\Gamma_{4+a}\,[S_a,\tilde \omega]\Bigr)+O(a) \cr&&
\eeq
where the index $a$ is summed $1,\ldots,6$.  We see that to
leading order in $a$
this correctly reproduces the fermionic part of the action for $\CN=4$
SYM in four dimensions, as given in \eq{target4}. Note that the target
theory has a full $SO(6)$ chiral symmetry that naturally emerges in
the continuum, even though the symmetry does not exist on the
lattice.  In a sense, the $SO(6)\times SO(4)$ symmetry of our theory
comes about much in the same way as the $SO(4)\times SU(4)$ symmetry
that emerges in the continuum with conventional staggered fermions in
four dimensions, even though
independent flavor and spacetime rotations symmetries do not exist at
finite lattice spacing.

Although not evident from the above analysis where we expanded about
smooth fields, one can show that there are no boson or fermion doublers in 
the theory living at the edge of the Brillouin zone. We show this
explicitly for the bosons in Appendix~\ref{sec:ap2}, and it follows by
supersymmetry for the fermions as well.  We also refer the reader to an earlier paper where
we worked through a similar example in detail for both bosons and
fermions \cite{Cohen:2003aa}.

Our conclusion for this section is that our construction of the
$A_4^*$ lattice with explicit $\CQ=1$ supersymmetry does indeed give
$\CN=4$ SYM theory in four dimensions in the continuum limit, at tree
level. In the concluding section we will make several remarks about
the renormalization of this theory.  In the next section we construct
the $A_3^*$
lattice for $\CQ=16$ SYM in three  dimensions.

\section{The three dimensional lattice}
\label{sec:4}

\subsection{The target theory}
 The sixteen supercharge theory in  three dimensions  can be obtained 
by dimensional reduction of $\CN=1$ $U(k)$ SYM theory in 
ten dimensions. The action  
possess a global $SO(3) \times SO(7)  $ symmetry, where $SO(3)$ is  the
Euclidean counterpart of the Lorentz symmetry and $SO(7)$ is the $R$-symmetry
of the theory. Under this symmetry the gauge bosons transform as
  $(3,1)$,  the scalars  as  $(1, 7)$, and
the fermions as $(1,8) \oplus (1,8)$.
The action  has a form form  similar to that in    \Eq{target4}, with
the ranges of the indices changed appropriately, with $\mu,\nu$ running from $1 \ldots 3$; 
$a,b$ from  $1 \ldots 7$ and  $\tilde \Gamma_{4+a}$  replaced 
by  $\tilde \Gamma_{3+a}$. Furthermore,  $g_4^2$
is replaced by $g_3^2$, which has mass dimension equal to one.

The low energy  theory is believed to 
be an interacting  conformal field theory.   
The  gauge boson in three dimensions 
is dual to a compact scalar.  In the 
infrared limit of the theory (at scales well below $g_3^2$) 
this scalar is thought to decompactify, joining the other 
seven scalars to form the $(1,8)$ representation of an enhanced
$SO(8)$ $R$-symmetry \cite{Seiberg:1998ax}. 


\subsection {The mother theory in $SU(4) \times U(1) \times U(1)$ 
multiplets}

To create a three dimensional lattice from $\CQ= 16$ mother 
theory \eq{act1}, we orbifold by $Z_N^3$ where the three $Z_N$ transformations are 
determined  by the  three-vector ${\bf r}$ charges  
defined in \eq{rcharges} with $d=3$. In the case of the four
dimensional lattice analyzed in the previous section we saw that the
four dimensional $\bfr$ vectors generated the Cartan subalgebra of the
$SU(5)$ subgroup of the $SO(10)$ symmetry of the mother theory;  in the present case of a three dimensional lattice, the
3-vectors $\bfr$ generate the Cartan subalgebra of the $SU(4)$ subgroup of
$SO(10)$.  Consequently, the assignment of the fields of the mother theory onto
links and sites follows from their $SU(4)$ weights.  
It is convenient to decompose the variables of the mother theory under
$SU(4) \times U(1) \times U(1)$.  The two $U(1)$ generators may be
taken to be
\beq
Q_0\equiv\sum_{m=1}^5 q_m\ ,\qquad Q_1=q_5\ ,
\eqn{u12}
\eeq 
where the $q_m$ are defined in \eq{qdef}.

With this definition of the $U(1)$ charges, it is a simple matter to
figure out the decomposition of the ${\bf 10}$ of bosons and ${\bf
  16}$ of fermions of the mother theory under $SU(4)\times
U(1)_0\times U(1)_1$.  One finds for the bosons
\beq
v \sim{\bf 10} \ \longrightarrow\ \left( z \oplus t\right)\oplus \left(\mybar z \oplus \mybar t\right)
\sim \left( {\bf 4_{1,0}}\oplus{\bf 1_{1,1}}\right)\oplus \left({\bf \mybar  4_{-1,0}} 
\oplus {\bf 1_{-1,-1}}\right)\ ,
\eeq
where we have grouped together the variables that had been
irreducible $SU(5)$ representations on the four dimensional lattice
construction of the previous section. 

 For the fermions one has
\beq
\omega\sim {\bf 16} \ 
\longrightarrow&\ & \lambda  \oplus \left(  \xi \oplus \chi\right)\oplus \left(\psi \
 \oplus \alpha\right)\cr
&\sim&{\bf 1_{\frac{5}{2},\frac{1}{2} }} \oplus  \left( {\bf  {6}_{\frac{1}{2} ,\frac{1}{2} }\oplus {\bf\mybar {4}_{\frac{1}{2} ,-\frac{1}{2} }}
  }\right)
   \oplus  \left( {\bf 4_{-\frac{3}{2},-\frac{1}{2} }}\oplus
  {\bf {1}_{-\frac{3}{2},\frac{1}{2} }}\right) \ .
\eeq
This decomposition can be effected by means of the ladder operators
defined in  \eq{adef} \footnote{Note that in this section
  the Latin indices $m,n,p...$  take the values $1,\ldots,4$ and
  repeated indices are summed.}:
\beq
\omega = \left(\lambda +  \xi_{mn}\,\frac{1}{2}
\hat A^m \hat A^n + \chi_m \hat A^m \hat A^5
-\psi^m \,\frac{\epsilon_{mnpq}}{6} \hat A^n \hat A^p \hat A^q \hat A^5
- \alpha \frac{\epsilon_{mnpq}}{24} \hat A^m \hat A^n \hat A^p \hat A^q 
  \right)\nu_+\cr
\eqn{omexp3}
\eeq
where $\nu_+$ is the highest weight spinor defined in \eq{nudef},
carrying $Q_0=5/2$ and $Q_1=1/2$, and   $\hat A^m$ carries charges
$Q_0=-1$, $Q_1=0$, while $\hat A^5$ carries charges $Q_0=Q_1=-1$.

Note that the above
expansion of $\omega$ is the same as \eq{omexp} in the previous
section, with the substitutions
\beq
\xi_{m5}\to \chi_m\ ,\qquad
\psi^5 \to \alpha\ .
\eeq
Together with the substitutions $z^5\to t$ and $\mybar z^5\to\mybar t$,
 the action of the mother theory \eq{act2} may be written in terms of the
 $SU(4)\times U(1)\times U(1)$ multiplets as
\beq
S &=& \frac{1}{g^2} \Tr 
\Biggl[ 
\frac{1}{2} \left( [\mybar z_m, z^m] + 
[\mybar t, t]\right)^2 + 
  \bigl|[z^m, z^n]\bigr|^2 + 2 \bigl|[t, z^m]\bigr|^2 \cr
\qquad\qquad\qquad
&&+ {\sqrt 2}\Bigl\{ 
\lambda\left( [\mybar z_m, \psi^m] + [ \mybar t, \alpha]\right)
- \xi_{mn}[z^m, \psi^n] - \chi_m\left([ z^m, \alpha] -  [t, \psi^m]\right)  
\cr\qquad\qquad\qquad&&+
\half \epsilon^{mnpq}\xi_{mn}\left(
 [\mybar z_p, \chi_{q}] + \frac{1}{4}[\mybar t, \xi_{pq}]\right) 
 \Bigr\} 
\Biggr]
 \eqn{act2SU4}
\eeq
where  $SU(4)\times U(1) \times U(1) $ symmetry is manifest. 

Following our treatment of the four dimensional lattice, we make the correspondence between the $\bfr$
charges and the $SU(4)$ tensor notation  explicit by defining
the four vectors
\beq
\bfmu_1  =  \{1,0,0\}\ ,\quad
\bfmu_2  =  \{0,1,0\}\, \quad
\bfmu_3  =  \{0,0,1\}\, \quad
\bfmu_4  =  \{-1,-1,-1\}\ . 
\eqn{mudef3}\eeq
The  $\bfmu_m$ vectors   specify the $\bfr$ charge directly in terms 
of the $SU(4)$ tensor indices: for each variable the $\bfr$ charge is 
given by a sum of
$\bfmu_m$ for each
upper  $SU(4)$ index $m$, and $-\bfmu_m$ for each lower index
$m$. Thus, as seen in Table~\ref{tab2},   $z^m$ and $\psi^m$ have
$\bfr=\bfmu_m$; while $\mybar z_m$ and  $\chi_m$ have
$\bfr=-\bfmu_m$; $\xi_{mn}$ has $\bfr=-(\bfmu_m + \bfmu_n)$; while the
$SU(4)$ singlets
$\lambda$, $\alpha$, $t$, and $\mybar t $ each have $\bfr=0$ and
become site variables. 

\begin{table}[t]
\begin{tabular}[t]
{|r||r|r||r|r|r|r||rcl|}
\hline
& $Q_0$\ & $Q_1$\ & $q_1\ $ & $q_2\ $ & $q_3\ $ & $q_4\ $ & 
&\bfr&
\\ \hline
$z_1$ &  $\, \ 1$ &  $\, \ 0$ &$\,\ 1$ & $\, \ 0$ & $\, \ 0$ & $\, \ 0$   & 
$\{1,0,0\}$&=&$\ \ \bfmu_1$\\ 
$ z_2$&  $\, \ 1$ &  $\, \ 0$ &$\, \ 0$ & $\,\ 1$ & $\, \ 0$ & $\, \ 0$ & 
$\{0,1,0\}$&=&$\ \ \bfmu_2$\\
 $z_3$ &$\, \ 1$  &  $\, \ 0$ & $\, \ 0$ & $\, \ 0$ & $\,\ 1$ & $\, \ 0$ &
$\{0,0,1\}$&=&$\ \ \bfmu_3$\\
$z_4$ & $\, \ 1$  &  $\, \ 0$ & $\, \ 0$ & $\, \ 0$ & $\, \ 0$ & $\,\ 1$ &
$\{-1,-1,-1\}$&=&$\ \ \bfmu_4 $\\ \hline
$t$ &  $\,\ 1$ &  $\, \ 1$ & $\, \ 0$ & $\, \ 0$ & $\, \ 0$ & $\, \ 0$ &
$\{0,0,0 \}$&= &\,\ \ {\bf 0}\\ \hline
$\mybar z_1 $ & $-1$ &  $\, \ 0$ &  $\,\ -1$  & $\, \ 0$ & $\, \ 0$ & $\, \ 0$
&
$\{-1,0,0\}$&=&$-\bfmu_1$\\ 
$\mybar z_2$& $-1$  &  $\, \ 0$ & $\, \ 0$ & $\, \ -1$ & $\, \ 0$ & $\, \ 0$ &
$\{0,-1,0 \}$&=&$-\bfmu_2$ \\
$\mybar z_3$ &  $-1$ &  $\, \ 0$ &$\, \ 0$ & $\, \ 0$ &  $\,\ -1$ & $\, \ 0$ &
$\{0,0,-1\}$&=&$-\bfmu_3$ \\
$\mybar z_4$ &  $-1$ &  $\, \ 0$ &$\, \ 0$ & $\, \ 0$ & $\, \ 0$ & $-1$ &
$\{1,1,1\}$&=&$-\bfmu_4$ \\
\hline
$ \mybar t $ &  $-1$ &  $\, \ -1$ &$\, \ 0$ & $\, \ 0$ & $\, \ 0$ & $\, \ 0$ &
$\{0,0,0\}$& =&\,\ \ {\bf 0}\\   \hline   \hline 
$\lambda$ & \, \ $\frac{5}{2}$ & $\half $ & $\, \ \half$ & $\, \ \half$ & $\, \ \half$ &
$\, \ \half$ &
$\{0,0,0 \}$&=&\,\ \ {\bf 0}\\ \hline 
$\psi_1$ &  $-\frac{3}{2}$ & $-\half $ & $\, \ \half$ & $-\half$ & $-\half$ & $-\half$ &
$\{1,0,0\}$&=&$\ \ \bfmu_1$
\\
$\psi_2$  & $-\frac{3}{2}$ & $-\half $ & $-\half$ & $\, \ \half$ & $-\half$ & $-\half$ & 
$\{0,1,0\}$&=&$\ \ \bfmu_2$ \\ 
$\psi_3$& $-\frac{3}{2}$ & $-\half $ & $-\half$ & $-\half$ & $\, \ \half$ & $-\half$ &
$\{0,0,1\}$&=&$\ \ \bfmu_3$\\
$\psi_4$ & $-\frac{3}{2}$ & $-\half $ & $-\half$ & $-\half$ & $-\half$ & $\, \ \half$ & 
$\{-1,-1,-1\}$&=&$\ \ \bfmu_4 $\\ \hline
$\alpha$ & $-\frac{3}{2}$ & $\, \ \half $ & $-\half$ & $-\half$ & $-\half$ & $-\half$ &
$\{0,0,0\}$&= &\,\ \ {\bf 0} \\  \hline 
$\xi_{12}$& \, \ $\half $ & $\,\ \half $ & $-\half$ & $-\half$ & $\, \ \half$ & $\, \ \half$ & 
$\{-1,-1,0 \}$&=&$-\bfmu_1 - \bfmu_2$\\
$\xi_{13}$ & \, \ $\half $ & $\, \ \half $ & $-\half$ & $\, \ \half$ & $-\half$ & $\, \ \half$ &
$\{-1,0,-1\}$&=&$-\bfmu_1-\bfmu_3$\\
$\xi_{14}$& \, \ $\half $ & $\, \ \half $ & $-\half$ & $\, \ \half$ & $\, \ \half$ &
$-\half$ & 
$\{0 ,1 ,1\}$&=&$-\bfmu_1-\bfmu_4$ \\
$\xi_{23}$ & \, \ $\half $ & $\, \ \half $ & $\, \ \half$ & $-\half$ & $-\half$ &
$\, \ \half$ &
$\{0,-1,-1\}$&=&$-\bfmu_2-\bfmu_3$\\
$\xi_{24}$& \, \ $\half $ & $\, \ \half $ & $\, \ \half$ & $-\half$ & $\, \ \half$ & $-\half$ & 
$\{1,0,1\}$&=&$-\bfmu_2-\bfmu_4$  \\
$\xi_{34}$ & \, \ $\half $ & $\, \ \half $ & $\, \ \half$ & $\, \ \half$ & $-\half$ & $-\half$ 
&
$\{1,1,0\}$&=&$-\bfmu_3-\bfmu_4$ \\
\hline
$\chi_{1}$& \, \ $\half $ & $-\half $ & $-\half$ & $\, \ \half$ & $\, \ \half$ &
$\, \ \half$ &
$\{-1,0,0\}$&=&$-\bfmu_1$ \\
$\chi_{2}$& \, \ $\half $ & $-\half $ & $\, \ \half$ & $-\half$ & $\, \ \half$ &
$\, \ \half$ &
$\{0,-1,0\}$&=&$-\bfmu_2 $ \\ 
$\chi_{3}$ & \, \ $\half $ & $-\half $ & $\, \ \half$ & $\, \ \half$ & $-\half$ & $\, \ \half$ &
$\{0,0,-1\}$&=&$-\bfmu_3$\\
$\chi_{4}$ & \, \ $\half $ & $-\half $ & $\, \ \half$ & $\, \ \half$ & $\, \ \half$ &
$-\half$ &
$\{1,1,1\}$&=&$-\bfmu_4$\\  \hline
 \end{tabular} 
\caption{ The $Q_0$,$Q_1$ ${q_m}$ and $r_\mu = (q_\mu-q_4)$ charges of the
  bosonic variables $v$
  and fermionic variables $\omega$  of the $\CQ=16$ mother  theory under the
  $ SO(10)\supset SU(4)$ decomposition 
$v=10\to 1 \oplus 4 \oplus \mybar 4  \oplus 1= t \oplus z^m   
\oplus \mybar z_m  \oplus \mybar t $, and 
$\omega= 16 \to 1\oplus 4 \oplus  6 \oplus \mybar 4 \oplus 1  = 
\lambda \oplus \psi^m \oplus \xi_{mn}  \oplus \chi_{m} \oplus \alpha  $. 
 }
\label{tab2}
\end{table}

\subsection {  Manifest  $\CQ=2$ supersymmetry }
After the orbifold projection by $Z_N^3$ of the mother theory \eq{act2SU4}, 
two out of  sixteen supersymmetries remain intact, corresponding to
the two components 
of the supersymmetry parameter $\kappa$ in \eq{mothersusy} which have
${\bf r}=0$.  To render this exact lattice supersymmetry
explicit, we express
the mother theory in terms of the unbroken $\CQ=2$ supersymmetric
multiplets \footnote{Many of the features of $\CQ=2$ supersymmetry are
  described in appendix of reference \cite{Kaplan:2002wv}. See also
  the discussion in \cite{Witten:1993yc}.}.  

The two surviving supersymmetries are parametrized by the independent
Grassmann numbers $\eta$ and
$\mybar\eta$ (analogous to $\lambda$ and $\alpha$ in
Table~\ref{tab2} respectively) where 
\beq
\kappa= \left( \eta - \mybar \eta \frac{\epsilon_{mnpq}}{24} 
\hat A^m  \hat A^n  
\hat A^p  \hat A^q  \right) \nu_{+}\ .
\eqn{restrict2}
\eeq
In component form, the (on-shell) $\CQ=2$ transformations are given by
\beq
\delta z^m &=& \sqrt{2}\,  i\eta\, \psi^m\ ,\cr
\delta\mybar z_m &=&-\sqrt{2}\, i \mybar \eta\, \chi_m \ ,\cr
\delta t &=& \sqrt{2}\,i\eta\, \alpha + \sqrt{2}\, 
i \mybar \eta\, \lambda\ ,\cr \delta\mybar t &=&0 \ ,\cr
\delta\lambda &=& -i\eta\,([\mybar z_m, z^m] + [\mybar t, t] )  \cr
\delta\alpha &=& i\mybar \eta\,([\mybar z_m, z^m] - [\mybar t, t] )  \cr
\delta\psi^m &=&  - 2 i \mybar \eta\, [\mybar t,  z^m] \cr
\delta\chi_m &=&   2 i \eta\, [\mybar t, \mybar z_m] \cr
\delta\xi_{mn}&=& - 2 i \eta\, [ \mybar z_m, \mybar z_n] - 2 i \mybar \eta\,
(\half  \epsilon_{mnpq})[z^p, z^q].
\eqn{q2onshell}\eeq
In order to introduce supermultiplets, we need an off-shell formulation, thus
we introduce the auxiliary fields $d$, $G^k$, and $\mybar
G_k$, where $d$ is real and $k=1,\ldots,3$.  Together $G^k$ and $\mybar
G_k$ form the ${\bf 6}$ representation of the $SU(4)$ symmetry, but as
we shall see, the superfield formalism only keeps the $SU(3)\subset
SU(4)$ symmetry manifest, under which the transform as $3 \oplus\mybar
3$.
It is convenient then to define new combinations of the $\xi$ fermions
as
\beq
\tilde\xi^k = \half \epsilon^{ijk} \xi_{ij}\ ,\qquad \tilde \xi_k =
\xi_{k4}\ ,\qquad i,j,k=1,\ldots,3\ .
\eqn{xidef}\eeq
The considerations  
similar to those found in \cite{Cohen:2003qw} lead us to the off-shell transformations of 
fermions 
\beq
\delta\lambda &=& -i\eta\,( [\mybar t, t] + i d)  \cr
\delta\alpha &=& -i\mybar \eta\,(  [\mybar t, t] - i d)  \cr
\delta\psi^m &=&  - 2 i \mybar \eta\, [\mybar t,  z^m] \cr
\delta\chi_m &=&   2 i \eta\, [\mybar t, \mybar z_m] \cr
\delta\tilde\xi^k&=& + \sqrt 2 i \eta\, G^k - 2 i \mybar \eta\,[z^k, z^4] \cr
\delta\tilde\xi_{k}&=& - 2 i \eta\, [ \mybar z_k, \mybar z_4] +  \sqrt{2} i \mybar \eta\,
 \mybar G_k \cr
\delta d &=& -\sqrt{2}\,\eta\,[\mybar t,\alpha] +
\sqrt{2}\,\mybar\eta\,[\mybar t,\lambda]\ ,\cr
\delta  G^k &=& -2i\mybar\eta\, \left(  [\mybar t,\tilde\xi^k]
  - [z^k,\psi^4] - [\psi^k,z^4]\right)\
,\cr
\delta \mybar G_k &=&-2i\eta\,\left([\mybar
  t,\tilde\xi_{k}] + [\mybar z_k,\chi_4]+ [\chi_k,\mybar z_4]\right)\ . 
\eqn{q2offshell}\eeq 
The supersymmetry transformations of the bosons $z$, $\mybar z$,
$t$ and $\mybar t$ remain as in \eq{q2onshell}.

A superfield notation is now possible, by introducing Grassmann
superspace coordinates $\theta$ and $\mybar \theta$.  The supercharges
are defined to be
\beq
\delta= i\left(\eta Q + \mybar \eta \mybar Q\right)\ ,\qquad Q = \partial_\theta + \sqrt{2}\, \mybar\theta\,[\mybar t, \,\cdot\, ]\
,\qquad \mybar Q = \partial_{\mybar\theta} + \sqrt{2} \theta\,
[\mybar t,\,\cdot\,]\ ,
\eeq
which are nilpotent, but which satisfy the nontrivial  anticommutation
relation $\{Q,\mybar Q\}= 2\sqrt{2}\,[\mybar
t,\,\cdot\,]$.  In addition, we define the chiral derivatives
\beq
\CD=\partial_\theta - \sqrt{2}\, \mybar\theta\,[\mybar t, \,\cdot\, ]\
,\qquad 
\mybar\CD= \partial_{\mybar\theta} - \sqrt{2} \theta\,
[\mybar t,\,\cdot\,]\ .
\eeq
Superfields annihilated by $\CD$ or by $\mybar\CD$ will be called
``anti-chiral'' and ``chiral'' superfields respectively.

The supersymmetry transformations of the components are then realized
by introducing the bosonic vector superfield \footnote{Note that $\bf
  T$ is not real, as would be a vector superfield in Minkowski space;
  however on analytic continuation back to Minkowski space, ${\bf T}$
  does satisfy a reality condition, and so warrants the moniker.}
\beq
{\bf T} = t + \sqrt{2}\,\theta \alpha + \sqrt{2}\, \mybar\theta 
\lambda
+\sqrt{2}\,\theta \mybar \theta \, ( i d) \ ,
\eeq
and the bosonic chiral and anti-chiral superfields
\beq
{\bf \Upsilon} &=& \frac{1}{\sqrt{2}}\mybar{{\CD}}{\bf T}= \lambda - \theta 
( + [\mybar t, t] + i d ) -
\sqrt2 \theta \mybar{\theta}[\mybar{t}, \lambda  ]\ , \cr
\mybar{\bf \Upsilon}&=& \frac{1}{\sqrt{2}}{\CD}{\bf T}= \alpha +  
\mybar{\theta}( - [\mybar t, t] + i d) + \sqrt2 \theta \mybar{\theta}
[\mybar{t},\alpha]\ ,\cr
{\bf Z}^m &=& z^m +\sqrt{2}\,\theta \psi^m -\sqrt{2}\, \theta\mybar\theta
[\mybar t ,z^m]\ ,\cr
{\bf \mybar Z}_m &=& \mybar z_m -  \sqrt{2}\,\mybar\theta \chi_{m} +
\sqrt{2}\, \theta\mybar\theta [\mybar t ,\mybar z_m]\ ,
\eeq
satisfying
\beq
 {\CD} \mybar {\bf \Upsilon} = {\mybar {\CD}} {\bf \Upsilon}= {\mybar\CD}{\bf Z}^m={\CD}\mybar {\bf Z}_m=0\ .
\eeq



In addition we have six so-called ``Fermi multiplets'' \footnote{For a
  discussion of the $\CQ=2$ supersymmetric  multiplet structure, see the appendix of
  ref. \cite{Kaplan:2002wv}.}, denoted by
${\bf \Xi}^k$ and $\mybar {\bf  \Xi}_{k}$
with $k=1,\ldots,3$.  There expansions into components are
\beq 
{\bf \Xi}^k &=& \tilde\xi^k +\sqrt{2}\, \theta \; G^k
  - \sqrt{2}\, \theta \mybar
\theta [\mybar t,\tilde\xi^k] - 2 \mybar \theta \,{\bf E}^k\ ,\\
\mybar {\bf \Xi}_{k} &=& \tilde\xi_{k} +\sqrt{2}\, \mybar \theta 
 \; \mybar G_k + \sqrt{2}\, \theta \mybar
\theta [\mybar t,\tilde\xi_{k}] - 2 \theta \,\mybar {\bf E}_{k}
\eeq
where we have introduced the holomorphic  functions  ${\bf E}^k$ and
antiholomorphic function  ${\mybar {\bf E}}_{k}$ 
\beq
{\bf E}^k =  [{\bf Z}^k, {\bf Z}^4 ] 
\ ,\qquad
{\mybar {\bf E}}_{k} = [ {\mybar {\bf Z} }_k, {\mybar {\bf Z} }_4 ]\
.
\eeq
The fermi multiplets satisfy the identities
\beq
{\mybar {\CD}} {\bf \Xi}^k =
-2 {\bf E}^k\ , \qquad 
 { {\CD}} \mybar {\bf \Xi}_{k}= -2\mybar {\bf E}_{k}\ ,
\eqn{dxi}\eeq
which are consistent with the identities
\beq
 \mybar \CD {\bf E}^k= \CD \mybar {\bf E}_{k}=0\ .
\eeq


Interactions for the Fermi multiplets are included by introducing the
holomorphic functions
\beq
{\bf J}_k = \half\epsilon_{ijk}[{\bf Z}^i,{\bf Z}^j]\ ,\qquad
\mybar {\bf J}^{k} = \half \epsilon^{ijk}[\mybar {\bf Z}_i,\mybar {\bf Z}_j]
\eeq
These functions satisfy
\beq
\Tr {\bf J}_k {\bf E}^k  = \Tr \mybar {\bf J}^{k} {\mybar {\bf E}}_{k}
= 0\ 
\eeq
due to the cyclic properties of the trace.
It follows then from \eq{dxi} that
\beq 
\mybar \CD \Tr  ( {\bf \Xi}^k {\bf J}_k)= 
2\,\Tr ( {\bf E}^k {\bf J}_k)= 0\ , \qquad           
\CD\Tr ( \mybar {\bf \Xi}_{k} \mybar {\bf J}^{k})=2\Tr \mybar {\bf J}^{k}
{\mybar {\bf E}}_{k}=0\ ,
\eeq
which implies that $ \Tr  ( {\bf \Xi}^k {\bf J}_k)$ and $\Tr ( \mybar
{\bf \Xi}_{k} \mybar {\bf J}^{k})$ are chiral and anti-chiral
superfields respectively.

>From the transformation properties of the ${\bf T}$, ${\bf Z}^m$ and
$\mybar{\bf Z}_m$ superfields, one sees that supersymmetric invariants
can be constructed from the trace of the $\theta\mybar\theta$
component of a vector superfield, the $\theta$ component of a chiral
superfield, or the $\mybar\theta$ component of an anti-chiral superfield.

Using these superfields which we have constructed, it is now possible
to write down the  mother theory action  in 
manifestly  $\CQ=2$  supersymmetric form as
\begin{equation}
S=\frac{1}{g^2} \Tr \left[\int d\theta d\mybar\theta\,  \left(
    \half \mybar{\bf \Upsilon}{\bf \Upsilon} 
+\frac{1}{\sqrt{2}} \mybar { {\bf Z}}_{m}[{\bf T}, {{\bf Z}}^m]
-\half {\bf \Xi}^k
\mybar{\bf \Xi}_k  \right) \right.
 +  \left.
\,\int d\theta\,
\, {\bf \Xi}^k{\bf J}_k   
+\,\int d\mybar\theta\,\mybar {\bf \Xi}_{k} \mybar {\bf J}^{k}    
\right]\ ,
\eqn{act16q2}
\end{equation}
summing $m$ over $1,\ldots,4$ and $k$ over $1,\ldots,3$.
When written in component form, the above action contains the
auxiliary field interactions
\begin{equation}
S_{\rm{aux}}= \frac{1}{g^2} \Tr \left[ \frac{d^2}{2}+ id
[\mybar z_m, z^m] + 
\mybar G_k G^k + \frac{1}{\sqrt{2}}\epsilon_{ijk} [z^i, z^j] G^k + \frac{1}{\sqrt {2}} \epsilon^{ijk}
[\mybar z_i, \mybar z_j]\mybar G_k    \right],
\end{equation}
One can verify that after replacing the auxiliary fields by the
solutions to their equations of motion
\begin{equation}
id = [\mybar z_m, z^m], \qquad
G^k= -\frac{1}{\sqrt 2} \epsilon^{ijk}[\mybar z_i, \mybar z_j], \qquad 
 \mybar G_k = -\frac{1}{\sqrt 2} \epsilon_{ijk}[ z^i, z^j] \, . 
\end{equation}
and makes use of the
definitions \eq{xidef}, the above action \eq{act16q2}
correctly reproduces the mother theory as written in \eq{act2SU4}.

\subsection{ The $D=3$, $ \CQ=2$ lattice action and its symmetries}
The charges given in Table~\ref{tab2} make it easy to write down
the action of the lattice theory that results from the orbifold
projection. In component form, the result is
\begin{equation}
\begin{aligned} 
S =&
  \frac{1}{g^2}\sum_{\bfn} \Tr \biggl[  \half\Bigl(\sum_{m=1}^4 
\left(\mybar z_m(\bfn-\bfmu_m)
    z^m(\bfn-\bfmu_m) - z^m(\bfn)\mybar z_m(\bfn)\right) + [\mybar t(\bfn), 
 t(\bfn)]\Bigr)^2\\
&  +\sum_{m,n=1}^4
\Bigl\vert\, z^m(\bfn) z^n(\bfn + \bfmu_m)-  z^n(\bfn) z^m(\bfn + \bfmu_n)
\Bigr\vert^2 +  2\sum_{n=1}^4
\Bigl\vert\, t(\bfn) z^n(\bfn)-  z^n(\bfn) t(\bfn + \bfmu_n)
\Bigr\vert^2 
  \\ & 
-\sqrt{2}\Bigl( \Delta_\bfn(\lambda,\mybar z_m, \psi^m)+ 
\Delta_\bfn(\lambda,\mybar t, \alpha) -
  \Delta_\bfn(\xi_{mn}, z^m,\psi^n) -  \Delta_\bfn(\chi_{m}, z^m,\alpha) +
\Delta_\bfn(\chi_{m}, t,\psi^m)  
 \\ & 
  +\half \epsilon^{mnpq}\Delta_\bfn(\xi_{mn},\mybar z_p,\chi_{q})
 + \frac{1}{8}\epsilon^{mnpq}\Delta_\bfn(\xi_{mn},\mybar t, \xi_{pq})
\Bigr) 
\biggr] 
\\ &
\eqn{d3lat}
\end{aligned} 
\end{equation} 
The function $\Delta$ is the same as defined in \eq{Deltadef} for the
four dimensional lattice.

We can also express the lattice action in a manifestly 
\CQ=2 supersymmetric form. The $\CQ=2$ superfields on the lattice 
may be written as 
\begin{equation}
\begin{aligned}
{\bf Z}^m(\bfn) &= z^m(\bfn) +\sqrt{2}\,\theta \psi^m(\bfn) -
\sqrt{2}\, \theta\mybar\theta
\Bigl(\mybar t({\bfn}) z^m(\bfn)- z^{m}(\bfn) \mybar t(\bfn + \bfmu_m) \Bigr)\ ,\\
\mybar {\bf Z}_{m}(\bfn) &= \mybar z_{m}(\bfn)
+\sqrt{2}\,\mybar\theta \chi_{m}(\bfn) +\sqrt{2}\, \theta\mybar\theta
\Bigl(\mybar t(\bfn+\bfmu_m) \mybar z_{m}(\bfn) -\mybar z_m(\bfn) \mybar 
t(\bfn)
\Bigr)   
\ , \\
{\bf \Xi}^k(\bfn) &= \tilde\xi^k(\bfn) +\sqrt{2}\, \theta
G^k(\bfn )   -\sqrt{2}\, \theta \mybar\theta
\Bigl(\mybar t(\bfn -\bfmu_4-\bfmu_k )\tilde\xi^k(\bfn) -\tilde\xi^k(\bfn) \mybar t(\bfn)
\Bigr) - 2 \mybar \theta {\bf E}^k(\bfn)
\ ,\\
\mybar{\bf \Xi}_{k}(\bfn) &= \tilde\xi_{k}(\bfn) -\sqrt{2}\,\mybar\theta 
\mybar G_k(\bfn)
  +\sqrt{2}\, \theta \mybar\theta
\Bigl(\mybar t(\bfn+ \bfmu_k + \bfmu_4) \tilde\xi_{k}( \bfn) - \tilde\xi_{k}(\bfn) 
\mybar t(\bfn) \Bigr) -2 \theta {\bf \mybar E}_{k}(\bfn) \ ,\\
{\bf T}(\bfn) &= t(\bfn) + \sqrt{2}\,\theta \alpha(\bfn) + \sqrt{2}\, \mybar
\theta \lambda (\bfn)
+\sqrt{2}\,\theta \mybar\theta 
+id (\bfn) \ ,\\
{\bf \Upsilon}(\bfn) &=  \lambda(\bfn) - 
\theta \Bigl(
+[\mybar t(\bfn), t(\bfn)] +i d (\bfn) \Bigr) 
-\sqrt2 \theta \mybar{\theta}[\mybar t (\bfn), \lambda (\bfn)]\ ,\\
\mybar {\bf \Upsilon}(\bfn)& = \alpha (\bfn) + \mybar{\theta} \Bigl(
-[\mybar t (\bfn), t (\bfn)] +i d(\bfn) \Bigr)
+ \sqrt2 \theta \mybar{\theta}[\mybar t (\bfn), \alpha (\bfn)]\ .
\end{aligned}
\eqn{sfields3d}
\end{equation}
 The ${\bf E}$  functions are given as
\beq
{\bf E}^k(\bfn) &=& \bfZ^{k}( \bfn - \bfmu_4 - \bfmu_k) {\bfZ}^{4}( \bfn-\bfmu_4) - 
  \bfZ^{4}( \bfn - \bfmu_4 - \bfmu_k) \bfZ^{k}( \bfn-\bfmu_k) \cr
{\bf \mybar E}_{k} (\bfn)&=& \mybar \bfZ_{k}( \bfn + \bfmu_4) \mybar \bfZ_{4}( \bfn) - 
 \mybar \bfZ_{4}( \bfn + \bfmu_k) \mybar \bfZ_{k}( \bfn)  \qquad
\eeq
The ${\bf J}$ functions can be written as 
\beq
{\bf J}_k(\bfn)&=& \half \epsilon_{ijk}\left 
( \bfZ^{i}( \bfn) \bfZ^{j}( \bfn+ \bfmu_i) - 
  \bfZ^{j}( \bfn ) \bfZ^{i}( \bfn +\bfmu_j)\right)  \cr
{\bf \mybar J}^{k}( \bfn)&=& \half \epsilon^{ijk} \left 
(\mybar \bfZ_{i}( \bfn - \bfmu_i) \mybar \bfZ_{j}( \bfn+ \bfmu_k + \bfmu_4 ) - 
 \mybar \bfZ_{j}( \bfn - \bfmu_j) \mybar \bfZ_{i}( \bfn + \bfmu_k + \bfmu_4 )\right)
\eeq


The lattice action, which is in fact orbifold projection of the  mother 
theory action in \Eq{act16q2}, may be written in terms of these 
lattice superfields as
\begin{equation}
\begin{aligned}
S=\frac{1}{g^2} \sum_{\bfn}\Tr &\Biggl[\int d\theta d\mybar\theta\,  \biggl(
    \half \mybar{\bf \Upsilon}({\bfn}){\bf \Upsilon}({\bfn})  -\half\,{\bf\Xi}^{k}(\bfn)
\mybar{\bf \Xi}_{k}(\bfn)   \cr 
&\qquad\qquad
+\frac{1}{\sqrt{2}}{\bft}({\bfn})\left({\bfZ}^{m}({\bfn}){{\mybar\bfZ}}_{m}(\bfn) - 
{{\mybar\bfZ}}_{m}(\bfn-
    \bfmu_m) {\bfZ}^{m}({\bfn-\bfmu_m})\right)\biggl)\\ &
  + \int d\theta\, {\bf \Xi}^k({\bfn }) {\bf J}_{k}(\bfn)  +
   \int d\mybar\theta \,\mybar{\bf \Xi}_k({\bfn }) \mybar{\bf
    J}^{k}(\bfn)   
\Biggr]
\end{aligned}
\eqn{lact3}
\end{equation}
By eliminating the auxiliary fields by their 
 equations of motion,  
\begin{eqnarray}
G^k(\bfn) &=& -\frac{1}{\sqrt{2}}\epsilon^{ijk}\,\,( \mybar z_{i}(\bfn+\bfmu_j)\mybar
z_{j}(\bfn) - \mybar z_{j}( \bfn + \bfmu_i) \mybar z_{i}( \bfn))  \nonumber \\
\mybar G_k(\bfn) &=& -\frac{1}{\sqrt{2}}\epsilon_{ijk}\,\, ( z^{i}(\bfn- \bfmu_i - \bfmu_j )   
z^{j}(\bfn - \bfmu_j) 
-  z^{j}( \bfn - \bfmu_i - \bfmu_j ) z^{i}( \bfn- \bfmu_i ))    \nonumber \\
id (\bfn) &=&
\; ( \mybar z_{m}(\bfn-\bfmu_m)  z^{m}(\bfn-\bfmu_m) -
z^{m}(\bfn)\mybar z_{m}(\bfn )
\end{eqnarray}
one can show that the action \eq{lact3} is equivalent to the lattice action 
\eq{d3lat} given  in component form.  

\subsection {The continuum limit for d=3 lattice: tree level} 
\subsubsection{A cubic lattice}
The expansion of link fields  around the configuration 
\beq
z^{m}( \bfn)= \mybar z_{m}( \bfn) = \frac{1}{a {\sqrt 2}} {\bf 1}_k
,\,\,\, m= 1 \ldots 3,  \qquad z^{4}( \bfn )= \mybar z_{4}( \bfn) =0
\eqn{cub}\eeq
generates  a cubic spacetime 
lattice. The superfields $\bf Z_4$   and $\mybar {\bf Z}_4$
are residing  on the body diagonal  and ${\bf \Xi}^{k} \ \text{and}
\ {\bf \mybar \Xi}_{k}$ are residing on the face 
diagonals of the  cube (see Fig.~\ref{fig:lat3dcube}). The
symmetry of the lattice is 
$S_3 \ltimes Z_2$, with twelve group elements.   
  
\subsubsection{The $A_3^{*}$  (bcc)  lattice}
We expand the action about the most symmetric solution of moduli equations
\beq
z_{m}( \bfn)= \mybar z^{m}( \bfn) = \frac{1}{a {\sqrt 2}} {\bf 1}_k
,\,\,\, m= 1 \ldots 4,
\eqn{s4}
\eeq
which treats all bosonic link fields on equal footing and preserves 
the octahedral symmetry of the action.  The symmetry group is 
$S_4 \ltimes Z_2$, where $S_4$ corresponds to the permutation of 
$SU(4)$ indices  and $Z_2$  is the charge conjugation symmetry 
 swapping chiral and antichiral multiplets.  

Similar to the four dimensional example, we introduce four three dimensional 
vectors to relate the point $\bfn$ to a spacetime point. These lattice 
vectors can be chosen as 
\beq
    {\bf e_1} &=& (\frac{1}{\sqrt 2},  \frac{1}{\sqrt 6}, 
\frac{1}{\sqrt{12}})  \cr
{\bf e_2} &=& (-\frac{1}{\sqrt 2},  \frac{1}{\sqrt 6},
\frac{1}{\sqrt{12}} )  \cr
{\bf e_3} &=& (0,  -\frac{2}{\sqrt 6},
\frac{1}{\sqrt{12}})  \cr
{\bf e_4} &=& (0, 0,
-\frac{3}{\sqrt{12}}).
\eqn{latvec3}
\eeq 
These vectors  are the $SU(4)$ weights of the $\bf 4$ representation, 
and they  form a three simplex (tetrahedron) in three dimensions.
They satisfy the relations
\beq
\sum_{m=1}^{4} {\bf e}_m = 0, \qquad
{\bf e}_m\cdot {\bf e}_n= \delta_{mn}- \fourth \qquad  
\sum_{m=1}^{4} ({\bf e}_m)_{\mu} ({\bf e}_m)_{\nu}  = \delta_{\mu \nu}
\eeq 
 The matrix 
${\bf e}_m\cdot{\bf e}_n$ is the Gram matrix of $A_3^{*}$ \cite{Conway:1991},
also known as body-centered cubic (bcc) lattice. 

\DOUBLEFIGURE[t]{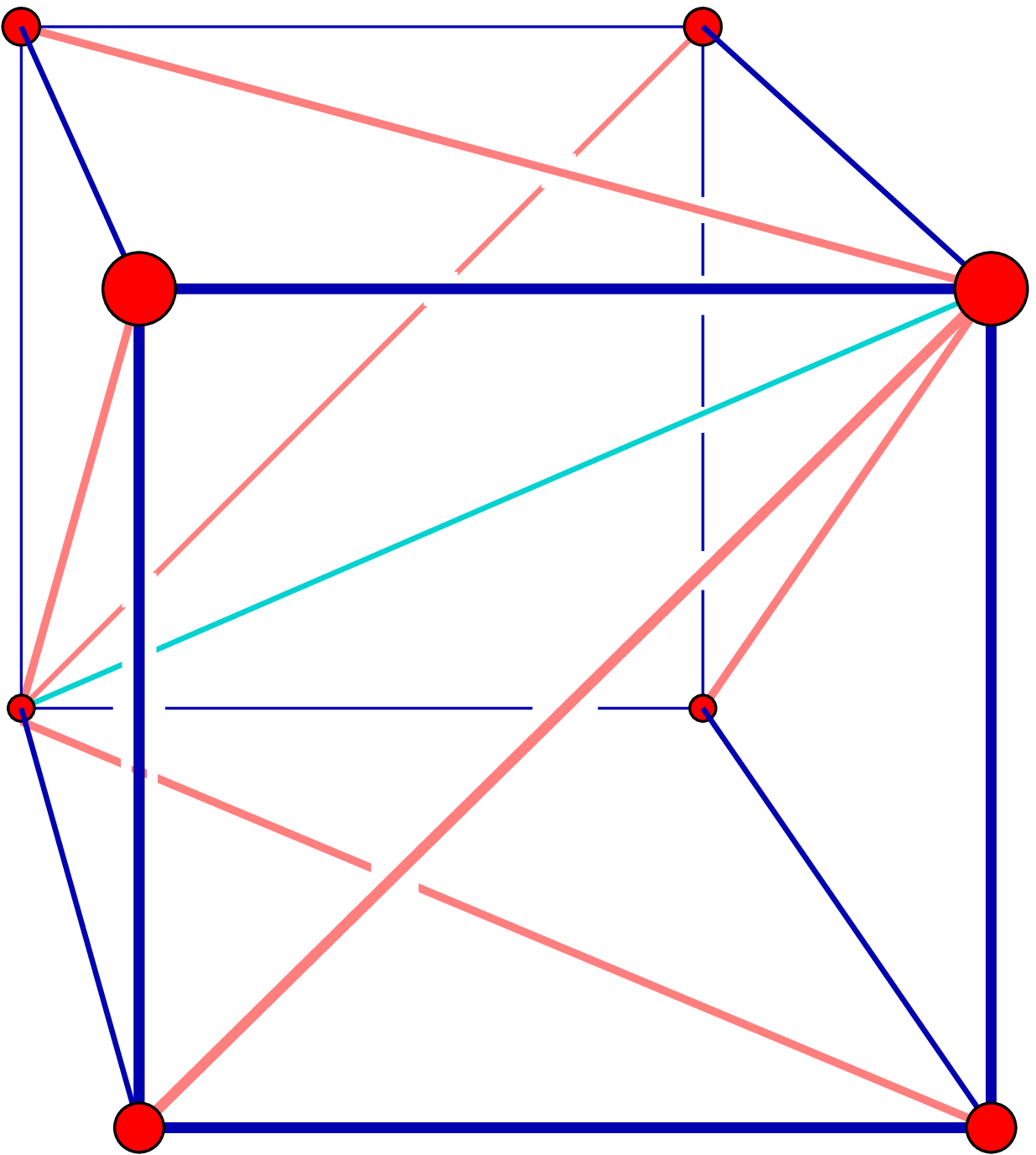,width=4.5cm}{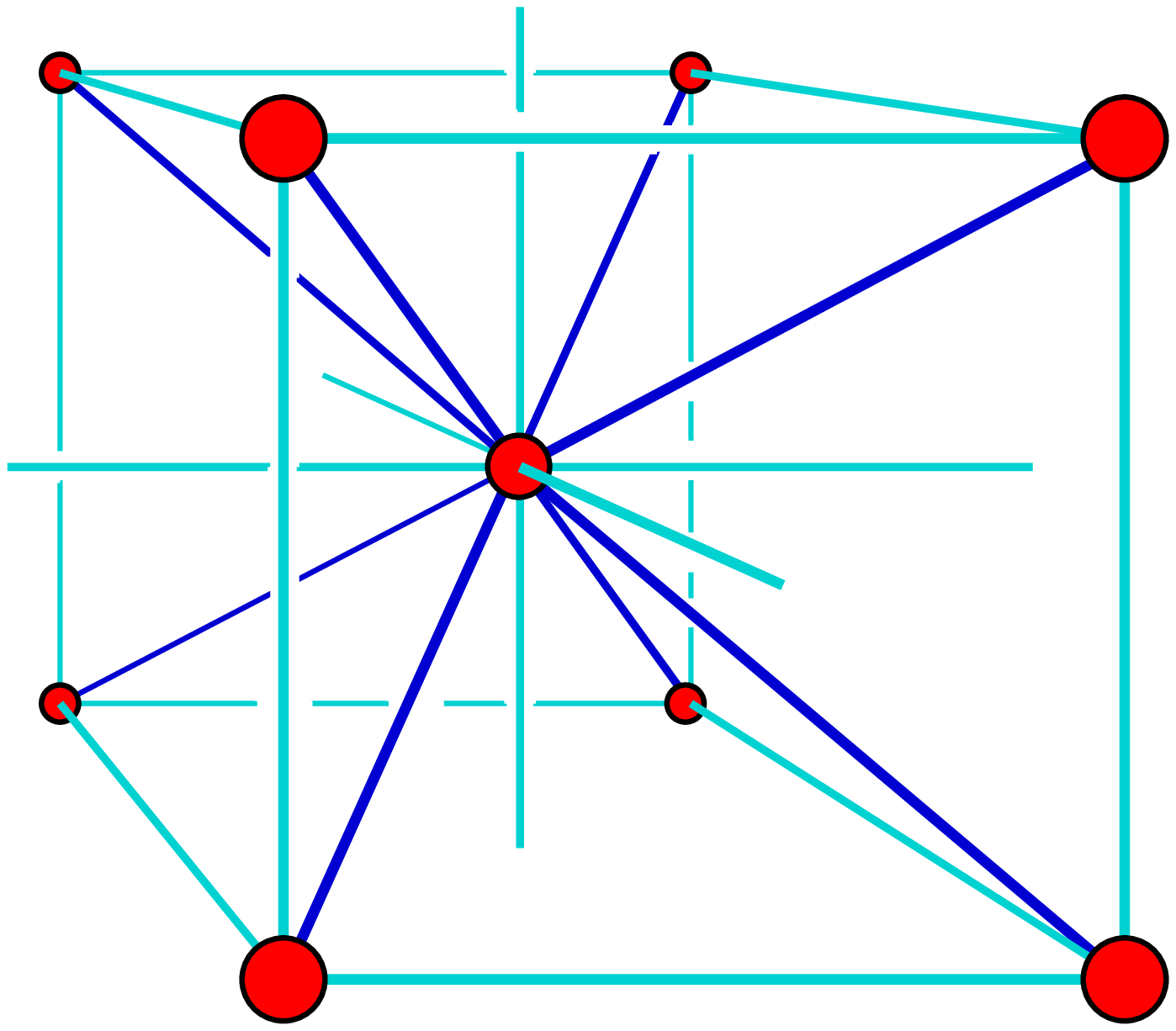,width=6cm}
{\sl The cubic lattice, corresponding to the trajectory in moduli space
  given in eq.~(4.34).\label{fig:lat3dcube}} 
{\sl  The $A_3^*$ lattice corresponding to the trajectory eq.~(4.35). The 
 eight nearest neighbor links (dark blue)
  emanating from the central site may be associated with the eight
 3-vectors $\pm \bfe_m$.  
 The six second-nearest-neighbor links (light blue) correspond to the six
  3-vectors $\pm \bfw_k$ in eq.~(4.39).\label{fig:lat3dBCC}}

The site $\bfn$ is identified with the spacetime location 
\beq
\bfR=a\sum_{\nu=1}^{3}({\bfmu}_{\nu}\cdot \bfn){\bf e}_{\nu} 
\eqn{ra3}
\eeq
and a lattice displacement of one unit in direction ${\bfmu}_m$ corresponds
to a spacetime translation $a{\bf e}_{m}$. It is easy to see that 
each of the four links occupied by four $z^m$ variables has length 
$|a{\bf e}_{m}|=\sqrt{\frac {3}{4}}\,a$, unlike the case of the less symmetric 
cubic lattice 
where  $z^4$ resides on a link $\sqrt 3 $ times longer then the ones occupied
by  the three $z^m$.

A picture of the lattice is shown in Fig.~\ref{fig:lat3dBCC}. The  dark
blue links between nearest neighbor sites correspond to the  eight
  3-vectors $\pm \bfe_m$. 
It is helpful to also define the three orthonormal vectors 
\beq
\bfw_k \equiv \bfe_4 + \bfe_k\ ,\qquad \bfw_i\cdot\bfw_j=\delta_{ij}\
,\qquad i,j,k=1,\ldots,3\ .
\eqn{w3Ddef}
\eeq
The six $\pm\bfw_k$ vectors correspond to the light blue links between
second nearest neighbors in Fig.~\ref{fig:lat3dBCC}.  
The $\bfZ^m$ and
  $\mybar\bfZ_m$ superfields reside on the dark blue links; the $\bfXi^k$ and $\mybar\bfXi_k$ superfields live on
the light blue links, and the $\bf T$ and $\bf\Upsilon$ superfields
live on the sites.  However one should note that the superfields are
not completely local, and contain terms looking like the square root
of a plaquette.

In order to relate the lattice fields  to continuum fields,  we expand 
the lattice action around the point \eq{s4}.  However, the structure of the
lattice is dictated by $SU(4) \times U(1) \times U(1)$ and the continuum 
fields transform under $SO(3)\times SO(7)$.  To make the connection between
lattice and continuum fields clear, 
we introduce a $4 \times 4$  real orthogonal matrix  
\beq
\CE = \begin{pmatrix}
 \ \ \frac{1}{\sqrt{2}} &  \ \ \frac{1}{\sqrt{6}} & 
\ \ \frac{1}{\sqrt{12}} &  \ \ \frac{1}{2} \cr
-\frac{1}{\sqrt 2} &  \ \ \frac{1}{\sqrt{6}} & \ \ \frac{1}{\sqrt{12}} & 
\ \  \frac{1}{2}   \cr
\ \ 0 &  -\frac{2}{\sqrt{6}} &  \ \ \frac{1}{\sqrt{12}} & 
\ \  \frac{1}{2}  \cr
\ \ 0 & \ \ 0 & -\frac{3}{\sqrt{12}} &  \ \ \frac{1}{2}  
\end{pmatrix} = \left(\CE^T\right)^{-1}
\eqn{cedef3}\eeq
Note that  $\CE_{m\mu}$, with $\mu=1,\ldots 3$, are the components of the vectors
$\bfe_m$ of \eq{latvec3}. This matrix has the property that
\beq
\left(\sum_{m=1}^{4} \bfe_m \,\CE_{mn}\right)_\mu
=\begin{cases}\delta_{ n\mu} & n=1,\ldots,3 \cr 0 & n=4 \ .\end{cases}
\eqn{ceprop3}
\eeq
which serves as a bridge between the $SU(4)$ tensors of the lattice
construction, and the $SO(3)$ representations of the continuum theory.
In terms of this matrix we then define the
expansion of $z^m$ about the point in moduli space \eq{s4} to be
\beq
z^m = \frac{1}{a\sqrt{2}} + \sum_{m=1}^4 \CE_{mn} \Phi_n =
\frac{1}{a\sqrt{2}} + \sum_{\mu=1}^3 (\bfe_m)_\mu \Phi_\mu +
\frac{1}{2} \Phi_4\ ,
\eqn{zexp3}\eeq
with
\beq
\Phi_\mu \equiv \left(\frac{S_\mu + i V_\mu}{\sqrt{2}}\right)\ ,\quad
\mu=1,\ldots,3\ ,\qquad \Phi_4 \equiv \left(\frac{S_4 + i
    S_5}{\sqrt{2}} \right)  \ , \qquad t \equiv \left(\frac{S_6 + i
    S_7}{\sqrt{2}} \right) \ \ 
\eeq
where $V_\mu$ and $S_i$ are hermitean $k\times k$ matrices, corresponding to 
gauge fields and scalars of the continuum theory. 

We now will expand the action \eq{d3lat} to leading order in powers of
the lattice spacing $a$, with the goal to show the equivalence in the
continuum limit  at tree
level between our lattice action and the target theory action.  
Since the Jacobian of the transformation
between lattice coordinates $\bfn$ and spacetime coordinates $\bfR$ in
\eq{ra3} equals $a^3/2$, we first must rescale our coupling $g$ such
that
\beq
\frac{1}{g^2} = \frac{a^3}{2\,g_3^2}  \ ,\qquad
\lim_{a\to 0} \frac{1}{g^2} \sum_{\bfn} = \frac{1}{g_3^2} \int d^3\bfR\ .
\eeq

The analysis of the bosonic action of the lattice theory follows 
similarly to \ref{sec:3e}. 
None of the three types of  bosonic terms  are
individually $SO(3)\times SO(7)$ invariant.  
 However, upon adding them
one gets the bosonic part of the target theory action,
\beq
S_\text{boson}= \frac{1}{g_3^2}\int d^3\bfR\, \Tr\,\left[
\frac{1}{2} (D_\mu S_a)^2 +\frac{1}{4} V_{\mu\nu}^2 -\frac{1}{4}
[S_a,S_b]^2\right] +O(a)\ ,
\eeq
where in this section the indices $a,b = 1,\ldots,7$ are $SO(7)$
indices, while $\mu,\nu=1,\ldots,3$ are spacetime indices.
Notice that  in this theory, two out of seven scalar arises from the site
fields, whereas the other 
five  scalar fields 
arise from the 
link variables transforming nontrivially under the lattice symmetries,
yet they become in scalars under the $SO(3)$ spacetime rotations,
transforming instead under the independent $SO(7)$ global $R$-symmetry that
emerges in the continuum.

Similar to the analysis of fermionic terms in \S~\ref{sec:3e}, 
 we can express the continuum limit of the fermion action 
\eq{d3lat} in terms of $\tilde\omega$ as
\beq
S_\text{f}&=&\frac{1}{g_3^2}\int d^3\bfR\, \Tr\,\frac{1}{2}\,
\tilde\omega^T \CC\, \bigl(\Gamma_m\, \CE_{m\mu} D_\mu \tilde\omega +
  i\,\Gamma_{m+5}\, \CE_{mn}[S_n,\tilde \omega]+ i  \Gamma_{m}\,
  \CE_{m4}[S_5,\tilde \omega]  \nonumber \\  \cr
&&+ i  \Gamma_{10}\,
  [S_6,\tilde \omega]  + i  \Gamma_{5}\,
  [S_7,\tilde \omega]  \bigr)\cr&&
\eeq
where the $\Gamma_\alpha$ are $SO(10)$ gamma matrices in the basis
used to define the mother theory, \eq{act1}. Since $\CE$ is an
orthogonal matrix, we can define a new gamma matrix basis for $SO(10)$
\beq
\tilde \Gamma_{\mu} = \Gamma_m\, \CE_{m\mu}\ ,\ \  \tilde \Gamma_{n+3} =
\Gamma_{m+5}\, \CE_{mn}\ ,\ \  \tilde \Gamma_{8}= \Gamma_m\, \CE_{m4} 
\ ,\ \    \tilde \Gamma_{9} =
\Gamma_{10} \ ,\ \  \tilde \Gamma_{10} =
\Gamma_{5}\,  \ .
\eeq
In the new basis the charge conjugation matrix is unchanged,
$\tilde\CC = \CC$, given that we specified in \eq{gamsym} that the
$\Gamma_m$ be antisymmetric for $m=1,\ldots,5$ and symmetric for
$m=6,\ldots,10$.  Therefore, the above continuum limit of the lattice
fermion action may be written as
\beq
S_\text{fermion}&=&\frac{1}{g_3^2}\int d^3\bfR\  \frac{1}{2}\, \Tr\,\Bigl(
\tilde\omega^T \tilde\CC\, \tilde \Gamma_\mu D_\mu \tilde\omega +
i\,\tilde\omega^T \,\tilde\CC \,\tilde\Gamma_{3+i}\,[S_i,\tilde 
\omega]\Bigr)+O(a) \cr&&
\eeq
where the index $\mu$ is over $1 \ldots 3$ and  $i$ is over $1 \ldots 7$. 
The chiral 
symmetry of the theory, $SO(7)$, which does not exist for any finite lattice 
spacing,  emerged naturally in the continuum.    

In conclusion, our construction of $A^{*}_3$ lattice with $\CQ=2$
supersymmetry  correctly reproduce the sixteen supercharge  ($\CN=8$) SYM
theory  in $d=3$ dimensions.  The discrete and continuous symmetries on the
lattice, $S_4 \ltimes Z_2 \times (U(1) \times U(1))$ enhances to 
$SO(3)\times  SO(7)$ symmetry in the continuum.    

\section {The two dimensional lattice}
\label{sec:5} 

\subsection{The mother theory with manifest $\CQ=4$ supersymmetry in
  $SU(3)\times U(1)\times SO(4)$ multiplets}
\label{sec:5a}

We now turn to the  sixteen supercharge target theory in two
dimensions, also known as $\CN=(8,8)$ supersymmetry, which possesses
an $SO(8)$ $R$-symmetry. In this case the
lattice possesses four exact supercharges, and the multiplet structure
is identical to that of the familiar $\CN=1$ supersymmetric gauge
theories in four dimensions, and we can use superfields reduced from
four dimensions to describe the theory.  Here we give an abbreviated version
of the analysis, trusting that familiarity with the previous two sections of this
paper will make it straight forward to fill in the missing details.

To create a two dimensional lattice,  we orbifold the mother theory \eq{act1} by a
$Z_N\times Z_N$ symmetry.  The two dimensional $\bfr$ charges generate
the Cartan algebra of an $SU(3)$ subgroup of $SO(10)$ embedded in the
natural way along the chain $SO(10)\to SO(4)\times SO(6) \to
SO(4)\times SU(3)\times U(1)$.  The $SO(4)\times U(1)\simeq
SU(2)\times SU(2)\times U(1)$ symmetry will remain exact on the
lattice, while the $SU(3)$ symmetry is broken to $U(1)\times U(1)$,
by the orbifold projection, with fields assigned to links and sites according to their $SU(3)$
weights.

The ten bosons and the sixteen fermions of the mother theory decompose
under the  $SU(3)\times SU(2)\times SU(2)\times U(1)$ subgroup of
$SO(10)$ as
\beq
v \sim{\bf 10} \ \longrightarrow\  z \oplus \mybar z \oplus  
\tilde v
\sim ({\bf 3,1,1})_{\bf 1} \oplus  ({\bf \mybar 3,1,1})_{- {\bf 1}} 
\oplus  ({\bf 1,2,2})_{\bf 0} \ .
\eeq
\beq
\omega \sim {\bf 16} \ 
\longrightarrow\   \psi \oplus \mybar \psi  \oplus \lambda   \oplus  
\mybar  \lambda \sim 
({\bf 3,2,1})_{-\frac {\bf 1}{ {\bf 2}}} \oplus 
({\bf \mybar 3,1,2})_{\frac {\bf 1}{ {\bf 2}}}
\oplus  ({\bf 1,2,1})_{\frac {\bf 3}{ {\bf 2}}}
\oplus  ({\bf 1,1,2})_{-\frac {\bf 3}{ {\bf 2}}}\ \ 
\eeq
The fermions are now doublets under the $SU(2)\times SU(2)$ symmetry,
and we adopt the conventions of Wess and Bagger \cite{Wess:1992cp}
adapted for Euclidean spacetime (see Appendix~\ref{sec:ap3}). For a
more explicit discussion of the above decomposition, see Appendix~\ref{sec:ap4}.

 After
orbifolding, the location on the lattice of 
the above variables is determined as before by the $\bfr$ charges, as
given in  Table~\ref{tab3}.  We see that $\tilde v$ and $\lambda$ are
site variables, while $z^m$, $\psi^m$, $\mybar z_m$ and $\mybar
\psi_m$ are link variables, where the links are designated by $\bfr$
equaling one of the three vectors
 \beq
\bfmu_1 &=& \{1,0\}\ ,\cr
\bfmu_2 &=& \{0,1\}\, \cr
\bfmu_3 &=& \{-1,-1\}\ .
\eqn{mudef2}\eeq

\begin{table}[t]
\begin{tabular}[t]
{|r||l||r|r|r||rcl|}
\hline
& $SU(2) \times SU(2) \times U(1) $\ & $q_1\ $ & $q_2\ $ & $q_3\ $& &\bfr&
\\ \hline
$z^1$ &  \qquad\qquad $ ({\bf 1,1})_{\bf 1}$ &$\,\ 1$ & $\, \ 0$ & $\, \ 0$ &  
$\{1,0\}$&=&$\ \ \bfmu_1$\\ 
$ z^2$& \qquad\qquad  $({\bf 1,1})_{\bf 1}  $ &$\, \ 0$ & $\,\ 1$ & $\, \ 0$ & 
$\{0,1\}$&=&$\ \ \bfmu_2$\\
 $z^3$ & \qquad\qquad$ ({\bf 1,1})_{\bf 1}$  & $\, \ 0$ & $\, \ 0$ & $\,\ 1$ & 
$\{-1,-1 \}$&=&$\ \ \bfmu_3$\\ \hline
$\mybar z_1 $ &  \qquad\qquad$ ({\bf 1,1})_{\bf -1} $ &  $\,\ -1$  & $\, \ 0$ & $\, \ 0$ &
$\{-1,0\}$&=&$-\bfmu_1$\\ 
$\mybar z_2$&  \qquad\qquad$ ({\bf 1,1})_{\bf -1}  $  & $\, \ 0$ & $\, \ -1$ & $\, \ 0$ & 
$\{0,-1 \}$&=&$-\bfmu_2$ \\
$\mybar z_3$ &  \qquad\qquad $({\bf 1,1})_{\bf -1}  $ &$\, \ 0$ & $\, \ 0$ &  $\,\ -1$ & 
$\{1,1\}$&=&$-\bfmu_3$ \\ \hline
%
%
$\tilde v$
    & \qquad\qquad $ ({\bf 2,2})_{\bf 0}$  & $\, \ 0$ & $\, \ 0$ & $\, \ 0$ &
$\{0,0\}$ &=&\ \ \,{\bf 0}\\ 

\hline   \hline 
$\lambda $ &  \qquad\qquad$ ({\bf 2,1})_{\bf \frac{3}{2}}$ & $\ \half$ & $\ \half$ & $\ \half$ &
$\{0,0 \}$&=&\ \ \,{\bf 0}\\ 
$\mybar \lambda$ &  \qquad\qquad$({\bf 1,2})_{\bf -\frac{3}{2}}  $ & $-\half$ & $-\half$ & $-\half$ &
$\{0,0\}$& =& \ \ \,{\bf 0}\\  \hline 
$\psi^1 $ &  \qquad\qquad $ ({\bf 2,1})_{\bf -\frac{1}{2}} $ & $\ \half$ & $-\half$ & $-\half$ & 
$\{1,0\}$&=&$\ \ \bfmu_1$
\\
$\psi^2 $  & \qquad\qquad $({\bf 2,1})_{\bf -\frac{1}{2}} $ & $-\half$ & $\ \half$ & $-\half$ & 
$\{0,1\}$&=&$\ \ \bfmu_2$ \\ 
$\psi^3 $& \qquad\qquad $({\bf 2,1})_{\bf -\frac{1}{2}} $ & $-\half$ & $-\half$ & $\ \half$ & 
$\{-1,-1\}$&=&$\ \ \bfmu_3$\\ \hline
$\mybar \psi_{1}$& \qquad\qquad  $ ({\bf 1,2})_{\bf \frac{1}{2}}$ & $-\half$ & $\ \half$ & $\ \half$ &
$\{-1,0 \}$&=&$-\bfmu_1$ \\
$\mybar \psi_{2}$&  \qquad\qquad$ ({\bf 1,2})_{\bf \frac{1}{2}} $ & $\ \half$ & $-\half$ & $\ \half$ & 
$\{0,-1\}$&=&$ -\bfmu_2$  \\
$\mybar \psi_{3}$ &  \qquad\qquad$({\bf 1,2})_{\bf \frac{1}{2}}  $ & $\ \half$ & $\ \half$ & $-\half$ &
$\{1,1\}$&=&$-\bfmu_3$ \\  
\hline
 \end{tabular} 
\caption{ The $SU(2)\times SU(2)\times U(1)$,  ${q_m}$ and $r_\mu = (q_\mu-q_3)$ representations of the
  bosonic variables $v$
  and fermionic variables $\omega$   
 of the $\CQ=16$ mother  theory under the
  $ SO(10)\supset SU(3)\times SU(2) \times SU(2) \times U(1) $ decomposition.
}

\label{tab3}
\end{table}

The mother theory may be most easily expressed in a manifestly $\CQ=4$
supersymmetric form by writing the $\CN=4$ SYM in four
dimensions using $\CN=1$ superfields, and then dimensionally reducing
to zero dimensions.  The result is the action
\beq
S &=& \frac{1}{g^2} 
\Tr \biggl[ 
\int d^2\theta \; d^2 \mybar \theta \;\; \mybar {\bfZ}_m 
e^{2{\bf V}} {\bfZ}^m  e^{-2 {\bf V}}+ \; 
\frac{1}{4}\int d^2  \theta \;{\bf W}^{\alpha}{\bf W}_{\alpha} +   
\frac{1}{4}\int d^2  {\mybar \theta} \;  \mybar {\bf W}_{{\dot \alpha}}
\mybar {\bf W}^{
{\dot \alpha}} \\ \nonumber 
&+& \frac{\sqrt 2 }{3!} \epsilon_{mnp}  \int d^2 \theta  \, {\bfZ}^m 
[\bfZ^n, \bfZ^p]  
-   \frac{\sqrt 2}{3!} \epsilon^{mnp} \int d^2 \mybar \theta  \, \mybar \bfZ_m[
\mybar \bfZ_n, \mybar \bfZ_p]  
\biggr]
\eeq
where $\bfZ^m$ and $\mybar \bfZ_m$ are chiral and anti-chiral
superfields respectively, and ${\bf V}$ is a vector multiplet, expanded in
components as
\beq
\bfZ^m&=& z^m + \sqrt 2 \theta \psi^m + \theta \theta F^m \cr
{\mybar \bfZ}_m&=& \mybar z_m + \sqrt 2 \mybar \theta \mybar \psi_m + 
\mybar \theta \mybar \theta \mybar F_m  \cr
{\bf V} &=& -\theta \sigma_a \tilde v_a \mybar \theta  
+ \theta \theta \mybar \theta
\mybar \lambda + \mybar \theta \mybar \theta  \theta
\lambda + \half \theta \theta \mybar \theta \mybar \theta\, d \cr
{\bf W}_{\alpha}&=& -\lambda_{\alpha} + ( -(\sigma_{ab}v_{ab})_{\alpha}^
{\ \  \beta}  + \delta_{\alpha}^{ \beta} d)\theta_{\beta} - \theta 
\theta (\sigma_a)_{\alpha \dot \beta}[v_a, \mybar \lambda^{\dot \beta}] \cr
{\bf W}^{\dot \alpha}&=& - \mybar \lambda^{\dot\alpha} + ( (\mybar 
\sigma_{ab}v_{ab})^{\dot \alpha}_
{\ \ \dot  \beta}  + \delta^{\dot \alpha}_{\dot  \beta} d)\mybar 
\theta^{\dot \beta} -  \mybar \theta 
\mybar \theta (\mybar \sigma_a)^{\dot \alpha  \beta}[v_a,  \lambda_{\beta}].
\eeq
The ${\bf W}_{\alpha}$ and $\mybar {\bf W}_{{\dot \alpha}}$ are the
usual spinorial field strength chiral superfields that give rise in four
dimensions to the kinetic terms for the gauge bosons and gauginos.

The off-shell supersymmetric variations of these components in terms
of the four Grassmann parameters $\zeta$ and $\mybar \zeta$
(transforming as $(2,1)$ and $(1,2)$ respectively under $SU(2)\times
SU(2)$) are given by\footnote{As before, we define $\delta = (i\zeta Q
  - i \mybar\zeta\mybar Q)$.  In four dimensions, one has $Q =
  \partial_\theta -i\sigma^\mu \mybar\theta\partial_\mu$, and so one
  might expect in the dimensionally reduced theory that $Q$ and
  $\mybar Q$ would be the nilpotent operators $\partial_\theta$ and
  $\partial_{\mybar\theta}$, which would lead to
  $[\delta_1,\delta_2]=0$, while \eq{susy2d} yields instead
  $[\delta_1,\delta_2]=-\epsilon_{ij}\zeta_i\sigma_a \mybar \zeta_j
  [\tilde v_a,\cdot]$.  This occurs because we have chosen Wess-Zumino
  gauge to eliminate the extraneous components of the ${\bf V}$
  supermultiplet.  The supersymmetry transformation must now include a
  field-dependent gauge transformation to maintain the WZ gauge
  condition.  A similiar phenomenon occured in the $\CQ=2$,
  $d=3$ lattice of the previous section. See
  ref. \cite{Sohnius:1985qm} for a discussion.}
\beq
\delta z^m &=&  \sqrt 2 i \zeta \psi^m,\qquad \cr
 \delta {\mybar z}_m &=&  -\sqrt 2 i {\mybar \zeta}  {\mybar \psi}_m \cr
\delta  \psi^m &=& + \sqrt 2 i (\sigma_a){\mybar \zeta}[\tilde v_a, z^m] +  
\sqrt 2 i 
\zeta  F^m \cr
\delta  {\mybar \psi}_m &=& - \sqrt 2 i  ({\mybar \sigma}_a ) {\zeta}
[\tilde v_a,{\mybar z_m}]- \sqrt 2 i \mybar \zeta {\mybar F}_m \cr
\delta \tilde v_a &=&  +i \mybar \lambda \mybar \sigma_a \zeta -  
i {\mybar \zeta}
{\mybar \sigma}_a \lambda \cr
\delta \lambda &=& -i ({\sigma}_{ab} \tilde v_{ab})\zeta  + 
i\zeta d  \cr
 \delta \mybar \lambda &=& -i({\mybar \sigma}_{ab}\tilde v_{ab}) \mybar \zeta  
-i \mybar 
\zeta d \cr
\delta d &=& i \zeta \sigma_a [\tilde v_a, \mybar \lambda] + i 
 [\tilde v_a, \lambda] \sigma_a \mybar \zeta \ \cr
\delta F^m &=& - i\sqrt 2  \mybar \zeta \mybar \sigma_a [ \tilde v_a , \psi^m]
+ 2i [z^m, \mybar \lambda ]\mybar \zeta
 \cr
\delta \mybar F_m &=&  
-i\sqrt 2   \zeta  \sigma_a [ \tilde v_a , \mybar \psi_m]
+ 2i [\mybar z_m,  \lambda ] \zeta \ .
\eeq 
where the auxiliary fields satisfy the equations of motion  
\beq
d &=&  - [\mybar z_m,z^m]\ , \cr
 F^m &=&   -\frac{1}{\sqrt 2}\epsilon^{mnp}[\mybar z_n,\mybar z_p]\cr
 \mybar F_m &=& +  \frac{1}{\sqrt 2} \epsilon_{mnp}[z^n,z^p]\ .
\eqn{susy2d}\eeq

\subsection{The $d=2$, $\CQ=4$ lattice theory}
\label{sec:5b}

The steps for constructing the $d=2$, $\CQ=4$ lattice  for the
$(8,8)$ target theory are similar to those followed in previous
sections.  After the $Z_N\times Z_N$ orbifold projection of the mother
theory, one obtains the lattice action, written with manifest $\CQ=4$
supersymmetry
\begin{equation} 
\begin{aligned}
S=
&\frac{1}{g^2}\sum_{\bfn} \Tr\biggl[
 \int  d^2\theta \; d^2 \mybar \theta \;\; \mybar {\bfZ}_{m}(\bfn) 
e^{2{\bf V}({\bfn})} {\bfZ}^{m}( \bfn) e^{-2{\bf V} (\bfn+ \bfmu_m)}  \\ &
+ \; 
\frac{1}{4} \int d^2  \theta \; 
{\bf W}^{\alpha}(\bfn){\bf W}_{\alpha}( \bfn)   +   
\frac{1}{4} \int d^2  {\mybar \theta} \;  \mybar {\bf W}_{{\dot \alpha}}( \bfn)
\mybar {\bf W}^{
{\dot \alpha}}({\bfn})  \\ &
+ \frac{\sqrt 2}{3!} \epsilon_{mnp}  \int d^2 \theta  \, {\bfZ}^{m}(\bfn)  
( \bfZ^n ( \bfn + \bfmu_m) \bfZ^{p}( \bfn - \bfmu_p) - \bfZ^{p}
( \bfn + \bfmu_m) 
\bfZ_{n}( \bfn-\bfmu_n) ) \\    
& +   \frac{\sqrt 2}{3!} \epsilon^{mnp} \int d^2 \mybar \theta  \,
\mybar \bfZ_{m}( \bfn) 
( \mybar \bfZ_{n}(\bfn - \bfmu_n) \mybar \bfZ_{p}( \bfn + \bfmu_m) -
\mybar \bfZ_{p}(\bfn - \bfmu_p) \mybar \bfZ_{n}( \bfn + \bfmu_m) ) 
\biggr]
\end{aligned}
\eqn{d2lat}
\end{equation}

One then expands the theory about a particular trajectory in moduli
space.  For a square lattice, the expansion is about
\beq
z^{m}( \bfn)= \mybar z_{m}( \bfn) = \frac{1}{a {\sqrt 2}} {\bf 1}_k
,\,\,\, m= 1 \ldots 2,  \qquad z^{3}( \bfn )= \mybar z_{3}( \bfn) =0.
\eqn{squ}\eeq
With this choice $\bfmu_1$, $\bfmu_2$ and $\bfmu_3$ get mapped to the
lattice vectors $\hat{\bf x}$, $\hat{\bf y}$ and $-(\hat{\bf x}+
\hat{\bf y})$ respectively, and the lattice appears as  in  Fig.~\ref{fig:lat2dsquare}.
A more symmetric alternative is the expansion about
\beq
z_{m}( \bfn)= \mybar z^{m}( \bfn) = \frac{1}{a {\sqrt 2}} {\bf 1}_k
,\,\,\, m= 1 \ldots 3,
\eqn{s3}
\eeq
which treats all bosonic link fields on equal footing and gives rise
to the $A_2^*$ (triangular) lattice  shown in  Fig.~\ref{fig:lat2dhex}.

\DOUBLEFIGURE[t]{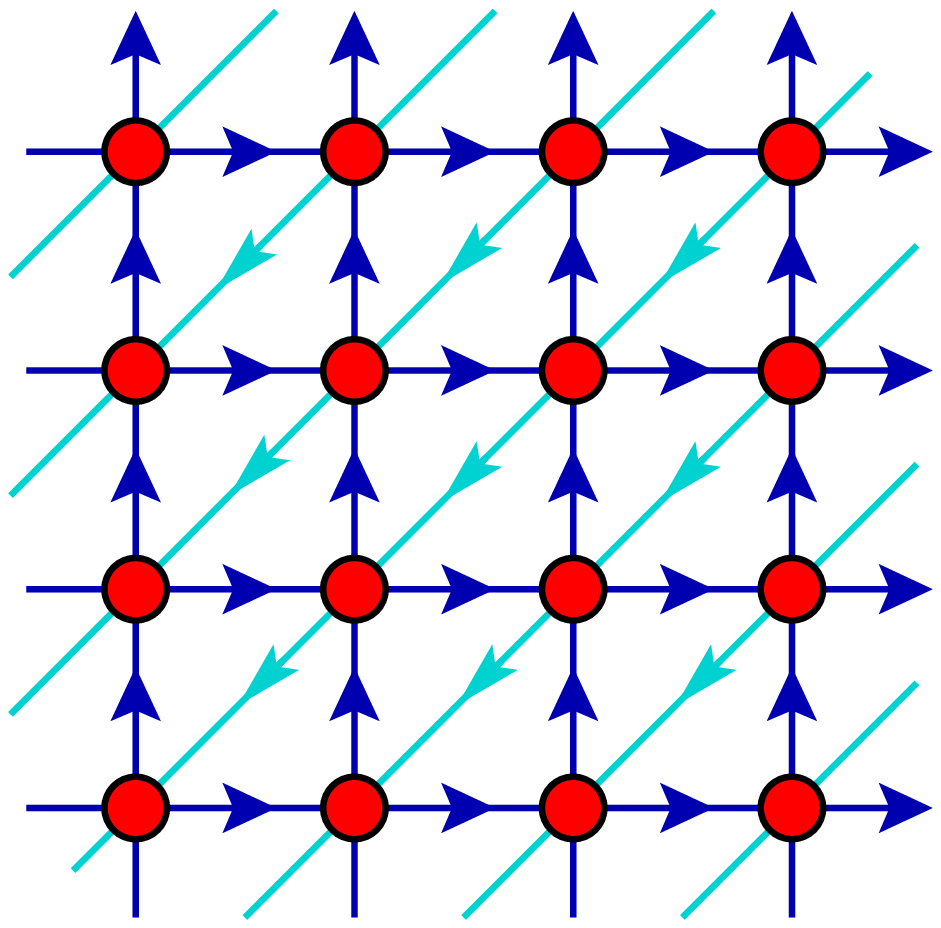,width=6cm}{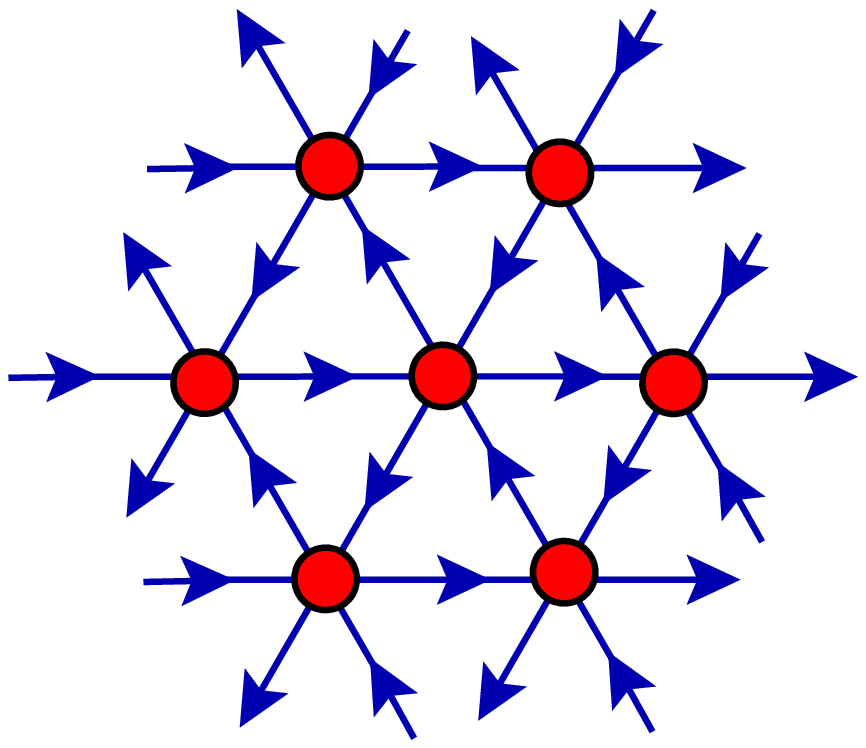,width=6cm}
{{\sl The square $d=2$ lattice following from the expansion eq.~(5.9). 
 The $\bfr=\bfmu_{1,2}$ variables in
  Table~3.
reside on the $\hat{\bf x}$ and  $\hat{\bf y}$
  links respectively, while charge $\bfr=\bfmu_3$ maps variables onto
  the diagonal links; $\tilde v$ and $\lambda$, $\mybar\lambda$ have
  $\bfr={\bf 0}$ and reside
at the sites.\label{fig:lat2dsquare}}}
{{\sl The triangular $d=2$ lattice ($A_2^*$) corresponding to expansion about the
  symmetric point in moduli space, eq.~(5.10).  The $\bfr=\pm\bfmu_{m}$ variables 
  $z^m$, $\mybar z_m$, $\psi^m$ and $\mybar \psi_m$  reside on the
  links, while $\tilde v$ and $\lambda$, $\mybar\lambda$ have 
  $\bfr={\bf 0}$ and reside 
at the sites. \label{fig:lat2dhex}}}

To analyze the continuum limit of the $A_2^*$ lattice, we introduce three two dimensional 
vectors to relate the point $\bfn$ to a spacetime point. These lattice 
vectors can be chosen as 
\beq
    {\bf e_1} &=& (\frac{1}{\sqrt 2},  \frac{1}{\sqrt 6} )\ ,  \cr
{\bf e_2} &=& (-\frac{1}{\sqrt 2},  \frac{1}{\sqrt 6} ) \ , \cr
{\bf e_3} &=& (0,  -\frac{2}{\sqrt 6} )  \ ,
\eqn{latvec2}
\eeq 
and satisfy  the relations  
\beq
\sum_{m=1}^{3} {\bf e}_m = 0, \qquad
{\bf e}_m\cdot{\bf e}_n= \delta_{mn}- {\textstyle{\frac{1}{3}}} \qquad  
\sum_{m=1}^{3} ({\bf e}_m)_{\mu} ({\bf e}_m)_{\nu}  = \delta_{\mu \nu}
\eeq 
The lattice vectors are the $SU(3)$ weights of the $\bf 3$ representation, 
and they  form a 2-simplex (equilateral triangle) in two dimensions. 
The matrix  ${\bf e}_m\cdot{\bf e}_n$ is 
the Gram matrix of $A_2^{*}$ \cite{Conway:1991},
also known as hexagonal lattice.         

The site $\bfn$ is identified with the spacetime location 
\beq
\bfR=a\sum_{\nu=1}^{2}({\bfmu}_{\nu}\cdot\bfn){\bf e}_{\nu} = 
\eqn{ra2}
\eeq
and a lattice displacement of one unit in direction ${\bfmu}_m$ corresponds
to a spacetime translation $a{\bf e}_{m}$.  
Each of the three links occupied by three $z^m$ variables has length 
$|a{\bf e}_{m}|=\sqrt{\frac {2}{3}}\,a$, unlike the case of the less symmetric 
square lattice where  $z^3$ resides on a link $\sqrt 2 $ times longer 
then the ones occupied by  the three $z^m, \ m=1,2$. 

In order to relate the lattice fields  to continuum fields,  we expand 
the lattice action around the point \eq{s4}.  However, the structure of the
lattice is dictated by $SU(3)$ and the continuum 
fields transform under $SO(2)\times SO(8)$ where $SO(2)$ Euclidean analog of
the Lorentz symmetry and  $SO(8)$ is the global R-symmetry.  
To make the connection between lattice and continuum fields, 
we introduce a $3 \times 3$  real orthogonal matrix  
\beq
\CE = \begin{pmatrix}
 \ \ \frac{1}{\sqrt{2}} &  \ \ \frac{1}{\sqrt{6}} & 
\ \ \frac{1}{\sqrt{3}}  \cr
-\frac{1}{\sqrt 2} &  \ \ \frac{1}{\sqrt{6}} & \ \ \frac{1}{\sqrt{3}} \cr
\ \ 0 &  -\frac{2}{\sqrt{6}} &  \ \ \frac{1}{\sqrt{3}} 
\end{pmatrix} = \left(\CE^T\right)^{-1}
\eqn{cedef2}\eeq
Note that  $\CE_{m\mu}$, with $\mu=1, 2$, are the components of the vectors
$\bfe_m$ of \eq{latvec3}. This matrix has the property that
\beq
\left(\sum_{m=1}^{3} \bfe_m \,\CE_{mn}\right)_\mu
=\begin{cases}\delta_{ n\mu} & n=1,\ldots,2 \cr 0 & n=3 \ .\end{cases}
\eqn{ceprop2}
\eeq
which serves as a bridge between the $SU(3)$ tensors of the lattice
construction, and the $SO(2)$ representations of the continuum theory.
In terms of this matrix we then define the
expansion of $z^m$ about the point in moduli space \eq{s3} to be
\beq
z^m = \frac{1}{a\sqrt{2}} + \sum_{m=1}^3 \CE_{mn} \Phi_n =
\frac{1}{a\sqrt{2}} + \sum_{\mu=1}^2 (\bfe_m)_\mu \Phi_\mu +
\frac{1}{\sqrt 3} \Phi_3\ ,
\eqn{zexp2}
\eeq
with
\beq
\Phi_\mu \equiv \left(\frac{S_\mu + i V_\mu}{\sqrt{2}}\right), \ 
\mu=1,2\ ,\qquad  \Phi_3 \equiv \left(\frac{S_3 + i
    S_4}{\sqrt{2}} \right), 
\eeq
and
\beq
\{\tilde v_0,\tilde v_1,\tilde v_2,\tilde v_3\} \equiv \,
\{S_8,S_5,S_6,S_7\} \ . 
\eeq
where $V_\mu$ and $S_a$ are hermitean $k\times k$ matrices, corresponding to the
two gauge fields and eight scalars of the continuum theory. 

We  expand the action \eq{d2lat} to leading order in powers of
the lattice spacing $a$ to obtain the continuum limit  at tree
level.
Since the Jacobian of the transformation
between lattice coordinates $\bfn$ and spacetime coordinates $\bfR$ in
\eq{ra2} equals $a^2/ \sqrt{3}$, we first must rescale our coupling $g$ such
that
\beq
\frac{1}{g^2} = \frac{a^2}{\sqrt{3}\,g_2^2}  \ ,\qquad
\lim_{a\to 0} \frac{1}{g^2} \sum_{\bfn} = \frac{1}{g_2^2} \int d^2 \bfR \ .
\eeq

The analysis of the bosonic action of the lattice theory gives  in the 
continuum  $SO(2)\times SO(8)$ invariant action.  
\beq
S_\text{boson}= \frac{1}{g_2^2}\int d^2\bfR\, \Tr\,\left[
\frac{1}{2} (D_\mu S_a)^2 +\frac{1}{4} V_{\mu\nu}^2 -\frac{1}{4}
[S_a,S_b]^2\right] +O(a)
\eeq
where $\mu, \nu=1,2$  and $a,b=1 \ldots 8$.

Similar to the analysis of fermionic terms in \S~\ref{sec:3e}, 
 we can express the continuum limit of the fermion action 
\eq{d2lat} in terms of $\tilde\omega$ as\footnote{$\tilde\omega$ for
  this theory may be obtained from the $\tilde \omega$ constructed in
  the $d=4$ lattice, followed by the substitutions given in
  Appendix~\ref{sec:ap4}.}
\beq
S_\text{fermion}&=&\frac{1}{g_2^2}\int d^2\bfR\, \Tr\,\frac{1}{2}\,
\tilde\omega^T \CC\, \Bigl(\Gamma_m\, \CE_{m\mu} D_\mu \tilde\omega +
  i\,\Gamma_{m+5}\, \CE_{mn}[S_n,\tilde \omega]+ i  \Gamma_{m}\,
  \CE_{m3}[S_4,\tilde \omega]  \nonumber \\ \cr
&&   
+  i \sum_{i=1}^2  \left(
\Gamma_{11-i}\,
  [S_{2i+3},\tilde \omega]  +   \Gamma_{6-i}\,
  [S_{2i +4},\tilde \omega]\right)
\Bigr)
\eeq
where the $\Gamma_\alpha$ are $SO(10)$ gamma matrices in the basis
used to define the mother theory, \eq{act1}\footnote{In this section
  $SU(3)$ tensor indices are denoted by  $m,n$ and are summed
from 1 to $3$; $SO(8)$ vector indices are denoted by $a,b$ and 
are summed from
1 to $8$ ; and $SO(2)$ vector indices are denoted by $\mu$,$\nu$ and 
are summed from $1$ to $2$.}. Since $\CE$ is an
orthogonal matrix, we can define a new gamma matrix basis for $SO(10)$
\begin{equation}
\begin{aligned}
&\tilde \Gamma_{\mu} = \Gamma_m\, \CE_{m\mu}\ ,\ \  \tilde \Gamma_{n+2} =
\Gamma_{m+5}\, \CE_{mn}\ ,\ \  \tilde \Gamma_{6}= \Gamma_m \, \CE_{m3}, \ \ \\ 
&\tilde \Gamma_{2i+5} = \Gamma_{11-i} \ ,\ \  
\tilde \Gamma_{2i+6} =\Gamma_{6-i} \ ,\qquad i=1,2\ . \  
\end{aligned}
\end{equation}
In the new basis the charge conjugation matrix is unchanged.
Therefore, the above continuum limit of the lattice
fermion action may be written as
\beq
S_\text{fermion}&=&\frac{1}{g_2^2}\int d^2\bfR\  \frac{1}{2}\, \Tr\,\Bigl(
\tilde\omega^T \tilde\CC\, \tilde \Gamma_\mu D_\mu \tilde\omega +
i\,\tilde\omega^T \,\tilde\CC \,\tilde\Gamma_{2+a}\,[S_a,\tilde 
\omega]\Bigr)+O(a)\ . \cr&&
\eeq
The chiral 
symmetry of the theory, $SO(8)$, which does not exist for any finite lattice 
spacing,  emerges naturally in the continuum. Combining
$S_{\text{fermion}}+S_{\text{boson}}$  correctly reproduce the sixteen
supercharge  $\CN=(8,8)$ SYM 
theory  in $d=2$ dimensions. 

\section{The one dimensional lattice; or Euclidean path 
integrals for M-theory} 
\label{sec:6}

The sixteen supercharge $U(N)$ matrix quantum mechanics is interesting
because it has been argued that the large $N$ limit corresponds to
$M$-theory \cite{Banks:1997vh}. Because of this limit, a Hamiltonian
approach to the theory is not very practical (as one would expect for
a theory that is supposed to contain higher dimensional physics), and
so a path integral approach may prove to be more promising.  Here we
construct a version of the theory on a one-dimensional lattice in the
Euclidean time direction, which
possesses eight exact supersymmetries.  The other eight appear in the
continuum limit.  

The theory continuum theory has a one-component gauge boson 
$V$  which is not dynamical, but which is rather a Lagrange multiplier. Integrating it out 
enforces the constraint on physical states that they be gauge
invariant. The $R$-symmetry of the theory is $SO(9)$, under which
the scalars transform as the ${\bf 9}$ dimensional  vector representation and 
the fermions as the ${\bf 16}$ dimensional spinor representation. 
The action of the target theory is 
\beq
S_{\text{target}} = \frac{1}{g_1^2} \,\int d{ \tau}\, \Tr
\Bigl( \frac{1}{2} (D_\tau S_a)^2 -\frac{1}{4}\left[S_a,S_b\right]^2
 + \frac{1}{2}\tilde \omega^T\,\tilde{\CC}\, D_\tau \, \tilde \Gamma_1\,\tilde \omega +
\frac{i}{2}\, \tilde\omega^T\, \tilde{\CC}\, \tilde\Gamma_{1+a}\, [S_a,\tilde \omega]
\Bigr)\cr
\eqn{target1}
\eeq
 where we introduced $SO(10)$ gamma matrices $\tilde \Gamma_{\alpha}$
 and $SO(9)$ indices $a,b=1,\ldots,9$. The 
$ SO(9)$ symmetry of the action is manifest.

\subsection {The lattice action}
\label{sec:6a}

To create the $d=1$ lattice we decompose $SO(10)$ multiplets along the chain
$SO(10)\supset SO(4)\times SO(6)\supset SU(2)\times
SO(6)\times U(1)$:
\beq
v \sim{\bf 10} \ \longrightarrow\  
({\bf 2,1})_{\bf 1} \oplus  ({\bf 2,1})_{- {\bf 1}} 
\oplus  ({\bf 1,6})_{\bf 0} \ .
\eeq
\beq
\omega \sim {\bf 16} \ 
\longrightarrow\   
({\bf 2,4})_{{\bf 0}} \oplus  ({\bf 1, \mybar 4})_{{\bf 1}} \oplus
({\bf 1, \mybar 4})_{-{\bf 1}} 
\eeq
We can then orbifold by the $Z_N$ contained within the above $U(1)$
symmetry, creating a one dimensional lattice, while leaving intact the
$SU(2)\times SO(6)$ global symmetry, and eight of the original
sixteen supercharges. 

To describe this $d=1$  theory it is convenient to use the $\CN=1$ superfield language of four
dimensions employed in  \S~\ref{sec:5} for the $d=2$ lattice,
at the price of only having manifest only four of the eight exact
supercharges, and an $SO(4)$ subgroup of the global $SO(6)$ symmetry.
The chiral superfields $Z^m$ and $\mybar Z_m$ with $m=1,2$ from the $d=2$
lattice are taken to have $\bfr=+\bfmu_m$ and $\bfr=-\bfmu_m$
respectively, where we define  
\beq
\bfmu_1 &=& +1\ ,\cr
\bfmu_2 &=& -1\ .
\eqn{mudef1}\eeq
We see then that $z^1$ and $\mybar z_2$ oriented along  the forward $\bfmu_1=-\bfmu_2$ link and
comprise the $(2,1)_1$ boson representation, while $z^2$ and $\mybar
z_1$ reside on the backward link $\bfmu_2=-\bfmu_1$ and form the
$(2,1)_{-1}$. The fermions $\psi^1$ and $\mybar \psi_2$  similarly live
on the forward link;  they each have two components and form the
$(1,4)_1$ representation, while $\psi^2$ and $\mybar\psi_1$ live on
the backward link and are the $(1,\mybar 4)_{-1}$.  In terms of
superfields, $Z^1$ and $\mybar Z_2$ live on the forward link and form
a hypermultiplet of the exact  $\CQ=8$ supersymmetry, while $Z^2$ and $\mybar Z_1$ form a hypermultiplet
along the backward link.

The site variables on our $d=1$ lattice are the vector superfield $V$
and the chiral superfield $Z^3$ from the $d=2$ lattice discussion of \S~\ref{sec:5}.  
Together the six real bosons in $z^3$, $\mybar z_3$ and $\tilde v_a$,
$a=0,\ldots,3$ form the $(1,6)_0$, while the eight fermion components in
$\lambda$, $\mybar\lambda$, $\psi^3$ and $\mybar\psi_3$ form the
$(2,4)_0$. Together the $V$ and $Z^3$, $\mybar Z_3$ superfields form
an extended vector multiplet of $\CQ=8$ supersymmetry.

The one dimensional lattice action with eight exact supersymmetry 
may be written in  manifestly 
$\CQ=4$  multiplets as    
\begin{equation} 
\begin{aligned}
S=&
\frac{\Tr}{g^2}\sum_{\bfn} \biggl[
 \int  d^2\theta \; d^2 \mybar \theta \;\; \mybar {\bfZ}_{3}(\bfn) 
e^{2{\bf V}(\bfn)} {\bfZ}^3( \bfn) e^{-2{\bf V}(\bfn)}   +  
\mybar {\bfZ}_{m}(\bfn) 
e^{2{\bf V}(\bfn)} {\bfZ}^{m}(\bfn) e^{-2{\bf V}(\bfn+ \bfmu_m)} \\
&
+ \; 
\frac{1}{4}\int d^2  \theta \; 
{\bf W}^{\alpha}({\bfn}){\bf W}_{\alpha}( \bfn) + {\rm a.h.} \\   
&+ \frac{\epsilon_{mn}}{2 \sqrt 2}  \int d^2 \theta  \, {\bfZ}^{3}(\bfn)  
( \bfZ^{m}( \bfn ) \bfZ^{n} (\bfn + \bfmu_m) - \bfZ^{n} (\bfn )
\bfZ^{m}( \bfn-\bfmu_n ))+  {\rm a.h.} 
\biggr]
\end{aligned}
\end{equation}
where a.h. stands for anti-holomorphic  integrals over antichiral 
superfields.  The $\CQ=4$ chiral superfields and the vector multiplet 
are  given by 
\beq
\bfZ^m&=& z^m + \sqrt 2 \theta \psi^m + \theta \theta F^m, \qquad   \cr
\bfZ^3&=& z^3 + \sqrt 2 \theta \psi^3 + \theta \theta F^3, \qquad   \cr
{\bf V} &=& -\theta \sigma_a \tilde v_a \mybar \theta  
+ \theta \theta \mybar \theta
\mybar \lambda + \mybar \theta \mybar \theta  \theta
\lambda + \half \theta \theta \mybar \theta \mybar \theta\, d 
\eeq

\subsection{The continuum limit for $d=1$ lattice}
\label{sec:6b}
We expand the  action about  the point 
\beq
z^{m}(\bfn)= \mybar z_{m}(\bfn) = \frac{1}{a {\sqrt 2}} {\bf 1}_k
,\,\,\, m= 1,2
\eqn{s2}
\eeq




To make the connection between lattice and continuum fields, 
we introduce a $2 \times 2$  real orthogonal matrix  
\beq
\CE = \begin{pmatrix}
 \ \ \frac{1}{\sqrt{2}} &  \ \ \frac{1}{\sqrt{2}} \cr 
-\frac{1}{\sqrt 2} &  \ \ \frac{1}{\sqrt{2}} 
\end{pmatrix} = \left(\CE^T\right)^{-1}
\eqn{cedef1}
\eeq
In terms of this matrix we then define the
expansion of the link bosons $z^m$ about the point in moduli space \eq{s2} to be
\beq
z^m = \frac{1}{a\sqrt{2}} + \sum_{n=1}^2 \CE_{mn} \Phi_n
\eqn{zexp1}
\eeq
with
\beq
\Phi_1 \equiv \left(\frac{S_1 + i V}{\sqrt{2}}\right)\ ,\quad  \Phi_2 \equiv \left(\frac{S_2 + i
    S_3}{\sqrt{2}} \right)\ 
\eeq
while the site bosons are rewritten as
\beq
z^3 = \left(\frac{S_8+iS_9}{\sqrt{2}}\right)\ ,\qquad \{\tilde v_0,\tilde
v_1,\tilde v_2,\tilde v_3\} = \{S_7, S_4,S_5,S_6\},
\eeq
where $V$ and $S_a$ ($a=1,\ldots,9$) are hermitean $k\times k$ matrices, corresponding to 
the nondynamical gauge field   and nine scalars of the continuum theory. 

We  expand the action \eq{d2lat} to leading order in powers of
the lattice spacing $a$ after performing the rescaling
\beq
\frac{1}{g^2} = \frac{a}{\sqrt{2}\,g_1^2}  \ ,\qquad
\lim_{a\to 0} \frac{1}{g^2} \sum_{\bfn} = \frac{1}{g_1^2} \int d \tau \ .
\eeq
to obtain the continuum limit  at tree
level.

Expanding the bosonic action of the lattice theory yields the 
continuum  $ SO(9)$ invariant action  
\beq
S_\text{boson}= \frac{1}{g_1^2}\int d \tau\, \Tr\,\left[
\frac{1}{2} (D_\tau S_a)^2  -\frac{1}{4}
[S_a,S_b]^2\right] +O(a)\ .
\eeq
We can express the continuum limit of the fermion action 
 in terms of the same $\tilde\omega$ as in \S\ref{sec:5}
\beq
S_\text{fermion}&=&\frac{1}{g_1^2}\int d\tau\, \Tr\,\frac{1}{2}\,
\tilde\omega^T \CC\, \Bigl(\Gamma_m\, \CE_{m1} D_\tau \tilde\omega +
  i\,\Gamma_{m+5}\, \CE_{mn}[S_n,\tilde \omega]+ i  \Gamma_{m}\,
  \CE_{m2}[S_3,\tilde \omega]  \nonumber \\ \cr
&&   
+  i\sum_{i=1}^3\left(  \Gamma_{11-i}\,
  [S_{2i+2},\tilde \omega]  +   \Gamma_{6-i}\,
  [S_{2i+3},\tilde \omega]\right)  \Bigr)
\eeq
where the $\Gamma_\alpha$ are $SO(10)$ gamma matrices in the basis
used to define the mother theory, \eq{act1}. Since $\CE$ is an
orthogonal matrix, we can define a new gamma matrix basis for $SO(10)$
\begin{equation}
\begin{aligned}
&\tilde \Gamma_{1} = \Gamma_m\, \CE_{m1}\ ,\ \  \tilde \Gamma_{n+1} =
\Gamma_{m+5}\, \CE_{mn}\ ,\ \  \tilde \Gamma_{4}= \Gamma_m \, \CE_{m2}, 
\ \ \ m,n=1,2
\ \ \\ 
&\tilde \Gamma_{2i+3} = \Gamma_{11-i} \ ,\ \  
\tilde \Gamma_{2i+4} =\Gamma_{6-i} \ , \  \  \ i=1,2,3
\end{aligned}
\end{equation}
In the new basis the charge conjugation matrix is unchanged.
Therefore, the above continuum limit of the lattice
fermion action may be written as
\beq
S_\text{fermion}&=&\frac{1}{g_1^2}\int d\tau\  \frac{1}{2}\, \Tr\,\Bigl(
\tilde\omega^T \tilde\CC\, \tilde \Gamma_1 D_\tau \tilde\omega +
i\,\tilde\omega^T \,\tilde\CC \,\tilde\Gamma_{1+i}\,[S_i,\tilde 
\omega]\Bigr)+O(a) \cr&&
\eeq
where $\mu=1$ is the continuous Euclidean time and  $i$ is over $1 \ldots 9$. 
The global R-symmetry $SO(9)$ is manifest  in the continuum action,
and we conclude that  our construction of one-dimensional lattice with $\CQ=8$
supersymmetry  correctly reproduce the Euclidean action for sixteen supercharge  matrix quantum 
mechanics. 

\section{Discussion and Prospects }
\label{sec:discussion}

We have exploited the technique of deconstruction
\cite{ArkaniHamed:2001ca,ArkaniHamed:2001ie} to create supersymmetric
lattices in Euclidean spacetime which serve as nonperturbative regulators for SYM theories
with sixteen supercharges in $d\le 4$ dimensions.  As argued in the
introduction, the target theories  are in many ways the most interesting quantum field
theories that have ever been constructed. Recently
the first  nonperturbative
construction of these theories was accomplished on a spatial lattice
(Relevant for a Hamiltonian formulation) \cite{Kaplan:2002wv};  in
this paper we provide a formulation of Euclidean spacetime lattices,
appropriate for a nonperturbative construction of the path integral
for these theories. Our lattices look very unconventional; the
structure is not the usual hypercubic lattice with scalars and
fermions living at sites and gauge fields on links.  In fact fermions
and scalars live on both sites and links, while the interactions are
most symmetrically described in $d$ dimensions by an $A_{d}^*$
lattice.  Despite their bizarre formulation, with spinless fields of
the continuum
represented by variables which transform nontrivially under the point
group of the lattice,  we have shown that  at tree level our lattices correctly reproduce
the desired target theories. 

An important problem not addressed here is whether  fine tuning  is
required when the effects of radiative
corrections are included, in order to attain the target theory in the
continuum limit.
  It is known from previous
work that the exact supersymmetry on the lattice greatly reduces or
entirely eliminates the
number of counterterms that may be required. In fact, it is expected that the combination of exact
supersymmetry and super-renormalizability will result in no
fine-tuning at all for the theories in $d\le 3$.  For the $d=4$
theory, $\CN=4$ SYM, standard power counting arguments used in
\cite{Kaplan:2002wv,Cohen:2003aa,Cohen:2003qw}  suggest that at worst
logarithmic fine tuning could be required.  However, whether or not
such fine-tuning is actually required requires a subtle analysis.  The
undesirable counterterms will violate the shift symmetry of the moduli
space.  The only possible source for this symmetry violation are those
terms that  must be added by hand at
finite volume in order to fix the lattice spacing, the vacuum value
for the trace of our
link variables $\bfz$ about which expand.  Such terms which fix the
trace of $\bfz$  are
analogous to the
 external ${\bf B}$ field  needed
to study magnetization in finite volume, and they can be removed in the
infinite volume limit.  Therefore any dangerous 
counterterm will have to depend on this source which lifts the vacuum
degeneracy, and therefore will involve IR physics in a nontrivial
way. It seems plausible to the authors that the continuum (UV) and
large volume (IR) limits of the lattice theory could be coordinated in
such a way as to obviate the need for any fine tuning.  Such an
analysis has yet to be done.

A alternative and  potentially fruitful line of inquiry would be to 
analyze the anomalous dimensions of 
the undesirable operators  (Lorentz violating, in general) in  the
gravitational dual to our lattices, as suggested in
ref.~\cite{Hellerman:2002qa}. 

Questions about the continuum limits of our lattices aside, the reader
might ask of what use are these lattices we have constructed?  There
is little prospect for their numerical simulation in the near
future, as they entail both massless fermions as well as a sign
problem\footnote{For example, the zero
  momentum sector of our lattices are equivalent to the matrix
  formulation of $M$-theory discussed in Ref.~\cite{Krauth:1998xh},
  where it was shown there is a sign problem.  Furthermore, there is
  no reason to expect the continuum target theories to have a real,
  positive fermion determinant since these SYM theories involve both
  gauge and Yukawa interactions (unlike the special case of $\CN=1$
  SYM in $d=4$ dimensions, where there are no scalars and positivity
  of the fermion Pfaffian can be proven).} from the
fermion determinant, both of which render current Monte Carlo
simulation methods impractical. 
 In the long run we hope of course
that such technical barriers can be surmounted, in which case the
lattices given here could provide a rigorous window not only onto
supersymmetric gauge dynamics, but into the behavior of quantum
gravity and string theory as well.  

In the meantime, we believe there is value in simply showing that such
a nonperturbative construction exists, in a formulation in which
supersymmetry plays a major role.  However, we have higher ambitions
for these constructions, namely that analytic study of the
supersymmetric lattices could provide valuable insights.  Beyond the analysis of radiative corrections
outlined above, several
topics  one might explore include:
\begin{itemize}
\item {\it Chiral symmetry.} One
interesting feature of our lattices is how global chiral symmetries
emerge without fine-tuning, and without resort to the standard
constructions of chiral lattice fermions
\cite{Kaplan:1992bt,Neuberger:1998fp}; it would be interesting to
understand whether fermion propagators on our lattice obey the
Ginsparg-Wilson relation \cite{Ginsparg:1981bj}, or whether some new
mechanism is at play. 

\item {\it Gauge duality.} It may be possible
 to analyze these lattices  along the lines of 
ref.~\cite{Ukawa:1979yv} in order to try to shed light on the fascinating dualities
present in $\CN=4$ SYM theory in four \cite{Montonen:1977sn}  and
three \cite{Kapustin:1999ha,Kitao:1998mf,Kitao:1999uj} dimensions.  

\item {\it The large $N_c$ limit and the gravity dual.} It may also be
  possible to  analyze  the large $N_c$ limit of these lattice gauge
theories with the hope of learning more about the gravity/string
theory  dual  \cite{Hellerman:2002qa}, exploiting the AdS/CFT
correspondence \cite{Maldacena:1998re}. 
\end{itemize}
We have no doubt that other interesting directions to explore exist
which have not occurred to us at present.

\acknowledgments
We are grateful for numerous conversations about this work with
 Andrew Cohen and Emannuel Katz, who were collaborators on earlier
 papers in the investigation of supersymmetric lattices.  This work was
 supported by DOE grants DE-FGO3-00ER41132 and  DE-FG02-91ER40676.

\bigskip
\noindent
{\it Note Added.}  During completion of this work a new paper on
latticizing $\CN=4$ SYM theory was posted by Simon Catterall
\cite{Catterall:2005fd}.
\bigskip

\appendix

\section{An explicit gamma matrix basis}
\label{sec:ap}

Here we give an explicit chiral basis for the $SO(10)$ gamma matrices used in
this paper, which can be useful for explicit computations.  They are
given in the form of a direct product of Pauli matrices, and have the
symmetry property \eq{gamsym} that the first five are antisymmetric,
and the second five are symmetric:
\beq
\Gamma_1 &=& \sigma_2\otimes \sigma_1\otimes 1 \otimes 1 \otimes 1\cr
\Gamma_2 &=& \sigma_2\otimes \sigma_3\otimes \sigma_1 \otimes 1 \otimes 1\cr
\Gamma_3 &=& \sigma_2\otimes \sigma_3\otimes \sigma_3 \otimes \sigma_1 \otimes 1\cr
\Gamma_4 &=& \sigma_2\otimes \sigma_3\otimes \sigma_3 \otimes \sigma_3 \otimes \sigma_1\cr
\Gamma_5 &=& \sigma_2\otimes \sigma_3\otimes \sigma_3 \otimes
\sigma_3\otimes \sigma_3\cr &&\cr
\Gamma_6 &=& \sigma_2\otimes \sigma_2\otimes 1 \otimes 1 \otimes 1\cr
\Gamma_7 &=& \sigma_2\otimes \sigma_3\otimes \sigma_2 \otimes 1 \otimes 1\cr
\Gamma_8 &=& \sigma_2\otimes \sigma_3\otimes \sigma_3 \otimes \sigma_2 \otimes 1\cr
\Gamma_9 &=& \sigma_2\otimes \sigma_3\otimes \sigma_3 \otimes \sigma_3 \otimes \sigma_2\cr
\Gamma_{10} &=& -\sigma_1\otimes 1 \otimes 1 \otimes 1 \otimes 1 \cr
&&\cr
\Gamma_{11} &=& \sigma_3\otimes 1 \otimes 1 \otimes 1 \otimes 1 \cr
&&\cr
C &=& -\sigma_2\otimes \sigma_1\otimes \sigma_2\otimes
\sigma_1\otimes \sigma_2\ .
\eqn{gambasis}\eeq
The five fermionic raising and lowering operators $\hat A_m=\half(\Gamma_m-i\Gamma_{m+5})$ defined
in \eq{adef} are given in this basis by
\beq
\hat A_1 &=& \sigma_2\otimes \sigma_- \otimes 1\otimes 1\otimes 1 \cr
\hat A_2 &=& \sigma_2\otimes \sigma_3 \otimes \sigma_-\otimes 1 \otimes 1\cr
\hat A_3 &=& \sigma_2\otimes \sigma_3 \otimes \sigma_3\otimes \sigma_-\otimes 1 \cr
\hat A_4 &=& \sigma_2\otimes \sigma_3 \otimes  \sigma_3\otimes\sigma_3\otimes \sigma_-\cr
\hat A_5 &=& -i\left(\sigma_-\otimes P_+ +\sigma_+\otimes P_-\right)\ , 
\eeq
where 
\beq
\sigma_\pm = \frac{\sigma_1\pm i\sigma_2}{2}\ ,\qquad 
P_\pm \equiv \frac{\left(1\otimes 1\otimes 1\otimes 1\right) \pm \left(\sigma_3\otimes
\sigma_3\otimes \sigma_3\otimes \sigma_3\right)}{2}\ .
\eeq

As described in \S~\ref{sec:2}, the $\hat A$ operators can be used
to decompose the spinor representation of $SO(10)$ under its $SU(5)$
subgroup.  In this basis the highest weight spinor $\nu+$ satisfying
\beq
\Gamma_{11} \nu_+ = \nu_+\ ,\qquad {\hat A}_m^\dagger\nu_+ = 0
\eeq
corresponds to the state
$\ket{\uparrow\uparrow\uparrow\uparrow\uparrow}$, or in a single index
notation takes the simple form $(\nu_+)_\alpha = \delta_{\alpha
  1}$. Using the latter form,
 the decomposition \eq{omexp} becomes in this basis
\beq
\omega = \left(\lambda +   \xi_{mn}\,\half 
\hat A^m \hat A^n 
-\psi^m \,\frac{\epsilon_{mnpqr}}{24} \hat A^n \hat A^p \hat A^q \hat A^r
  \right)\nu_+ = \begin{pmatrix}
\lambda\cr
\xi_{45}\cr
\xi_{35}\cr
\xi_{34}\cr
\xi_{25}\cr
\xi_{24}\cr
\xi_{23}\cr
-\psi^1\cr
\xi_{15}\cr
\xi_{14}\cr
\xi_{13}\cr
\psi^2\cr
\xi_{12}\cr
-\psi^3\cr
\psi^4\cr
-\psi^5\cr
{\bf 0}
\end{pmatrix}
\eqn{omexpap}
\eeq
where the bold ${\bf 0}$ at the bottom of the spinor represents a
column of sixteen zeros.

\section{Absence of doublers on the $d=4$ lattice}
\label{sec:ap2}

In this appendix we examine  the free boson spectrum of the  lattice action for the 
four dimensional $A_4^{*}$ lattice and show 
that the formulation does not have any boson doublers at corners of the
Brillouin zone.  It  then follows from supersymmetry that there are no
fermion doublers either, saving one a somewhat more tedious calculation.  The generalization to 
other dimensions is straightforward.
  
To find the spectrum, we use the decomposition 
\beq
z^{m}= \frac{1}{\sqrt 2}\left(\frac{1}{ a}+   x_{m} + i y_{m}\right)\ ,\quad 
\mybar z_{m}=  \frac{1}{\sqrt 2}\left(\frac{1}{ a}+   x_{m} - i y_{m}\right)\
, \qquad    
m=1, \ldots, 5\ .
\eeq 
These are related to the continuum variables   $S$  and $V$
(respectively the  six scalars and four gauge fields of the 
$\CN=4$ SYM theory in four dimensions)  via the orthogonal matrix 
$\CE$ of \eq{cedef} as 
\beq
x_m= {\CE}_{mn}S_n \ ,   \qquad
y_m =   {\CE}_{mn}  \begin{pmatrix}
                           V_{\mu} \cr
                           S_6  
                          \end{pmatrix}_n 
\ .
\eeq
 At quadratic order in $x$ and 
$y$, the lattice
action \eq{d4lat} is
\beq
&&\frac{1}{2g^2}\sum_{\bfR}\sum_{m,n=1}^5\,\Tr \left[ \left( \frac{ x_{m} 
(\bfR)  -  
x_{m} (\bfR - {{\bfe}_{n}})}{a}\right)^2  +\half \left(\frac{y_{m}  
(\bfR   ) - 
y_{m} ( \bfR-  {\bfe}_{n})}{a} - 
m\leftrightarrow n \right)^2 \right]\cr&&
\eeq
where $\bfR=a\,\sum_{\nu=1}^4\, n_\nu \bfe_\nu$ is the  coordinate of an
 lattice site, the $n_\nu$ being integers and the $\bfe_\nu$ being
the  $A_4^*$ lattice vectors of \eq{latvec}.
We compute the spectrum by means of a Fourier transform, 
\beq 
\phi(\bfR) = \frac{1}{N^2} \sum_{{\bf p}}
e^{i {\bf p\cdot\bfR }}\phi({\bf p})\ .  
\eeq
It is convenient to expand the momenta as
\beq
{\bf p} = \sum_{\nu=1}^4 p_\nu {\bf g_\nu}\ ,
\eeq
in terms of the reciprocal lattice vectors ${\bf g}_\nu$
defined by
 \beq 
  {\bf e}_{\mu}\cdot  {\bf g}_{\nu} = \delta_{\mu\nu},  \qquad
  \mu,\nu=1, \ldots, 4\ 
\eeq
so that $p_\nu = {\bf p}\cdot\bfe_\nu$. The ${\bf g}_\nu$ generate an
$A_4$ lattice and are given by the 
simple roots of $SU(5)$. The coefficients of $\bf p$ take on the discrete
values in the Brillouin zone, $p_\nu=2\pi {\hat p}_\nu/Na$, with
${\hat p}_\nu$ being an integer in the interval 
$(-N/2,N/2]$. Note that since the lattice vectors $\bfe_\nu$ are
not orthonormal, $p^2={\bf p}\cdot{\bf p}\ne\sum_{\nu=1}^4 p_\nu p_\nu$;  rather
if we define $p_5 = {\bf p}\cdot\bfe_5=-\sum_{\nu=1}^4 p_\nu$ then 
\beq
\sum_{m=1}^5 p_m^2 = \sum_{m=1}^5 ({\bf p}\cdot \bfe_m)^2 = p^2
\ ,
\eeq
where we used the property \eq{eprop} that
$\sum_{m=1}^5(\bfe_m)_\mu(\bfe_m)_\nu = \delta_{\mu\nu}$.



The kinetic terms for the bosonic action then take the form 
\beq 
\frac{1}{2g^2}\Tr \sum_{\bf p}  \big[ x_m ({\bf p})\,  M({\bf p})^2 \,   
x_m( {- \bf p}) +   y_m ({\bf p}) \, 
G_{m n}( {\bf p}) \,   
 y_n ({\bf p}) \big]
\eeq
where we have defined 
\beq
M({\bf p})^2 = \sum_{m=1}^{5}  {\CP}_{m}^2 \ ,\qquad
 G_{m n}({\bf p}) =\left( M ({\bf p})^2 \delta_{m n} -   {\CP}_{m}
    {\CP}_{n} e^{-i a ( {\bf p}\cdot\bfe _{m} - {\bf p}\cdot\bfe_{n})/2}\right) \ , 
\eeq
with 
\beq
\CP_m\equiv \left(\frac{2}{a}\right)\,\sin \frac{a\, p_m}{2}\ ,\quad m=1,\ldots,5\ .
\eeq
It is important for our investigation of doublers  that $\CP_\nu\ne 0$ at the edge of
the Brillouin zone,  $p_\nu=\pi/a$. 

Note that the second term in the definition of $G$ is a rank one
matrix (as it is a product of vectors)  with eigenvalue $-M({\bf
  p})^2$ (as easily computed from the trace of the matrix). Therefore $G$ is rank four
with four degenerate eigenvalues equal to  $M({\bf p})^2$.  The zero
eigenmode of $G$ is proportional to $\CP_n$  and is a consequence of
gauge invariance.  The remaining nine 
bosonic modes
are degenerate for a given momentum, with eigenvalue $M({\bf p})^2$,
which is seen to vanish only at ${\bf p}=0$ and not at the corners of the
Brillouin zone.  Therefore we have shown that there are no bosonic doublers,
and that our procedure in the body of this paper of finding the continuum theory by expanding
about ${\bf p}=0$ is justified. A similar analysis for the fermions is
possible but unnecessary, as the exact supersymmetry precludes fermion
doublers in the absence of their bosonic counterparts.

It is a matter of a few lines to show that the continuum limit of
action we found above is 
\beq
V\int \frac{{\rm d}^4{\bf p}}{(2\pi)^4}\, \frac{1}{2g^2} \Tr
\left[\sum_{a=1}^6  S_a({\bf p}) p^2 S_a(-{\bf p}) +
  \sum_{\mu,\nu=1}^4 V_\mu\left(p^2\delta_{\mu\nu}-p_\mu
    p_\nu\right) V_\nu(-{\bf p})\right]\ .
\eeq 
as one would expect.
 
\section{Spinor notation in Euclidean space}
\label{sec:ap3}

In this appendix we give our spinor notation for the  $\CQ=4$ exact
supersymmetries in Euclidean space, which possesses an $SO(4)\simeq
SU(2)_L\times SU(2)_R$ symmetry.

Spinors in the $(\half,0)$  representation of
$SU(2)_L\times SU(2)_R$ are unbarred and carry undotted indices;  the
$(0,\half)$ representation is barred and carries dotted indices.  In
Euclidean space, complex conjugation does not take one representation
into the other.   Indices are raised and lowered with the $\epsilon$
tensor: 
\beq
\epsilon^{ij} = -\epsilon^{ji} = -\epsilon_{ij} = \epsilon_{ji}\
,\qquad
\epsilon^{12}=-\epsilon_{12}=1\ ,
\eeq
with
\beq
\psi^\alpha &= \epsilon^{\alpha\beta}\psi_\beta\ ,\qquad
\psi_\alpha=\epsilon_{\alpha\beta} \psi^\beta \cr
\mybar\psi^{\dot\alpha} &= \epsilon^{\dot\alpha\dot\beta}\mybar\psi_{\dot\beta}\ ,\qquad
\mybar\psi_{\dot\alpha}=\epsilon_{\dot\alpha\dot\beta}
\mybar\psi^{\dot\beta} \ .
\eeq
Singlets composed of two spinors are then represented by
\beq
\psi \chi= \psi^{\alpha}\chi_{\alpha}= (\psi_1\chi_2-\psi_2\chi_1)=- \psi_{\alpha}\chi^{\alpha}
= \chi_{\alpha}\psi^{\alpha}=  \chi \psi  \\
\mybar \psi \mybar \chi = \mybar \psi_{\dot \alpha} \mybar \chi^{\dot \alpha}
= (\mybar\psi_{\dot 2} \mybar\chi_{\dot 1} - \mybar\psi_{\dot 1}
\mybar \chi_{\dot 2})=  - \mybar \chi^{\dot \alpha}  \mybar \psi_{\dot \alpha}
=  \mybar \chi_{\dot \alpha}  \mybar \psi^{\dot \alpha} =  
\mybar \chi   \mybar \psi 
\eeq

The 4-vector representation is the $(\half,\half)$ which
can be represented as a matrix with one dotted and one undotted
index. The invariant tensors are 
\beq
(\sigma_a)_{\alpha\dot\beta} = (1,i\vec\sigma)_{\alpha\dot\beta}\ ,
\qquad
(\mybar\sigma_a)^{\dot \alpha \beta}  = (1,-i\vec\sigma)^{\dot \alpha
  \beta} \ ,
\eeq
where the vector index $a=0,\ldots,3$ are never raised in Euclidean
space.  These matrices satisfy the relations
\beq
(\mybar \sigma_a)^{\dot \alpha \alpha} 
= \epsilon^{\dot \alpha \dot \beta} \epsilon^{ \alpha \beta}
 (\sigma_a)_{\beta \dot \beta}\ , \qquad  
\Tr 
 \sigma_a \mybar \sigma_b= 2 \delta_{ab},
 \qquad   (\sigma_a)^{\alpha \dot \alpha}   
(\mybar \sigma_a)_{\dot \beta \beta}= 2 \delta_{\alpha}^{\ \beta} 
  \delta_{\dot \alpha}^{\ \dot \beta}
\eeq

The $SU(2)_L$ and $SU(2)_R$ generators respectively are given by
\beq
(\sigma_{ab})_\alpha^{\ \ \beta} = \frac{i}{4} \left(\sigma_a\mybar\sigma_b -
  \sigma_b \mybar\sigma_a\right)_\alpha^{\ \ \beta}\ ,\qquad
(\mybar\sigma_{ab})^{\dot\alpha}_{\ \ \dot\beta} = \frac{i}{4} \left(\mybar\sigma_a \sigma_b -
  \mybar\sigma_a \sigma_b\right)^{\dot\alpha}_{\ \ \dot\beta}\ .
\eeq

\section{Fermion decomposition for the $d=2$ and $d=1$ lattices}
\label{sec:ap4}

 To be
explicit on how we define the $SO(10)\supset SO(4)\times SO(6)$
decomposition used in the $d=2$ lattice construction (whose structure
is inherited by the $d=1$ lattice as well), we define the $\gamma$ matrices of the global $SO(4)$
symmetry in terms of the $SO(10)$ $\Gamma_\alpha$ matrices of the
mother theory to be
\beq
\gamma_0 = -\Gamma_4\ ,\quad \gamma_1=\Gamma_{10}\ ,\quad
\gamma_2=\Gamma_5\ ,\quad \gamma_3=\Gamma_9\ ,
\eeq
and the $SO(4)$ generators
\beq
J_{\alpha\beta} = \frac{1}{4i}\left[\gamma_\alpha,\gamma_\beta\right]\
.
\eeq
Then $SO(4)\simeq SU(2)_L\times SU(2)_R$ is defined by the $SU(2)$
generators $L_i$ and $R_i$, $i=1,2,3$ defined by
\beq
L_i = \half\left(\half \epsilon_{ijk} J_{jk} + J_{0i}\right)\ , R_i =
\half\left(\half \epsilon_{ijk} J_{jk} - J_{0i}\right) \ ,
\eeq
which satisfy the $SU(2)_L\times SU(2)_R$ commutation relations
\beq
[L_i,L_j]=i\epsilon_{ijk}L_k\ ,\quad
[R_i,R_j]=i\epsilon_{ijk}R_k\ ,\quad
[L_i,R_j]=0\ .
\eeq

With these conventions, it is possible then to relate 
 we can express them
in terms of the $\lambda$, $\xi_{mn}$ and $\psi^m$ variables defined
for the $d=4$ lattice in \eq{omexp}.  The $SU(3)$ triplet fermions are
\beq
\psi^m_\alpha = (3,2,1) = 
\begin{pmatrix}
\psi^m \cr
\half \epsilon^{mnp}\xi_{np}
\end{pmatrix}_\alpha\ ,\quad
\mybar\psi_m^{\dot\alpha} = (\mybar 3,1,2)=
\begin{pmatrix}
\ \xi_{m5}\cr
-\xi_{m4}
\end{pmatrix}^{\dot\alpha}\ ,\quad m=1,2,3\ ,
\eqn{fsub1}\eeq
where the $\psi$ variables on the left side of the equation are those
used in the $d=1,2$ lattices, while those on the right are the
variables of \eq{omexp}.  Similarly, the $SU(3)$ singlet fermions are
given by
\beq
\lambda_\alpha = (1,2,1) = 
\begin{pmatrix}
\ \xi_{45}\cr
-\lambda
\end{pmatrix}_\alpha\ ,\qquad
\mybar\lambda^{\dot\alpha} = (1,1,2) =
\begin{pmatrix}
\psi^4 \cr
\psi^5
\end{pmatrix}^{\dot\alpha}\ .
\eqn{fsub2}\eeq
Thus, for example, in the particular basis of \S\ref{sec:ap}, the ${\bf
  16}$ of the mother theory is given by the spinor in \eq{omexpap},
followed by the above substitutions to express it in terms of
variables appropriate to the $d=2,1$ lattices.

The decomposition of the bosons is simpler, with
\beq
z^m=(3,1,1) \ ,\quad \mybar z_m=(\mybar 3,1,1)\ ,
\eeq
and
\beq
 (\tilde v_a
\sigma_a)_{\alpha\dot\beta} =(1,2,2)=i\sqrt{2} \,\begin{pmatrix}
\mybar z_4 & \ \mybar z_5 \cr
z_5 & - z_4
\end{pmatrix} 
=
\begin{pmatrix}
 \ \ (v_4+i v_9) &  (v_5+i v_{10}) \cr
(- v_5+i v_{10}) & ( v_4-i v_9)
\end{pmatrix}\ ,
\eqn{bsub}\eeq
with $(\psi^m)^\alpha (\tilde v_a \sigma_a)_{\alpha\dot\beta} \mybar
\psi_m^{\dot\beta}$ being an $SU(3)\times SU(2)_L\times SU(2)_R$
singlet.

\bibliography{latticeSUSY3}
\bibliographystyle{JHEP} 
\end{document}